\documentclass[final,5p,times,twocolumn]{elsarticle}

\usepackage{graphicx}
\usepackage{amssymb}
\usepackage{amsthm}
\usepackage{amsmath}
\usepackage{bm}
\usepackage{multirow}
\usepackage{color}
\usepackage{ upgreek }

\def\etal{{\it et~al.\/}}
\def\ie{{\it i.~e.\/}}
\def\MX{$\mathrm{MX_2}$}
\def\MoS{$\mathrm{MoS_2}$}

\def\WS{$\mathrm{WS_2}$}
\def\WSe{$\mathrm{WSe_2}$}
\newcommand{\kvec}{\mathbf{k}}
\newcommand{\abinitio}{\textit{ab initio}}

\journal{Surface Science Reports}

\begin{document}

\begin{frontmatter}



\title{Vibrational and optical properties of MoS$_2$: from monolayer to bulk}


\author[label1]{Alejandro Molina-S\'{a}nchez}
\author[label2]{Kerstin Hummer}
\author[label1]{Ludger Wirtz}
\address[label1]{Physics and Materials Science Research Unit, University of Luxembourg, 162a avenue de la Fa\"iencerie, L-1511 Luxembourg, Luxembourg}
\address[label2]{University of Vienna, Faculty of Physics, Department of Computational Materials Physics, Sensengasse 8/12, 1090 Vienna, Austria}

\begin{abstract}
Molybdenum disulfide, MoS$_2$, has recently gained considerable attention as a layered material where neighboring layers are only weakly
interacting and can easily slide against each other. Therefore, mechanical exfoliation allows the fabrication of single and multi-layers and opens
the possibility to generate atomically thin crystals with outstanding properties. In contrast to graphene, it has an optical gap of 1.9 eV. This
makes it a prominent candidate for transistor and opto-electronic applications.
Single-layer MoS$_2$ exhibits remarkably different physical
properties compared to bulk MoS$_2$ due to the absence of interlayer hybridization.
For instance, while the band gap of bulk and multi-layer MoS$_2$
is indirect, it becomes direct with decreasing number of layers.
In this review, we analyze from a theoretical point of view the electronic, optical, and vibrational properties of single-layer, few-layer and bulk
MoS$_2$. In particular, we focus on the effects of spin–orbit interaction, number of layers, and applied tensile strain on the vibrational and optical
properties. We examine the results obtained by different methodologies, mainly ab initio approaches. We also discuss which approximations are
suitable for MoS$_2$ and layered materials. The effect of external strain on the band gap of single-layer MoS$_2$ and the crossover from indirect to
direct band gap is investigated. We analyze the excitonic effects on the absorption spectra. The main features, such as the double peak at the
absorption threshold and the high-energy exciton are presented. Furthermore, we report on the phonon dispersion relations of single-layer,
few-layer and bulk MoS$_2$. Based on the latter, we explain the behavior of the
Raman-active $A_{1g}$ and $E^1_{2g}$ modes as a function of the number of
layers. Finally, we compare theoretical and experimental results of Raman, photoluminescence, and optical-absorption spectroscopy.
\end{abstract}




\end{frontmatter}

\section{Introduction}
\label{intro}

For many layered materials, it has been established that the few-layer or 
mono-layer phases have distinct properties with respect to their bulk 
counterparts. Often these properties are even changing between
the mono-, bi- and, tri-layer phases. 
Within the layers, the atoms are held together by strong covalent bonds
while the inter-layer bonds are rather weak and mostly due to 
van der Waals interaction. As a consequence, the layers can easily be separated
by mechanical exfoliation and single, quasi two-dimensional (2D), and 
few-layer systems of various
materials can easily be produced.\cite{Novoselov2005}
Some examples are graphene, hexagonal boron nitride (BN), semiconducting 
transition metal dichalcogenides \MX~ (M = Mo, W, Ta, and X = S, Se, Te), \cite{Wilson1969} the superconducting metal $\mathrm{NbSe_2}$, or the elemental 2D systems silicene, germanene, and phosphorene \cite{Jiang2015}. 

Many of these materials have potential for novel technological functionalities. 
Graphene is the most prominent single-layer material \cite{Novoselov2004}.
It does not only have outstanding physical properties such as high 
conductivity, flexibility, and hardness \cite{Geim2009}, but it is also a 
benchmark for fundamental physics. E.g., it displays an anomalous 
half-integer Quantum 
Hall effect due to the quasi-relativistic behavior (linear crossing in the
band-structure) of the $\pi$-electrons\cite{Katsnelson2006,Katsnelson2012}.
The fascinating properties of graphene have paved the way for intense investigations of alternative layered materials.\cite{Novoselov2005}

Electronics and optical applications often require materials with a sizeable 
band gap. For instance, the channel material in field-effect transistors must 
have a sufficient band gap to achieve high on/off ratios \cite{Lembke2015}. 
In this respect, the semiconducting transition metal 
dichalcogenides (TMDs) can complement or substitute the zero-band gap material
graphene\cite{RadisavljevicB.2011}. 
Single-layer \MoS~ is an appealing alternative for opto-electronic applications
with an optical gap of 1.8-1.9 eV, high quantum efficiency\cite{Mak2010,Splendiani2010}, an acceptable value for the electron 
mobility\cite{Lembke2012,Baugher2013},
and a low power of dissipation\cite{Radisavljevic2011}. It has potential application in nanoscale transitors \cite{RadisavljevicB.2011,Zhang2012,Lembke2012,Lembke2015}, photodetectors \cite{Lopez-Sanchez2013,Zhang2015}, and photovoltaics applications \cite{Fontana2013,Furchi2014,Pospischil2015}. Other TMDs such as single-layer \WS~also exhibit
high photoluminescence yield \cite{Gutierrez2013}. 

In this stimulating scenario, TMDs are being intensively investigated. 
Fabrication techniques such as the mechanical exfoliation \cite{Li2012c,Castellanos-Gomez2012b} and the liquid exfoliation \cite{Coleman2011}
produce single- and multi-layer crystals with high crystalline quality at 
low cost. This has increased notably
the amount of research groups working in both fundamental and applied aspects of
TMDs.
Concerning the electrical and optical properties of single-layer, multi-layer and bulk \MoS, extensive experimental 
investigations have been carried out within the last few years. 
The most important techniques are
photoluminescence, optical absorption, and electroluminescence spectroscopy
\cite{Mak2010,Splendiani2010,Korn2011,Sundaram2013}. It is widely accepted that single-layer \MoS~ has a 
direct band gap that transforms into an indirect gap with increasing number of layers. Similarly, bandgap 
engineering is possible by applying strain. The application of strain drives a 
direct-to-indirect band gap transition in single-layer \MoS~\cite{Scalise2012,Scalise2014,He2013,Conley2013,Dong2014,Guzman2014}. Moreover, suitable hydrostatic pressure reduces
the band gap of single- and multi-layer \MoS~resulting in a phase transition from semiconductor 
to metal \cite{Nayak2014,Nayak2015}. The group symmetry and the spin-orbit interaction in \MoS~also raise
interesting properties. The control of 
the valley polarization of the photo-generated electron-hole pairs paves the way for using \MoS~ in applications
related to next-generation  spin- and valleytronics \cite{Xiao2012,Mak2012,Zeng2012,Cao2012,Kumar2014}.
Further studies dealing with charged exciton complexes (trions)\cite{Mak2013,Plechinger2015} or with second harmonic generation have also been published \cite{Kumar2013,Malard2013}.

Many challenges remain to be solved in the field of TMDs. The problem of obtaining
high hole mobility in single-layer \MoS~ hinders the realization of p-n diodes. A proposed solution is using a
monolayer \WSe~ diode, in which the p-n junction is created
electrostatically by means of two independent 
gate voltages \cite{Pospischil2015,Baugher2015,Ross2015}. Another active research field is the 
design of Van der Waals heterostructures. Assembling atomically thin layers of
distinct 2D materials allows to enrich the physical properties \cite{geim2013}. Techniques like
chemical vapor deposition and wet chemical approaches are triggering
the fabrication of heterostructures \cite{Huang2014,Gong2014}. For example, flexible photovoltaic devices of TMDs/graphene layers exhibit
quantum efficiency above 30\% \cite{Britnell2013}. A photovoltaic effect has also been achieved using a 
\MoS$/$\WSe ~p-n heterojunction \cite{Furchi2014}. The different stacking 
configurations and the band alignments are important aspects 
in bilayer heterostructures\cite{he:prb:89,debbichi:prb:89,liu:natcommun:5}.

Another important activity in the field of TMDs is the characterization of the vibrational properties of \MX. Earlier studies of bulk \MoS~ using Raman and infrared spectroscopy \cite{Verble1970,Wieting1971} as well as Neutron scattering\cite{Wakabayashi1975} 
and electron-energy-loss spectroscopy\cite{Bertrand1991}
had already well characterized the phonons at $\Gamma$ and the phonon 
dispersion. In the recent years, a large number of Raman studies on mono- and
few-layer systems has emerged
\cite{Lee2010,Windom2011,Li2012,Plechinger2012,Boukhicha2013,Luo2013,Zhao2013,Zhang2013,ZhangX2015}. 
The Raman frequencies are correlated with the number of
layers which allows their unequivocal identification. 
The trend of the Raman modes 
$E_{2g}$ (in-plane mode) and $A_{1g}$ (out-of-plane mode) with the number of layers has been intensively discussed, both
theoretically \cite{Ataca2011,Molina-Sanchez2011,Luo2013,Terrones2014} and experimentally \cite{Lee2010,Rice2013,Terrones2014}. 
The $A_{1g}$ mode follows a predictable behavior. Its frequency grows with increasing number of layers, due to the interlayer interaction. 
The $E^1_{2g}$ mode shows the opposite trend, \ie, decreasing in frequency for an increasing number of layers.


The experimental findings are accompanied by a vast theoretical literature. The characteristic stacking
of ultra-thin layers of \MoS~ adds new challenges to the theoretical approaches. For example, the layer thickness 
influences the dielectric constant which becomes 
strongly anisotropic. This enhances the Coulomb interaction between carriers. The calculation of the
excitations has to include these dimensional effect for applying accurately the GW method and
the Bethe-Salpeter equation. For instance, one consequence is an
exciton binding energy in some layered materials of hundreds of meV,
much larger than in bulk semiconductors. 
Also, new models have been developed to explain the
interplay of the spin-orbit interaction and the crystal symmetry (which is layer dependent), and
its consequences, like the valley-Hall effect. Moreover, the interlayer 
interaction has demanded the improvement of the modelling of the van der Waals interaction in 
extended systems. The precision of the Raman spectroscopy has allowed to evaluate the accuracy
of \textit{ab initio} methods for calculating phonons, and it proves how useful is
the simulation of the vibrational properties for understanding the interlayer interaction and the
chemical bonding. Therefore, the research on layered materials
has contributed to the appeareance of new methods and to the reformulation of existing ones.
In this review, we give an overview of the challenges in the modelling of
the spectroscopic properties of \MoS~ and the solutions proposed. 
The discussion of the literature results is complemented by additional calculations. We believe
the topics discussed here will be also useful in the modelling and understanding of other
two-dimensional materials.

\section{Structural properties}
\label{structure}

\begin{figure}
\includegraphics[width=7.6 cm]{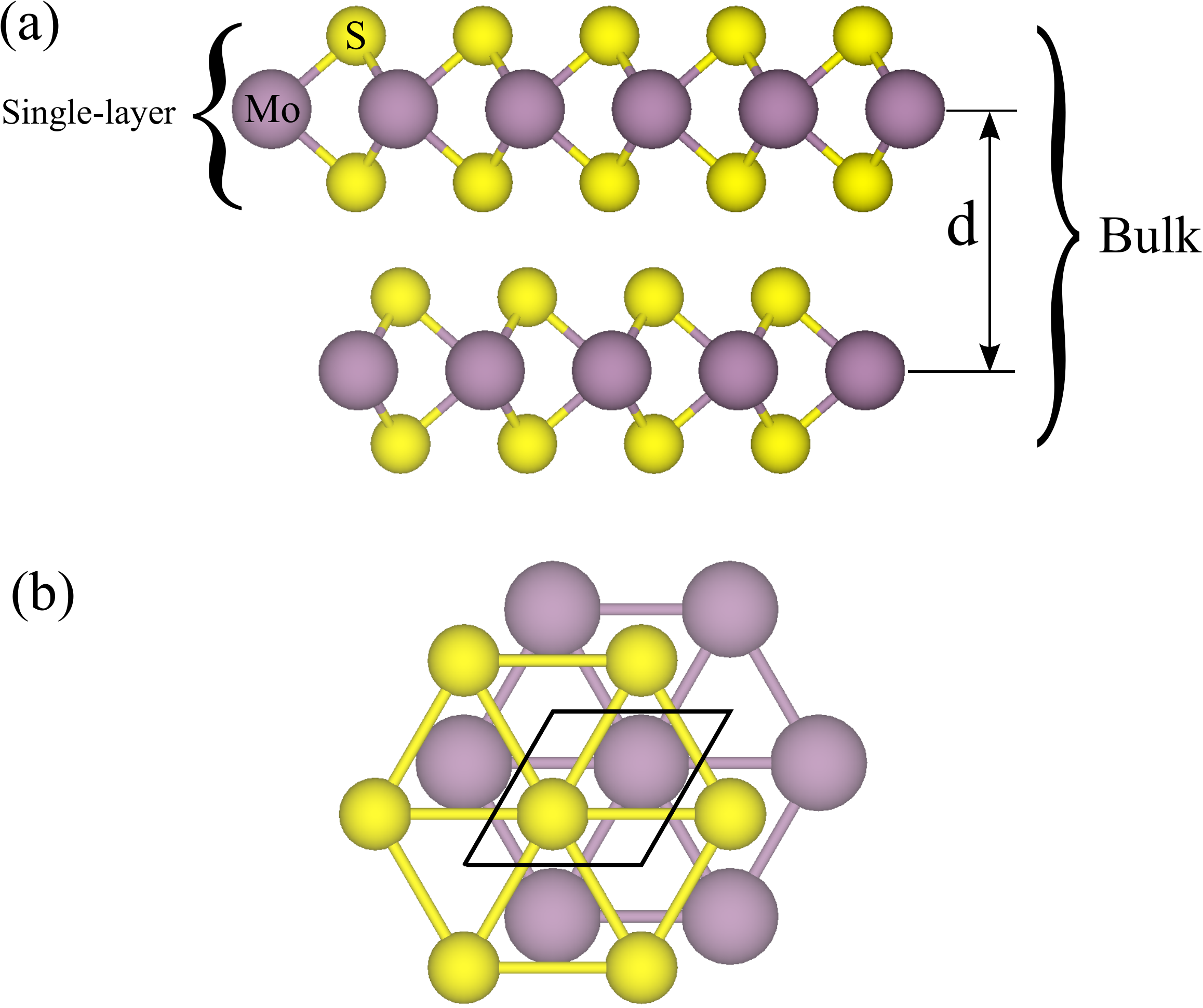}
\caption{(Color online) (a) \MoS bulk ~and single-layer.
The interlayer distance is denoted by $d$ (distance between Mo atoms
of different layers). (b) Top view
of the \MoS ~single-layer unit cell.}
\label{supercell} 
\end{figure}

Bulk molybdenum disulfide (\MoS) belongs to the class of transition metal 
dichalcogenides (TMDs) that crystallize in the characteristic 2H polytype.
The corresponding Bravais lattice is hexagonal and the space
group of the crystal is $P6_3/mmc$ ($D_{6h}$ non-symmorphic group). 
The unit cell is characterized by the lattice parameters $a$ (in-plane
lattice constant) and $c$ (out-of-plane lattice constant). 
The basis vectors are
\begin{equation}
\begin{array}{ccc}
 \bm{a}_1 & = & (\frac{1}{2}a, -\frac{\sqrt{3}}{2}a, 0),  \\
 \bm{a}_2 & = & (\frac{1}{2}a,  \frac{\sqrt{3}}{2}a, 0),  \\
 \bm{a}_3 & = & (0, 0, c). \\
\end{array}
\label{eq:basisvec}
\end{equation}
The unit cell contains 6 atoms,
two Mo atoms are located at the Wyckoff $2c$ sites and four S atoms at 
the Wyckoff $4f$ sites. With the internal parameter $z$, the positions,
expressed in fractional coordinates, are $\pm$(1/3, 2/3,1/4) for the Mo atoms
and $\pm$(2/3, 1/3,$z$) as well as $\pm$(2/3, 1/3,1/2-$z$) for the S atoms.

The single-layer contains one Mo and two S atoms. In this case, 
the inversion symmetry is broken and the space group (more precisely, layer group) is $P\bar{6}m2$ ($D^1_{3h}$ symmorphic group). The double-layer is constructed by adding another S-Mo-S layer, having now the layer group $P\bar{3}m1$ ($D^3_{3d}$ symmorphic group). Consequently, an odd number of layers has the 
same symmetry as the single-layer (absence of inversion symmetry),
whereas an even number has the symmetry 
of a double-layer (with inversion symmetry).

\begin{table*}
\begin{center}
\caption{Bulk 
\MoS~ experimental lattice parameters $a$, $c$, internal parameter $z$, and bulk modulus $B$.}
\begin{tabular}{lccccc}
\hline
\hline
   & $a$ (\AA) &  $c$ (\AA) & $z$   &  $c/a$  & $B$ (GPa)\\
\hline
Ref. \cite{dickinson:jacs:45}  & 3.160 & 12.294  & 0.621  & 3.890        &  \\
Ref. \cite{schoenfeld:acb:39} & 3.161 & 12.295  & 0.627(5)  & 3.890 &  \\
Ref. \cite{alhilli:jcg:15}       & 3.140 & 12.327  &        & 3.926 & 53.4$\pm$1.0 \\
Ref. \cite{petkov:prb:65}      & 3.168(1) & 12.322(1)  & 0.625  & 3.890  & \\
\hline
\hline
\label{tab:lattice}
\end{tabular}
\end{center}
\end{table*}

The crystal structure of \MoS can be specified as a stacking of
quasi-two-dimensional (2D) S-Mo-S layers along the 
$c$ direction. Within each layer, Mo atoms are surrounded by 6 S atoms in a
trigonal prismatic geometry as illustrated in Fig. \ref{supercell}.  
The bonding type is predominantly covalent within the atomically thin 
S-Mo-S layers,
whereas the layers themselves are weakly bound by Van der Waals (VdW) 
forces in the crystal. The inherent weakness of the 
interlayer interactions can result in different stacking sequences and therefore in different 
polytypisms as shown in Ref. \cite{he:submitted}.

Defining the optimized geometry is the first step for any calculation of the phonon spectra and/or the band structure. 
Most of the previous investigations used density-functional theory (DFT) on the level of the local-density approximation (LDA) or the generalized-gradient approximation (GGA) \cite{Ataca2011}.
We want to emphasize that in DFT the accuracy of the 
calculated quantities is determined by the treatment of
the exchange correlation (XC) energy given by the XC functional.
However, the standard local (LDA) and semilocal (GGA) XC functionals do not
account  for the long-range van der
Waals interactions, which are responsible for the stable stacking of the layers
and thus particularly relevant in two-dimensional 
materials. Nevertheless, the well-known LDA overestimates the (weak) covalent 
part of the interlayer bonding and compensates thus the missing vdW forces 
yielding a bound ground state for most layered materials. This explains
the success of LDA in obtaining the geometry of many
layered materials such as
graphite \cite{Wirtz2004}, boron nitride \cite{Kern1999,Serrano2007} or 
graphene on different substrates \cite{Allard2010,Fromm2013,Endlich2013}.
The good performance of LDA in layered materials (although fortuitous)
has made this approximation widely used in the calculations 
of structural properties.

In the present work, the calculations were partly performed with the Vienna \abinitio ~simulation package (VASP)\cite{kresse:cms:6,kresse:prb:54}
utilizing the projector augmented plane wave (PAW) method \cite{bloechl:prb:50,kresse:prb:59} to describe the core-valence interaction. 
PAW potentials with non-local projectors for the molybdenum (Mo) 4$s$, 4$p$, 4$d$, 5$s$ as well as sulfur (S) 3$s$, and 3$p$ valence states were generated
to minimize errors arising from the frozen core approximation. The valence electrons were treated by
a scalar-relativistic Hamiltonian and spin orbit coupling (SOC) was self-consistently included in all VASP calculations as described elsewhere \cite{kim:prb:80}.
VASP uses DFT with a variety of XC functionals ranging from LDA to different types of GGAs, to
hybrid functionals, and VdW density functionals. Furthermore, VASP has an 
implementation of many body perturbation theory such as the $GW$ approximation 
ranging from the single-shot $G_0W_0$ \cite{shishkin:prb:74} to a selfconsistent $GW$ (sc$GW$) approximation\cite{shishkin:prb:75,shishkin:prl:99}. 
Concerning standard DFT results presented in this work, the XC energy was treated
within the LDA\cite{LDA:prl:45} and the GGA. 
For the latter, the parametrization of Perdew, Burke, 
and Ernzerhof (PBE), in particular the PBEsol 
functional \cite{PBEsol:prl:100} was used. 

In order to improve the theoretical lattice parameters
calculated within DFT-LDA/GGA\cite{Ataca2011},
we have also studied the structural properties including VdW interactions 
starting
from the experimentally observed structural parameters summarized in Tab.~\ref{tab:lattice}.
For this purpose we used the
optB86b-VdW functional, recently implemented in VASP 
\cite{klimes:prb:83}. The optB86b functional is a non-local
correlation functional that approximately accounts for dispersion 
interactions. It is based on the VdW-DF proposed by Dion \etal \cite{dion:prl:92}, but employs an accurate exchange functional particularly optimized for the correlation part. \cite{klimes:prb:83} 

The structural optimization of hexagonal bulk \MoS ~requires 
a 2D energy minimization, since the 
ground state energy depends on two degrees of
freedom, \ie, the volume and the $c/a$ ratio. The experimentally observed 
structural parameters summarized in Tab. \ref{tab:lattice} 
have been used as starting point for the
calculation of the electronic properties of bulk as well 
as single layer (1L) \MoS ~within DFT and methods that go
beyond as described in detail below. This minimization was performed manually using LDA 
and PBEsol as well as the optB86b-VdW functional,
for comparison by varying the unit cell volume of bulk \MoS ~within $\pm 10\%$ of 
the experimentally observed equilibrium volume ($V_0$).
For each chosen volume the $c/a$ ratio was first optimized by fitting 
the energy dependence on $c/a$ to the Murnaghan equation of state (EOS). \cite{murnaghan:pnas:30}
The final DFT-optimized unit cell volume was obtained 
by subsequently fitting $E(V)$ to the Murnaghan EOS.
Note that in each single optimization step the atomic positions were also
relaxed by minimizing the total forces on the atoms until they
were converged to 0.05 eV/\AA.

In order to avoid effects from the changes in size of the basis set due to changes in the unit cell volume $V$,
the kinetic energy cutoff $E_{cut}$ has been increased to 350 eV. Convergence with respect to $\kvec$ sampling within the Brillouin zone was 
reached with $16\times16\times16$ $\Gamma$-centered meshes in case of optB86b-VdW and with $12\times12\times12$ $\Gamma$-centered meshes 
for LDA and PBEsol. The manually performed structural optimization was cross
checked with VASP calculations employing 
minimization algorithms parallel for the atomic positions and the $c/a$ ratio for selected volumes
in the range $\pm 5\%$ of $V_0$ and one subsequent Murnaghan EOS fit. From these calculations
the Bulk modulus is obtained from
\begin{equation}
B = \left. -\frac{1}{V}\frac{\partial^2E}{\partial V^2}\right|_{V=V_0}.
\end{equation}

\begin{figure}
\includegraphics[width=7.6 cm]{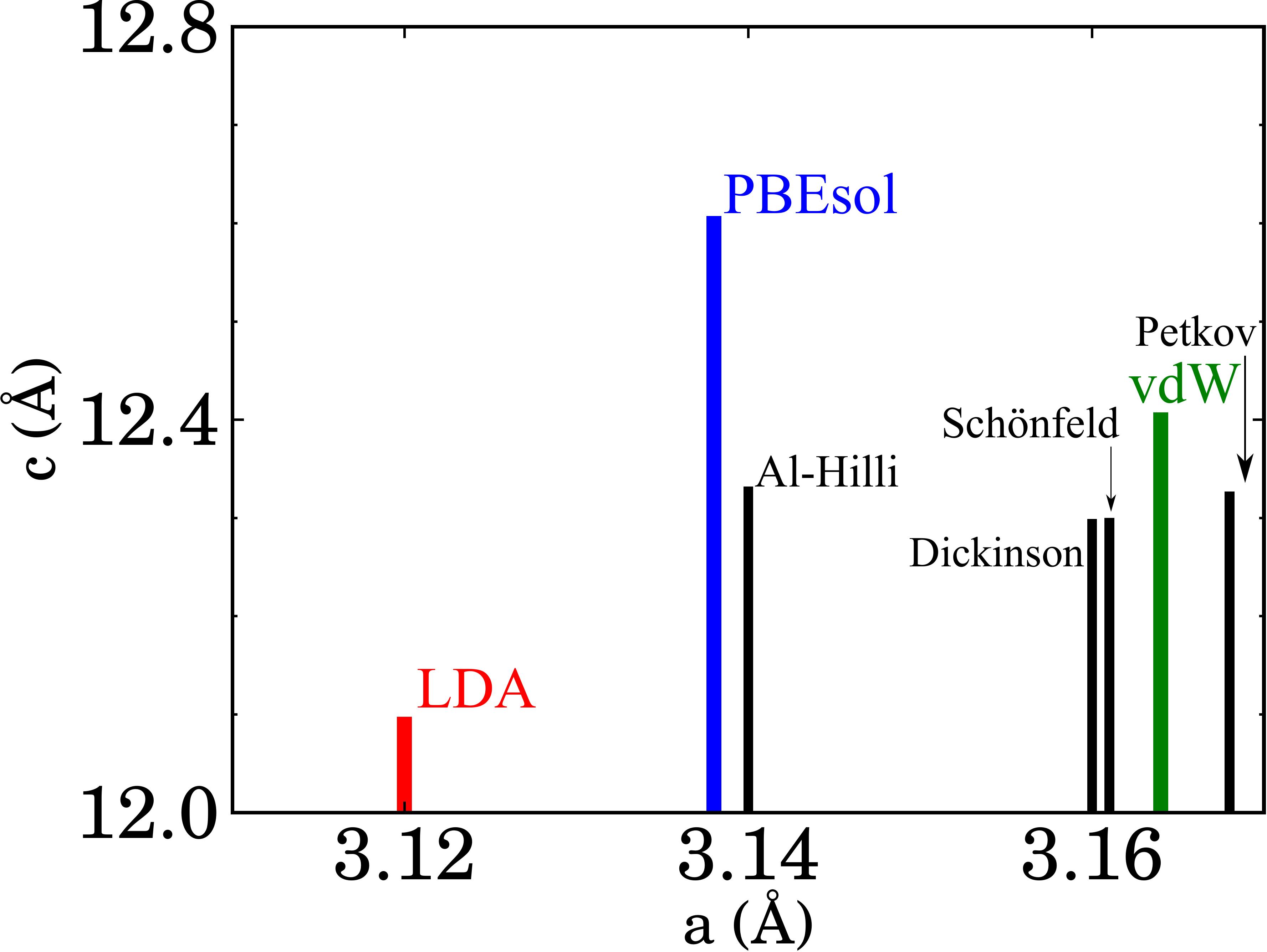}
\caption{Comparison of the theoretical $a$ and $c$ lattice parameters, 
obtained by LDA, PBEsol and vdW-DF,
with the experimental values of Refs. \cite{dickinson:jacs:45,
schoenfeld:acb:39, alhilli:jcg:15, petkov:prb:65}.}
\label{lattice-comparison} 
\end{figure}

In Table \ref{tab:structopt} the results of the structure optimization corresponding to different functionals
are summarized.
\begin{table*}
\begin{center}
\caption{Structural parameters of bulk \MoS ~obtained by minimizing $E(V,c/a)$ with different XC functionals.
$a$ and $c$ denote the lattice constants, $z$ the internal parameter specifying the atomic positions, $B$ the bulk modulus,
and $d$ the interlayer distance defined according to Fig. \ref{supercell}. The uncertainty
stemming mainly from the EOS fitting in $a$, $c$, and $c/a$ is $\pm$0.001 \AA, $\pm$0.01 \AA, and $\pm$0.01 \AA, respectively.}
\begin{tabular}{cccccccc}
\hline
\hline
        & $a$ (\AA) &  $c$ (\AA) & $z$ &  $c/a$ & $B$ (GPa) & $d$ (\AA) & VdW gap (\AA)\\
\hline
LDA     & 3.120 & 12.09  & 0.1214  & 3.87 & 40-43 & 6.039 &  2.933\\
PBEsol  & 3.138 & 12.60  & 0.1264  & 4.01  & 18-21 & 6.305 & 3.188\\
optB86b-VdW & 3.164 & 12.40 &  0.1236  & 3.92  & 39-40 & 6.203 & 3.068\\
\hline
\hline
\end{tabular}
\end{center}
\label{tab:structopt}
\end{table*}
When using LDA or optB86b-VdW functionals, the theoretical values of $B$ for bulk \MoS ~agree well with the experimental value given in Table \ref{tab:lattice}.
Concerning lattice parameters, we observe a small
underestimation of the in-plane parameter $a$, both in LDA (1.3 \%) and PBEsol (0.7 \%). 
Contrary, the $c$ parameter is underestimated by 1.6\% and overestimated by 2.4\% in LDA
and PBEsol, respectively. Compared to LDA, the PBEsol functional yields
the larger deviation of the resulting $c/a$ ratio. 
The improvement after including the VdW interactions is
substantial, with an error of only 0.09 \% (see Fig. 
\ref{lattice-comparison} for a comparison with the 
experimental results).

In conclusion, van der Waals functionals give the most accurate
results for lattice parameters and the bulk modulus. LDA tends 
to underestimate the interlayer distance and the $c$ parameter, but in average
gives acceptable results and it can be trusted in the 
prediction of structural properties.

Based on the ground state structures summarized in 
 the bulk \MoS ~charge density was 
calculated on a $\Gamma$-centered 12$\times$12$\times$3 
$\kvec$ mesh by converging the total energy to 0.1 meV using 
a kinetic cutoff energy of 350 eV and a Gaussian smearing with a smearing width of 50 meV. Tests with 18$\times$18$\times$3 $\kvec$ point 
grids have shown that the electronic band gaps are converged within 
20 meV compared to the results obtained with 
the 12$\times$12$\times$3 grid. 

The single-layer \MoS ~structure has been constructed from the optimized 
bulk structure (Tab. \ref{tab:structopt}) by selecting only the bottom  S-Mo-S layer and adding
20 \AA ~vacuum along $c$ direction. The atomic positions in the 
slab geometry have again been relaxed (force 
convergence criterion of 0.05 eV/\AA) before
calculating the band structure on $\Gamma$-centered 12$\times$12$\times$1 $\kvec$-point grids. 
Convergence tests of the eigenvalues
as a function of the vacuum space between repeated layers has 
been performed up to an accuracy of 15 meV,  
establishing a convergence distance of 20 \AA. 

As we will demonstrate later, the main 
features of the band structure of \MoS ~critically depend on the 
lattice optimization and even small differences can induce significant
changes. In the case of the phonon band structure, deviations are 
reflected in a rigid shift of the phonon frequencies but the main trends
are less affected than in the case of the electronic band structure.

\section{Lattice dynamics of MoS$_2$}
\label{phonon}

The knowledge of the phonon dispersion in a material is indispensable for the understanding of 
a large number of macroscopic properties such as the heat capacity,
thermal conductivity, (phonon-limited) electric conductivity, etc.
Vibrational spectroscopy (Raman spectroscopy and Infrared absorption spectroscopy)\cite{Wieting1971} give access to the phonons at the Brillouin zone center ($\Gamma$ point). Inelastic neutron scattering cite{Wakabayashi1975} allows to 
measure (almost) the full phonon dispersion.
Precise semi-empirical modeling of the phonon dispersion and {\it ab-initio} calculations in comparison to experimental data are a challenge by itself. However, precise modeling is also required because details in the vibrational spectra may also carry some information about the number of layers and the underlying substrate. For graphene, this has been widely explored: the so-called 2D line in the spectra splits into sub-peaks when going from the single to the multi-layer case\cite{Ferrari2006,Graf2007}. 
Last but not least, the 2D-line also changes position as a function of the underlying substrate\cite{Berciaud2009,Forster2013,Starodub2011,Endlich2013}. All these features are related to the double-resonant nature\cite{Thomsen2000} of Raman scattering in graphene and on the dependence of the highest optical mode on the screening.
For MoS$_2$ (and related semiconducting transition-metal dichalcogenides), the layer-dependence of the vibrational spectra is less spectacular than for graphene. Nevertheless a clear trend in the lower frequency inter-layer shear and breathing modes with increasing layer number can be observed\cite{Plechinger2012,Zhang2013,Boukhicha2013} (similar to graphene \cite{Tan2012}). Also in the high-frequency optical modes at $\Gamma$, a clear trend from single to multi-layer MoS$_2$ has been observed\cite{Lee2010} and reproduced in other dichalcogenides\cite{Berkdemir2013}. 

Interestingly, the behaviour of the phonon frequencies as a function of the number of layers does not always follow 
the intuitive trend. For instance, the frequency of the Raman active in-plane $E^{1}_{2g}$ phonon decreases with the increment of the number of layers. This is contrary to the expectation that the weak interlayer forces should increase the restoring forces and consequently result
in an increase of the frequency. The Raman active $A_{1g}$ does follow the
expected behaviour, increasing the 
frequency with the number of layers. Several attempts have been done to explain this trend \cite{Lee2010,Molina-Sanchez2011,Luo2013}. 
This will be critically reviewed in this section.

\subsection*{General Theory}
Before discussing the phonons of bulk and few-layer MoS$_2$, we briefly review the {\it ab-initio} calculation of phonons. 
(A complete discussion of the theory of phonons can be found, e.g., in Ref. \cite{Bruesch1982}.)
Starting from the equilibrium geometry of the system (see details in Section \ref{structure}), one obtains
the phonon frequencies from the solution of the secular equation 

\begin{equation}
\left| \frac{1}{\sqrt{M_IM_J}}\tilde{C}_{I\alpha,J\beta}(\bm{q}) 
- \omega^2(\bm{q})\right| = 0,
\label{secular}
\end{equation}
where $\bm{q}$ is the phonon wave-vector, and $M_I$ 
and $M_J$ are the atomic masses of 
atoms $I$ and $J$. The dynamical matrix is defined as

\begin{equation}
\tilde{C}_{I\alpha,J\beta}(\bm{q})
=\frac{\partial^2 E}{\partial u^{*\alpha}_I(\bm{q})
\partial u^{\beta}_J(\bm{q})},
\label{sec_der}
\end{equation} 
where $u^{\alpha}_I(\bm{q})$ denotes the displacement 
of atom $I$ in direction $\alpha$. The second
derivative of the energy in Eq.~\ref{sec_der} corresponds 
to the change of the force acting on atom $I$
in direction $\alpha$ with respect 
to a displacement of atom $J$ in 
direction $\beta$ \cite{Bruesch1982}. The elements of the dynamical matrix at a given wave-vector $\bm{q}$ can be obtained from
an {\it ab-initio} total energy calculation with displaced atoms in a correspondingly chosen supercell (that needs to be commensurate with $\bm{q}$).
Another approach consists in the use of density functional perturbation theory (DFPT)\cite{Baroni2001,Gonze1997}, where
atomic displacements are taken as a perturbation potential and the resulting changes in electron density and energy are calculated
self-consistently through a system of Kohn-Sham like equations. Within this approach the phonon frequency
can be obtained for any $\bm{q}$ while using the primitive unit-cell. Another way to obtain the dynamical 
matrix is to use empirical interatomic potentials or force constants. A decent fit of the MoS$_2$ phonon dispersion
from Stillinger-Weber potentials has recently been suggested by Jiang et al.\cite{Jiang2013}. Such a fit allows to study
much larger systems such as nanoribbons, nanotubes or heterostructures. The advantage of DFPT is, however,
the higher accuracy and the automatic inclusion of mid and long-range interactions.
For later reference, we note that real-space force constants can be obtained from the dynamical matrix
by discrete Fourier Transform:
\begin{equation}
C_{I\alpha,J\beta}(\bm{R}) = \frac{1}{N_c} \sum_{\bm{q}} e^{i \bm{q}\cdot\bm{R}} \tilde{C}_{I\alpha,J\beta}(\bm{q}),
\label{realforceconst}
\end{equation}
where $N_c$ is the number of unit-cells (related to the density of the $\bm{q}$-point sampling). The physical meaning
of the $C_{I\alpha,J\beta}(\bm{R})$ is the force acting on atom $I$ in direction $\alpha$ as atom $J$ in a unit cell 
at $\bm{R}$ is displaced in direction $\beta$ and all other atoms in the crystal are kept constant.

It is important to note that for polar systems, the phonons in the long-wavelength limit can couple to macroscopic electric fields.
In mathematical terms, this means that the dynamical matrix in the limit $(\mathbf{q}\rightarrow \bm{0})$
can be written as the sum of the dynamical matrix at zero external field and a ``non-analytic'' part that takes into account the coupling
to the electric field and depends on the direction in which the limit $\mathbf{q}\rightarrow \bm{0}$ is taken:
\begin{equation}
\tilde{C}_{I\alpha,J\beta}(\mathbf{q}\rightarrow \mathbf{0})
=\tilde{C}_{I\alpha,J\beta}(\mathbf{q}=\mathbf{0})+\tilde{C}^{NA}_{I\alpha,J\beta}(\mathbf{q}\rightarrow \mathbf{0}).
\label{srlr}
\end{equation} 
The non-analytic part contains the effect of the long-range Coulomb forces and is responsible for the splitting of some of the
longitudinal optical (LO) and transverse optical (TO) modes. \cite{Baroni2001,Gonze1997}:
Its general form is as follows:
\begin{equation}
\widetilde{C}^{NA}_{I\alpha,J\beta}({\bf q}) = \frac{4\pi}{\Omega}
e^2 \frac{({\bf q} \cdot {\bf Z^*}_I)_\alpha 
({\bf q} \cdot {\bf Z^*}_J)_\beta}{\bf q \cdot \epsilon \cdot q},
\label{nonana}
\end{equation}
where $\Omega$ is the volume of the unit cell, ${\bf Z^*}_s$ stands for the Born effective charge tensor of atom $s$ and $\epsilon$ is the dielectric tensor.
Since the dielectric tensor is fairly large in bulk MoS$_2$ ($\epsilon_{xx}=\epsilon_{yy}=15.4, \epsilon_{zz}=7.43$, the effect of LO-TO splitting is visible, but not very pronounced ($<$ 2.6 cm$^-1$).

We will discuss in the following first the phonon dispersions of bulk and single-layer MoS$_2$. Afterwards, we will discuss in detail the symmetry of bulk
and single-layer phonons at $\Gamma$. We will single our the Raman and infrared (IR) active modes and discuss how their frequencies evolve with the number of layers.
\subsection*{Phonon dispersion}

\begin{figure*}
\begin{center}
\includegraphics[width=16 cm]{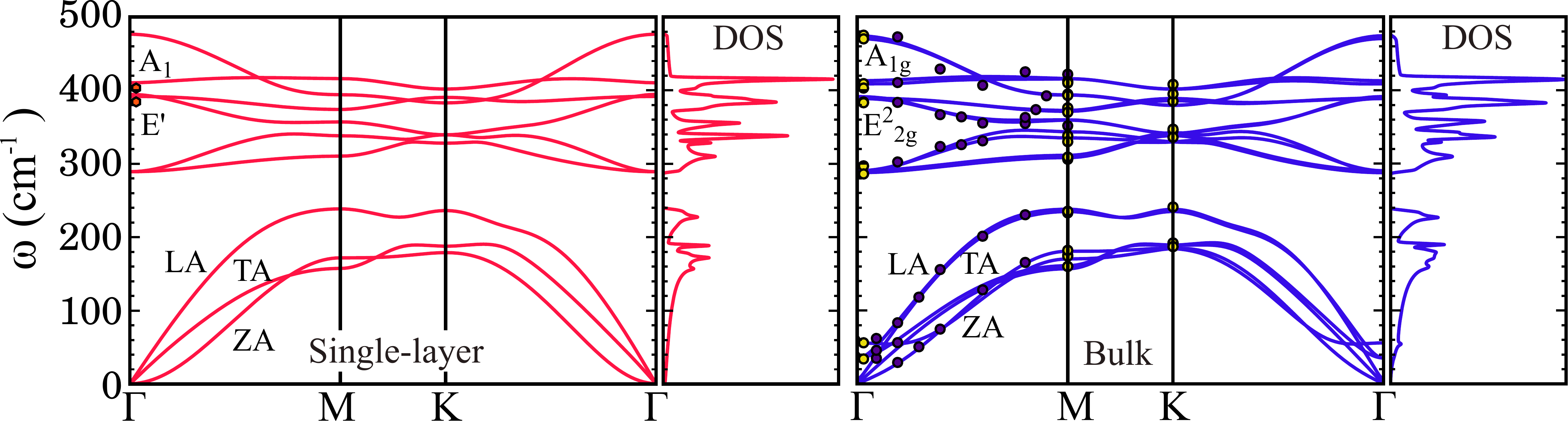}
\caption{Phonon dispersion curves and density of states of (a) single-layer and
(b) bulk \MoS . The symbols denote experimental data. Blue circles: neutron scattering \cite{Wakabayashi1975},
yellow circles: first and higher-order Raman scattering of bulk MoS$_2$ \cite{Livneh2010,Livneh2014},
red dots Raman scattering of single-layer MoS$_2$ \cite{Lee2010}.}
\label{phonon-sl-bulk} 
\end{center}
\end{figure*}

The phonon dispersions of single-layer and bulk MoS$_2$ are shown in Figure~\ref{phonon-sl-bulk}.
Overall, the single-layer and bulk
phonon dispersions have a remarkable resemblance. In the bulk, all single-layer
modes are split into two branches but since the inter-layer interaction
is weak, the splitting is very low (similar to the situation in graphite
and graphene.\cite{Wirtz2004} The only notable exception from this is the
splitting of the acoustic modes of bulk MoS$_2$ around $\Gamma$. 

We have also depicted the experimental data obtained with neutron inelastic scattering spectroscopy for 
bulk \MoS \cite{Wakabayashi1975} as well as the result of IR absorption and Raman scattering at $\Gamma$.
The overall agreement between theory and experiment is rather good, even for the
inter-layer modes. This confirms our expectation that the LDA describes reasonably well
the inter-layer interaction (even though not describing the proper physics
of the inter-layer forces, as discussed in Section \ref{structure}).

The bulk phonon dispersion has three acoustic modes. Those that
vibrate in-plane (longitudinal acoustic, LA, and transverse acoustic, TA) 
have a linear dispersion and higher energy than the out-of-plane acoustic (ZA) 
mode. The latter displays a $q^2$-dependence analogously 
to that of the ZA mode in graphene (which is a consequence of the
point-group symmetry \cite{Saito1998}). The low frequency optical modes
are found at 35.2 and 57.7 cm$^{-1}$ and correspond to rigid-layer shear mode
and layer-breathing mode (LBM) respectively (see left panels of Fig.~\ref{bulk-modes})
in analogy with the low frequency optical modes
in graphite\cite{Wirtz2004}). 
It is worth to mention the absence
of degeneracies at the high symmetry points $\bf{M}$ and $\bf{K}$ and the two
crossings of the LA and TA branches just before and after the $\bf{M}$ point. 
The high frequency 
optical modes are separated from the low frequency modes by a gap of 49 cm$^{-1}$.

The single-layer phonon dispersion is very similar to the bulk one. 
The number of phonon branches is reduced to nine. At low frequencies,
the shear mode and layer-breathing mode are absent.
At higher energies, very little difference between bulk and single-layer
dispersion can be seen. This is due to the fact that the inter-layer interaction is very weak. 
The subtle splitting and frequency shifts of zoner-center modes in gerade and ungerade modes 
(as going from single layer to the bulk) will be discussed below.

The densities of states (DOS) of single-layer and bulk are represented 
in the right panels of Fig.~\ref{phonon-sl-bulk}. 
The differences between single-layer and bulk DOS are minimal, 
except a little shoulder 
around 60 cm$^{-1}$ in the bulk DOS due to the low frequency optical modes. 
In both cases the highest
peaks are located close to the Raman active modes $E_{2g}^1$ and $A_{1g}$ and they are
due to the flatness of the bands around $\Gamma$.

\begin{figure}
\begin{center}
\caption{Raman spectrum from Ref. \cite{Windom2011}, recorded using
623.8 nm excitation (red dots). One-phonon density of states (blue lines) and two-phonon
density of states (green lines).}
\includegraphics[width=7.6 cm]{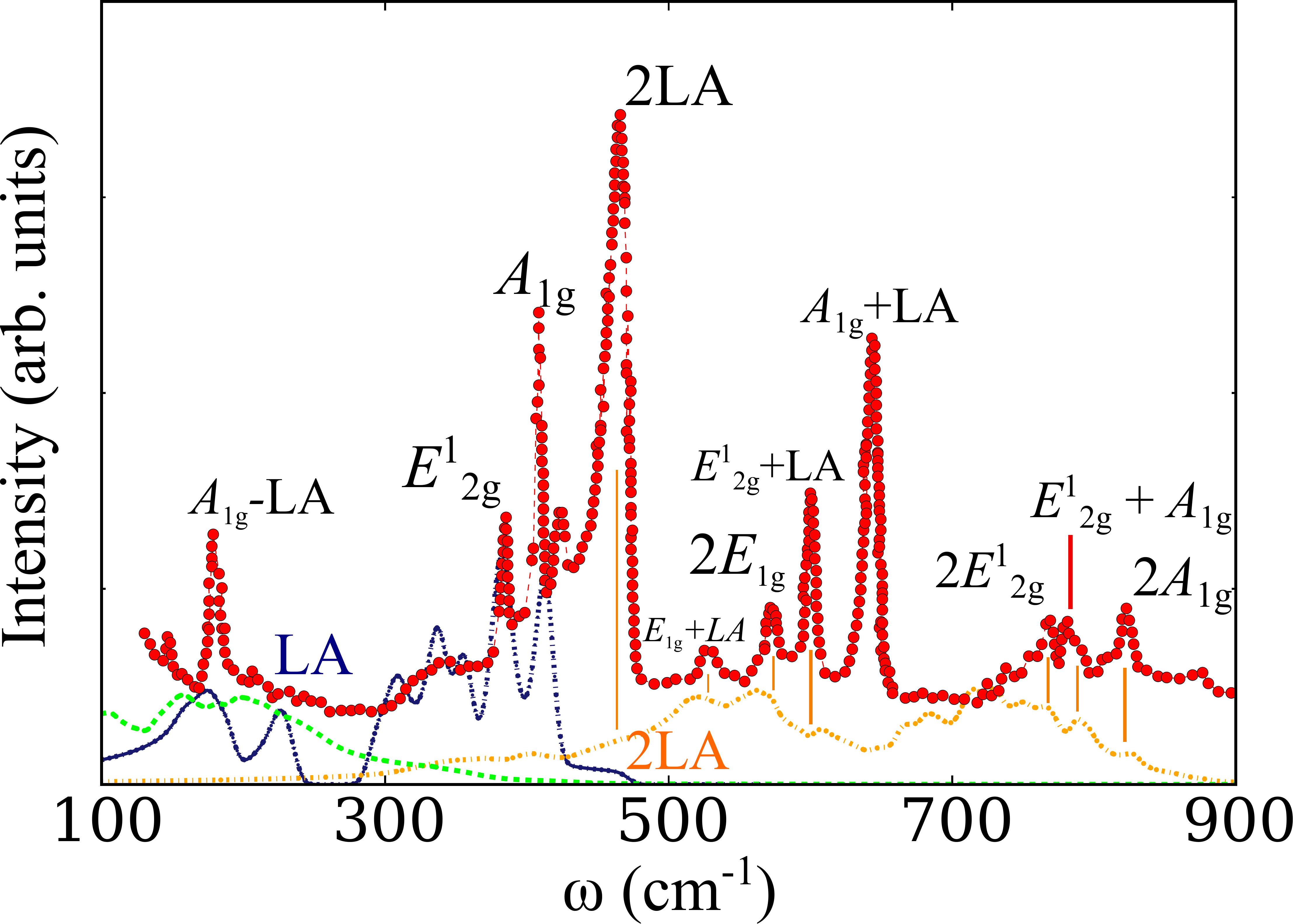}
\end{center}
\label{raman-exp}
\end{figure}

The density of states can be partially measured in 2nd and higher-order Raman spectra. We have represented in Fig. \ref{raman-exp} the Raman 
spectrum of MoS$_2$ bulk of Ref. \cite{Windom2011}, obtained by exciting the sample with a laser
frequency of 632 nm. Similar spectra can be found in older studies \cite{Chen1974, Stacy1985} and other recent studies \cite{Livneh2010,Livneh2014}. First order Raman peaks are due to the excitation of zero-momentum phonons. We can identify in the spectrum the modes $E^1_{2g}$ and $A_{1g}$. Moreover, working in
conditions of resonant Raman scattering, second-order Raman modes can be obtained. These modes come 
from the addition or subtraction of modes with opposite momentum, $\omega_i(\bm{q})\pm\omega_j(\bm{q})$, together with
the resonance of an intermediate excited electronic state\cite{Berkdemir2013}.
The result is a rich combination of Raman modes, as shown in Fig. \ref{raman-exp}. The identification can be done with the help of 
the density of states. We have calculated the 1-phonon and the 2-phonon density of states for MoS$_2$ bulk, represented by the
solid blue line and the dashed green line, respectively. 

A careful examination of the 2-phonon density of states tells us which 
phonons are participating in the Raman spectra. Thus, the 
longitudinal acoustic branch at M, denoted as $LA(M)$, couples 
to the modes $E^1_{1g}$, $E^1_{2g}$, and $A_{1g}$, resulting in overtones in the Raman spectrum. Other combinations
include $2\times E^1_{2g}$ or $2\times A_{1g}$, always with momentum $\bf{M}$. The second-order Raman modes are much more restrictive than first-order modes. The concurrence of phonon modes and electronic levels is needed. Such alignment
depends strongly on the electronic structure. Consequently, the second-order Raman spectrum has revealed useful to establish the fingerprints 
of single-layer systems with respect to the bulk, as discussed in Ref. \cite{Berkdemir2013}.

\subsection*{Symmetry analysis of phonon modes}

\begin{figure*}
\begin{center}
\includegraphics[width= 15 cm]{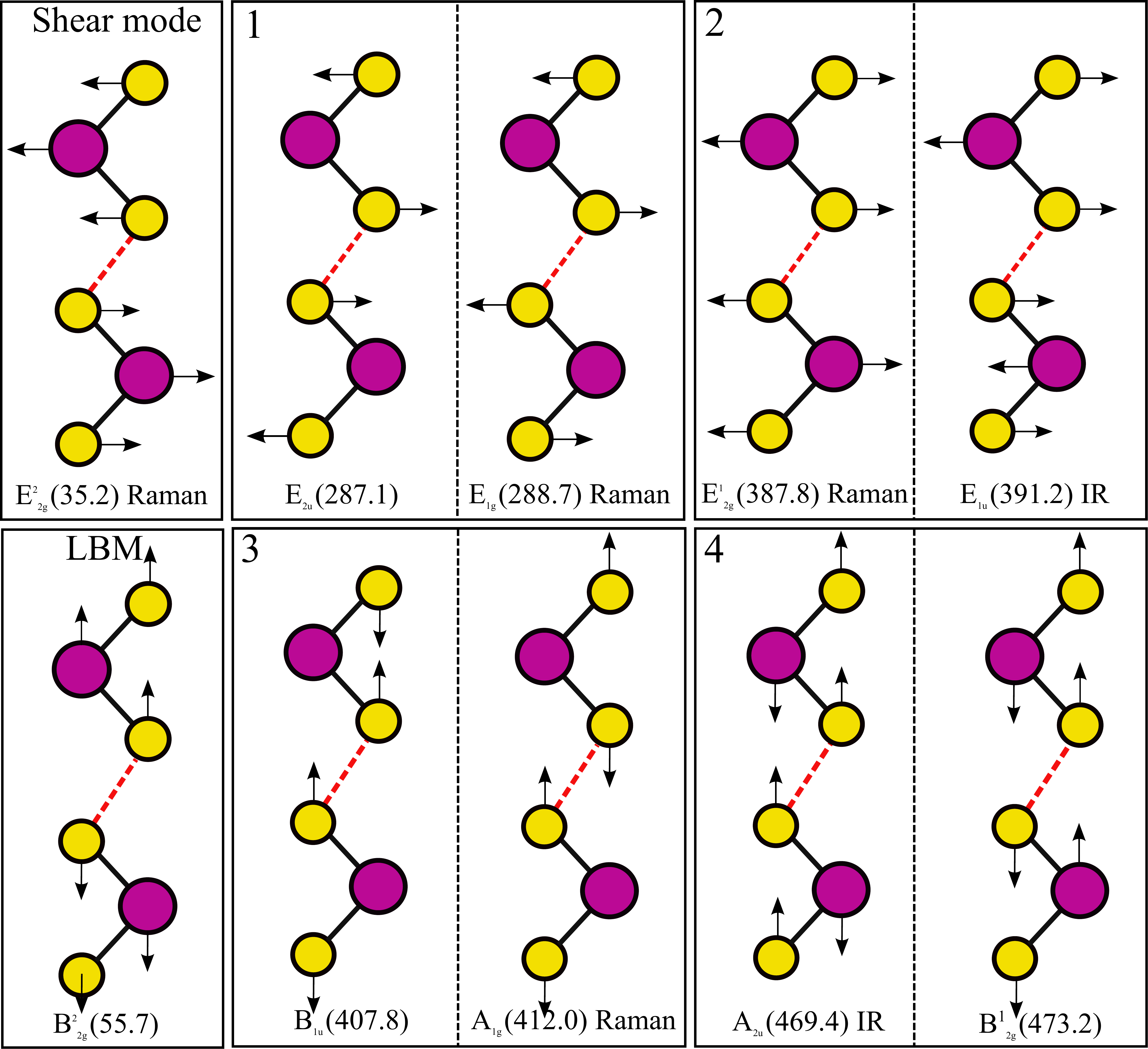}
\end{center}
\caption{Sketch of the optical phonon modes of bulk MoS$_2$. In the first row, the modes with polarization (atom-movement)
parallel to the layers are plotted in ascending order. In the second row, the perpendicular modes are shown. ``Davydov pairs'' of phonon modes are plotted in one box. The phonon frequencies (in cm$^{-1}$) are the calculated values of Ref.~\cite{Molina-Sanchez2011}.}
\label{bulk-modes}
\end{figure*}

\begin{figure*}
\begin{center}
\includegraphics[width= 15 cm]{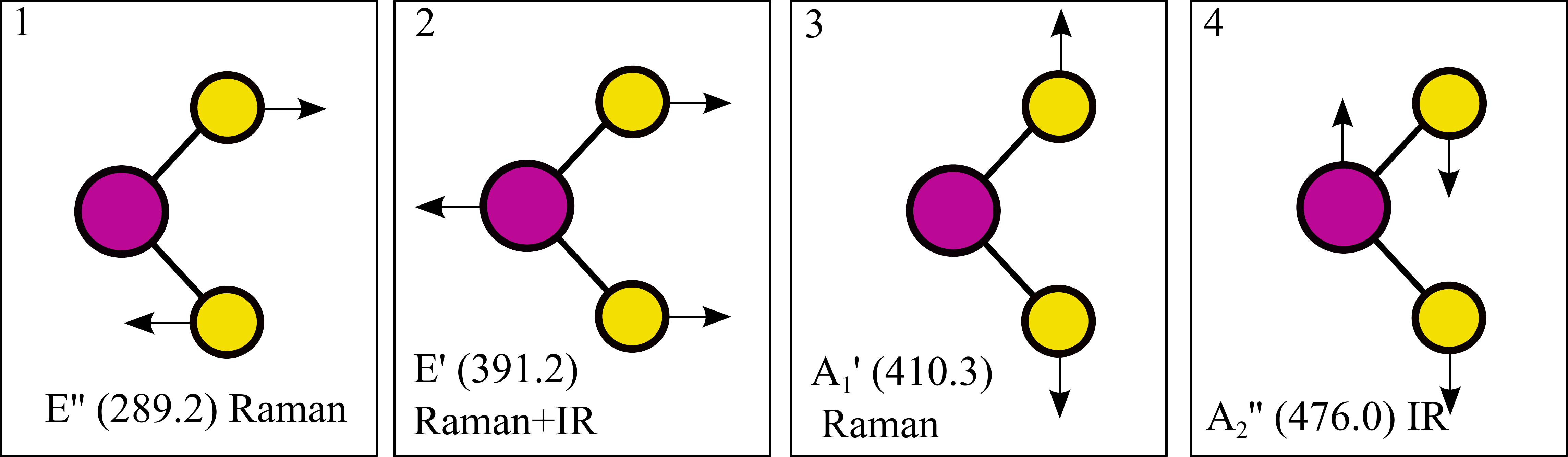}
\end{center}
\caption{Sketch of the optical phonon modes of single-layer MoS$_2$.
The phonon frequencies (in cm$^{-1}$) are the calculated values of Ref.~\cite{Molina-Sanchez2011}.}
\label{sl-modes}
\end{figure*}

We have drawn in Figs.~\ref{bulk-modes} and \ref{sl-modes} the atomic displacements (eigenvectors)
of optical phonon modes of bulk and single-layer MoS$_2$ at $\Gamma$.
Group theoretical analysis\cite{Verble1970,Wieting1971,Molina-Sanchez2011,Luo2013,Ribeiro-Soares2014} yields for the 15 optical modes of bulk MoS$_2$ (D$_{6h}$ symmetry)
the following decomposition in irreducible representations:
$A_{1g} \oplus A_{2u} \oplus 2B_{2g} \oplus B_{1u} \oplus E_{1g} \oplus E_{1u} \oplus 2E_{2g} \oplus E_{2u}$.
The $A_{1g}$, $E_{1g}$ and $E_{2g}$ modes are Raman active and the $A_{2u}$ and $E_{1u}$ modes
are infrared active.
For the 15 optical modes of double-layer MoS$_2$ (D$_{3d}$ symmetry), the decomposition is:
$3A_{1g} \oplus 2A_{2u} \oplus 3E_g \oplus 2 E_u$ where the gerade modes are Raman active and the ungerade ones
are IR active.
For the 6 optical modes of the single-layer, one obtains the following irreducible representations:
$A'_1 \oplus A''_2 \oplus E' \oplus E''$. The $A'_1$ and $E''$ modes are Raman active, the $E'$ mode
is both IR and Raman active.

The attribution of the different symmetries to the phonon modes in Figs.~\ref{bulk-modes} and \ref{sl-modes} can be understood
quite intuitively (taking into account that the drawings are simplified 2D versions of the real 3D modes).
All $E$ modes are doubly degenerate and correspond thus to in-plane vibrations of Mo and/or S atoms because
a rotation by any angle yields another phonon-mode with the same frequency.
The non-degenerate $A$ and $B$ modes must therefore correspond to perpendicular movement of the atoms.
For bulk MoS$_2$, (space group $P6_3/mmc$, point group $D_{6h}$), there is an inversion center half-way between two S atoms 
of neighboring layers.
We can thus distinguish between gerade (g) and ungerade (u) modes which are symmetric/anti-symmetric
with respect to inversion. The gerade modes are those where atoms in neighboring layers move with a phase shift of $\pi$
while the ungerade modes correspond to in-phase movement. All phonon modes of bulk MoS$_2$ thus come in pairs
of gerade and ungerade modes which are close in frequency. This can be clearly seen for the modes in panels 1 to 4
of Fig.~\ref{bulk-modes}. Furthermore, the ``shear mode'' at 35.2 cm$^{-1}$ is the gerade 
counterpart of the in-plane acoustic mode (not shown) and the ``layer-breathing mode'' (LBM) at 55.7 cm$^{-1}$ is the gerade counterpart
of the out-of-plane acoustic mode. In almost all cases, the gerade mode is higher in frequency than the ungerade mode. This is because the
weak (Van-der-Waals like) bond between S atoms of neighboring sheets is elongated and squeezed in the gerade mode (thus gives rise to an 
additional restoring force) but kept constant in the ungerade mode. The notable exception is the case of the modes in panel 2 where
the ungerade mode is higher in energy. We will come back to this important case in the next subsection.
One can easily see that only ungerade modes can be IR active: for a mode to be IR active, a net dipole must be formed through the displacement 
of positive charges in one direction and negative charges in the opposite direction. However, in gerade modes, the dipoles formed on one layer
 are canceled out by the oppositely oriented dipoles on the neighboring layer.
Since in systems with inversion symmetry, a phonon mode cannot be both IR and Raman active, only the gerade modes 
can be Raman active in bulk MoS$_2$.

The distinction between $A$ and $B$ modes is made by rotating the crystal by $2\pi/6$ around the principal rotation axis. This rotation is a non-symmorphic symmetry,
i. e., it has to be accompanied by a translation normal to the layer-plane in order to map the crystal into itself. In our reduced 2D representation of the
vibrational modes this corresponds to a translation of the 3 atoms of the upper layer onto the 3 atoms of the lower layer. If the arrows change direction, 
the mode is $B$, otherwise $A$. Finally, for the singly degenerate modes, the subscript $1$ ($2$) stand for modes that are symmetric (antisymmetric) with respect to rotation around
a C$_2$ axis crossing an Mo atom perpendicularly to the 2D plane of projection. For the doubly degenerate $E$ modes, it is the other way around. 

For even N-layers of MoS$_2$, the space-group symmetry is $P\bar{3}m1$ and the assignment of the phonon-mode symmetries
has to be done according to the $D_{3d}$ point-group symmetry. Since inversion symmetry is present, the mode assignment is 
very similar to the one of bulk MoS$_2$. For the doubly degenerate $E$ modes (see Fig.~\ref{bulk-modes}), the subscripts 1 and 2 
are dropped. All $E_u$ modes are IR active and all $E_g$ modes are Raman active. Out of the perpendicularly polarized modes,
the inactive $B_{1u}$ mode turns into an IR active $A_{2u}$ mode, the inactive $B_{2g}^1$ modes turns into a Raman active 
$A_{1g}$ modes. Notably, the layered breathing mode (LBM) is, in principle, Raman active. Indeed, for double and 4-layer MoS$_2$,
this mode has been detected in Raman measurements, albeit with small amplitude\cite{Zhang2013}.

For the single layer and for add-numbered multi-layers, the space group is $P\bar{6}m2$ and the corresponding point-symmetry group is D$_{3h}$. Since inversion symmetry is absent in this group, there is no distinction between gerade and ungerade modes. Instead, modes that are 
symmetric under $\sigma_h$ (reflection at the xy-plane) are labeled with a prime and anti-symmetric modes with a double prime (Fig.~\ref{sl-modes}).

The experimental and theoretical frequencies of all phonon modes of single-layer and bulk MoS$_2$ at $\Gamma$ are summarized 
in Table~\ref{symmetry}.
For the IR active modes of bulk MoS$_2$, we give both the values for longitudinal-optical (LO) and transverse-optical (TO) 
modes. The LO-TO splitting is calculated from the non-analytic part of the dynamical matrix (Eq.~\ref{nonana}) which only 
affects the IR active modes. 

\begin{table*}
\begin{center}
\caption{Phonon modes at $\Gamma$ of bulk and single-layer MoS$_2$ (inspired by Table II of Ref.~\cite{Wieting1971}). 
The polarization of the modes is in-plane ($xy$) or perpendicular ($z$). The irreducible representation (Irrep.) of each 
mode is calculated from the corresponding point-symmetry group ($D_{6h}$ for bulk, $D_{3h}$ for single-layer).
For the character of the modes, we distinguish between Raman active (R), infrared active (IR), 
acoustic modes (a), and inactive modes (i).}
\begin{tabular}{|cc|lcrr|lcrr|}
\hline
& & \multicolumn{4}{|c|}{single layer ($D_{3h}$ symmetry)}
& \multicolumn{4}{c|}{bulk ($D_{6h}$ symmetry)} \\
\hline
\multirow{2}{*}{Pol.} & \multirow{2}{*}{Atoms}   & \multirow{2}{*}{Irrep.} & 
\multirow{2}{*}{Char.} & \multicolumn{2}{c|}{Freq. (cm$^{-1}$)} & 
\multirow{2}{*}{Irrep.} & 
\multirow{2}{*}{Char.} & \multicolumn{2}{c|}{Freq. (cm$^{-1}$)} \\
     &          &        &       &   Calc.\cite{Molina-Sanchez2011} & Exp.\cite{Lee2010}  &   &    &  Calc.\cite{Molina-Sanchez2011} & Exp.\cite{Wieting1971} \\
\hline
\multirow{2}{*}{$xy$} &  \multirow{2}{*}{Mo+S}  &\multirow{2}{*}{$E'$}       &\multirow{2}{*}{a}   &  \multirow{2}{*}{0.0}  &                                     & $E_{1u}$   & a &  0.0      &                 \\
                      &                         &                            &                     &                        &                                     & $E_{2g}^2$ & R &  35.2     & 33 \\
\hline
\multirow{2}{*}{$z$}  &  \multirow{2}{*}{Mo+S}  &\multirow{2}{*}{$A_2^{''}$} &\multirow{2}{*}{a}   &  \multirow{2}{*}{0.0}  &                                     & $A_{2u}$   & a &  0.0      &      \\
                      &                         &                            &                     &                        &                                     & $B^2_{2g}$ & i &  55.7     &      \\
\hline
\multirow{2}{*}{$xy$} &  \multirow{2}{*}{   S}  &\multirow{2}{*}{$E''$}      &\multirow{2}{*}{R}   &  \multirow{2}{*}{289.2}&                                     & $E_{2u}$   & i &  287.1    &         \\
                      &                         &                            &                     &                        &                                     & $E_{1g}$   & R &  288.7    &     287    \\
\hline
\multirow{3}{*}{$xy$} &  \multirow{3}{*}{Mo+S}  &\multirow{3}{*}{$E'$}       &\multirow{3}{*}{R+IR($\bm{E}\bot\bm{c}$)}& \multirow{3}{*}{391.7} & \multirow{3}{*}{384.3} & $E_{2g}^1$  &R  & 387.8     & 383    \\
                      &                         &                            &                     &                        &           & \multirow{2}{*}{$E_{1u}$}  & \multirow{2}{*}{IR($\bm{E}\bot\bm{c}$)} & TO: 388.3 &   384           \\
                      &                         &                            &                     &                        &                                     &            &                       & LO: 391.2 &      387           \\
\hline
\multirow{2}{*}{$z$}  &  \multirow{2}{*}{   S}  &\multirow{2}{*}{$A'_1$}     &\multirow{2}{*}{R}   & \multirow{2}{*}{410.3} & \multirow{2}{*}{403.1}& $B_{1u}$   & i & 407.8     &             \\
                      &                         &                            &                     &                        &                                     & $A_{1g}$   &R  & 412.0     &     409                   \\
\hline
\multirow{3}{*}{$z$}  &  \multirow{3}{*}{Mo+S}  &\multirow{3}{*}{$A_2^{''}$} &\multirow{3}{*}{IR($\bm{E}||\bm{c}$)}  & \multirow{3}{*}{476.0} &                   &  \multirow{2}{*}{$A_{2u}$}   & \multirow{2}{*}{IR($\bm{E}||\bm{c}$)} &TO: 469.4          & 470       \\
                      &                         &                            &                     &                        &                                     &            &   &LO: 472.2      &   472                        \\
                      &                         &                            &                     &                        &                                     & $B^1_{2g}$ & i &473.2      &                           \\
\hline
\end{tabular}
\label{symmetry}
\end{center}
\end{table*}

\subsection*{Anomalous Davydov splitting}
As mentioned above, in bulk MoS$_2$, all modes appear as pairs of gerade and ungerade modes (Fig.~\ref{bulk-modes} and 
Table~\ref{symmetry}). This is because the unit cell of bulk MoS$_2$ contains 6 atoms while the single-layer unit cell 
only contains 3.
The frequency splitting of gerade and ungerade modes in layered materials and molecular crystals is known as 
``Davydov splitting'' or ``factor group splitting'' \cite{Davydov1969,Dawson1975}.
The model of Davydov has been used in particular to explain the splitting of modes that are IR and R active for the single-layer
into a Raman active mode and an IR active mode of the bulk (mode No.~2 in Figs.~\ref{bulk-modes} and \ref{sl-modes}). 
It was recognized early \cite{Wieting1971,Kuroda1979} that neither the weak Van-der-Waals coupling between neighboring
layers nor a simple model of dipolar couplings matches the experimental observation that 
$\nu_{\mbox{Raman}} < \nu_{\mbox{IR,TO}} < \nu_{\mbox{IR,LO}}$ for some layered materials and, in particular, MoS$_2$.
A model involving quadrupole interaction was proposed by Ghosh et al.\cite{Ghosh1976,Ghosh1983} but could not be underpinned
by numerical calculations. 

The explanation of the ``normal'' Davydov splitting in van-der-Waals bonded layered materials is straightforward.
As can be seen in Fig.~\ref{bulk-modes}, the weak inter-layer bonding can be viewed as an additional (weak) spring
constant acting between sulfur atoms from neighboring layers (red dashed lines). For the ungerade modes, the S-atoms
are moving in phase and the additional spring thus is not ``used''. However, for the gerade modes,
where the S-atoms are vibrating with a phase shift of $\pi$, the additional spring is elongated and compressed and thus
yields an additional restoring force. This leads, in general, to an upshift of the frequencies of the gerade modes. 
Since the interaction is weak, the frequency shift is small ($<$ 5cm$^{-1}$). Furthermore, the effect is more
pronounced for the perpendicularly polarized modes than for the in-plane modes (Fig.~\ref{bulk-modes} and Table~\ref{symmetry}).
The only exception from the ``normal'' Davydov splitting is thus the mode No.~2. One might argue that this case is
exceptional, because the LO-shift of the $E^1_{2g}$ mode makes its frequency higher then the one of the $E_{1u}$ mode.
However, even without the LO-shift, experiments\cite{Wieting1971} and calculations\cite{Molina-Sanchez2011} agree that
$\nu(E^1_{2g}) < \nu(E_{1u,\mbox{TO}})$. 

\begin{figure*}
\begin{center}
\includegraphics[width=15.0 cm]{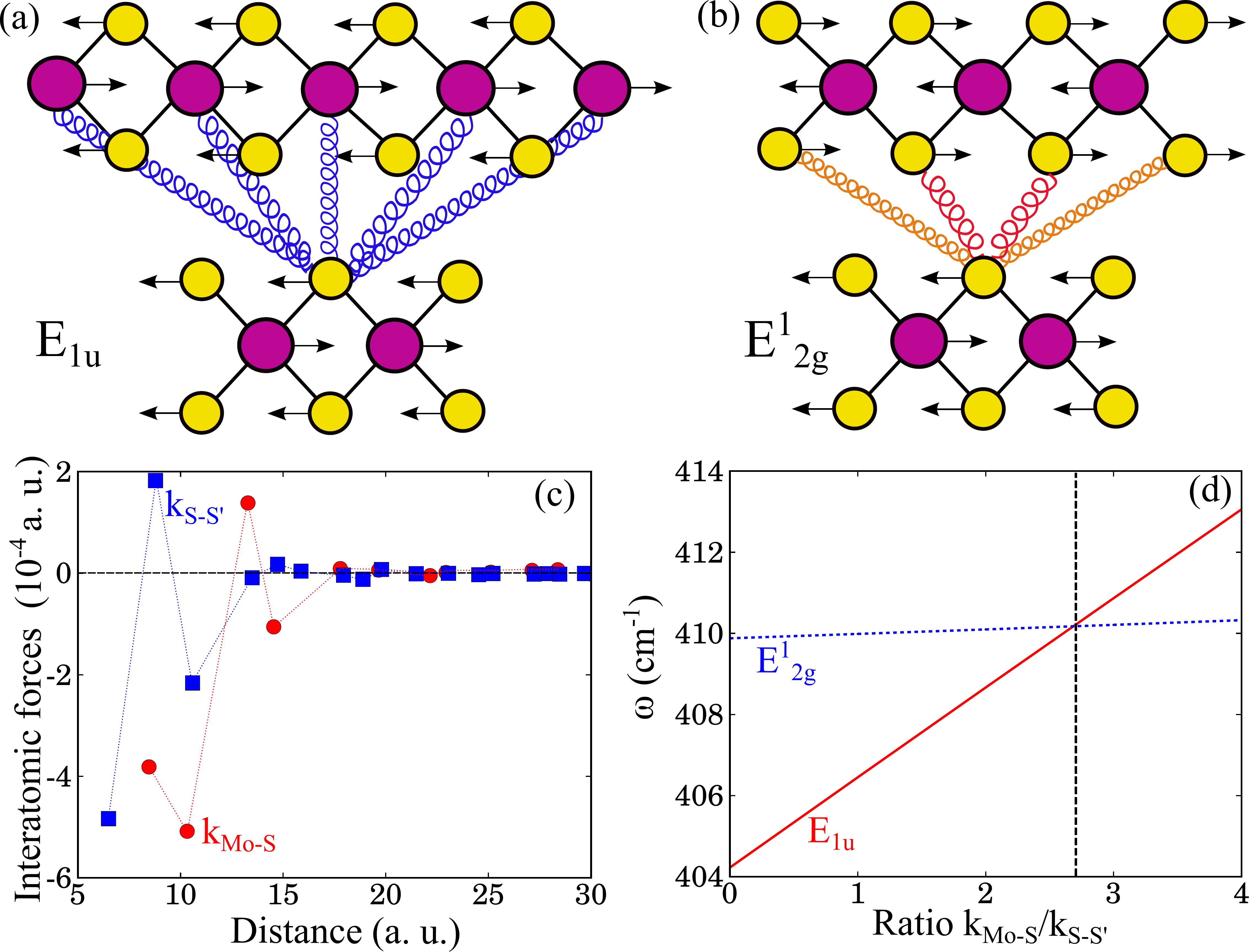}
\end{center}
\caption{(a) and (b) Sketch of the $E_{1u}$ and $E_{2g}^1$ modes.
(c) Sketch of in-plane force constants $k_{S-S'}$ and $k_{Mo-S}$.
(d) Phonon frequencies of $E_{1u}$ (red) and $E^1_{2g}$ 
(blue) modes as a function of the ratio $k_{Mo-S}/k_{S-S'}$. }
\label{forces}
\end{figure*}

Since our {\it ab-initio} calculations reproduce the anomalous Davydov splitting for mode No.~2, 
we can use the interatomic force constants, derived from the calculations, in order to find the
origin of this seemingly anomalous behaviour. The situation is depicted in Fig.~\ref{forces} (a) and (b).
We analyze in detail the force constants between S atoms of neighboring layers (blue springs) and also the force 
constants between S atoms on one layer with Mo atoms on the neighboring layer (red springs).
In the $E^1_{2g}$ mode it is the sum of all the S--S spring constants that leads
to an additional restoring
force and thus to an up-shift (with respect to the same mode in the fictitious isolated layer).
However, for the $E_{1u}$ mode, an additional restoring force arises as well, this time due to the Mo--S interactions. As it turns
out, their effect is stronger than the ones of the S--S springs. This follows from the numerical
values of the horizontal components of the S--S and Mo--S force constants. We present the values in
Fig.~\ref{forces} (c). For a given S atom, we calculate the sum of the (horizontal) force constants over all
nearest, next-nearest, ..., S and Mo atoms of the adjacent layer. Negative sign implies restoring force (the
S atoms is pushed back to the left when displaced to the right). In Fig.~\ref{forces} (c) one 
can see that the interaction of S with the three next-nearest S atoms of the adjacent layer is stronger than the
interaction of S with the closest Mo atom. However, the \textit{cumulative} effect of the S--Mo interactions
is larger than the one of the S--S interaction. This explains why for this mode pair the sign of the
Davydov splitting is negative. 
The dominance of the inter-layer S--Mo interaction over the inter-layer S--S interaction was already invoked in the
force-constant model of Luo et al.\cite{Luo2013} In that model, all the interaction was ascribed to the closest atom pairs of adjacent layers. 
This renders the semi-empirical model simple and quantitatively successful. 
In semi-empirical calculations using the code GULP\cite{Gale1997,Gale2003}, we have verified that it is possible to inverse the frequencies of the $E^1_{2g}$ and $E_{1u}$ modes by changing 
the relative strength of the cross-layer S--S and S--Mo interaction (see Fig.~\ref{forces} (d)). However,
the {\it ab-initio} calculations demonstrate that the physical reality is more complex. The 2nd-nearest neighbor interaction
between sulfur atoms across layers is even repulsive. Thus, the correct balance between S--S and S--Mo interaction is only found by summing over all interactions. It turns out that the force constants decrease rather quickly with increasing distance and the Coulomb contribution (from the effective charges) is rather small.

Note that the situation is different for mode No.~1 in which the Mo atoms
do not move and the inter-layer S--Mo interaction thus plays a negligible role for the Davydov splitting.
In this case, the Raman active $E_{1g}$ mode is slightly higher in frequency than the inactive $E_{2u}$ mode, following the intuitive expectation and yielding the ``normal'' sign of the Davydov splitting.

We will see in the following that also the dependence of the frequencies of the
$E^1_{2g}$ on the number of layers follows an unexpected trend which can be used for the determination of the number of layers via Raman spectroscopy.

\subsection*{Dependence of Raman active modes on number of layers}

Since the beginning of the research on \MoS flakes, the Raman modes have been used to identify 
the number of layers\cite{Lee2010,Plechinger2012,Zhao2013,Zhang2013}. The correspondence between frequency and 
number of layers has been done by comparing with other techniques such as atomic force microscopy or optical 
contrast. The phonon modes used to characterize the 
thickness are usually the high-frequency Raman modes $E^1_{2g}$ 
and $A_{1g}$ (see Fig. \ref{bulk-modes}) and the breathing and shearing modes at 
low-frequency \cite{Zhang2013}. 
We will summarize in the following the results and analyze the physics of the
frequency trends.

\subsubsection*{High-frequency modes}

In the single layer, the high frequency $\Gamma$ modes
$E_{2g}^1$ and $E_{1u}$ collapse into the mode $E'$.
(From Fig.~\ref{sl-modes} it is evident that 
with increasing inter-layer distance, the modes
$E_{2g}^1$ and $E_{1u}$ acquire the same frequency.) 
Interestingly, as measured in Ref.~\cite{Lee2010} and indicated
in Table~\ref{symmetry} (see also Figs.~\ref{bulk-modes} and \ref{sl-modes}), 
the bulk $E_{2g}^1$ mode is lower in frequency than the single-layer $E'$ mode.
This contradicts the expectation
that the additional inter-layer interaction should increase the frequency from
single-layer to bulk.
Even the bulk $E_{1u}$ mode (which is higher in frequency than the $E^1_{2g}$ 
mode due to the anomalous Davydov splitting) is slightly lower
than the single layer $E'$ mode.
The same behaviour (that the bulk modes are lower in frequency than the single-layer
mode) can be observed for the in-plane mode No.~1 ($E_{2u}$ and $E_{1g}$ in bulk versus $E''$ 
in the single layer) and for the out-of-plane mode No.~4 ($A_{2u}$ and $B^1_{2g}$ in bulk
versus $A_2''$ in the single layer).
Only the out-of-plane mode $A_{1g}$ (No.~3) follows the expected trend that the 
inter-layer interaction increases the frequency with respect to   
the single-layer mode $A_1'$.

\begin{figure}
\begin{center}
\includegraphics[width=7.6 cm]{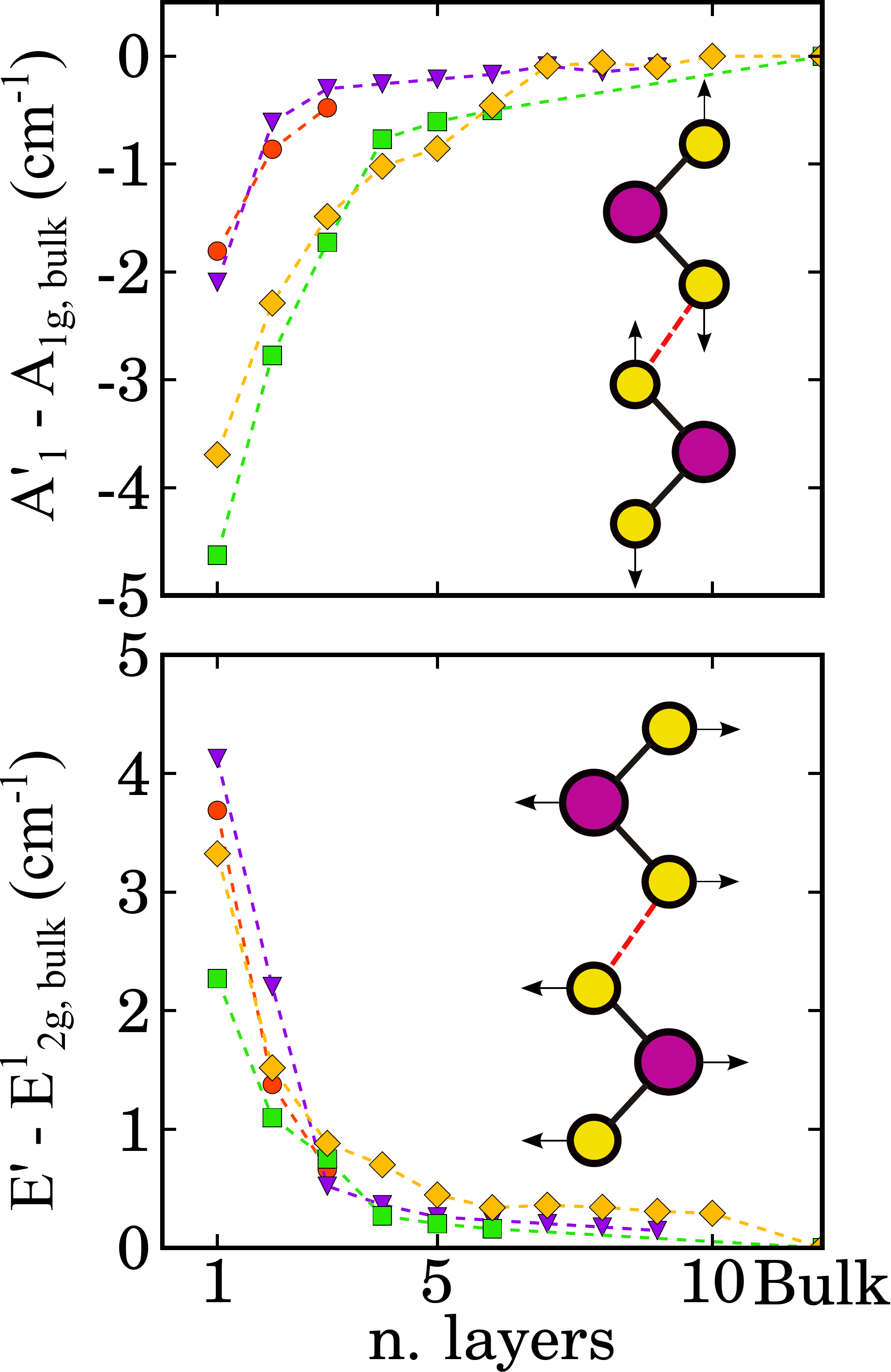}
\end{center}
\caption{Phonon frequencies of $A_{1g}$ and $E^1_{2g}$ modes as a 
function of MoS$_2$ layer thickness. The symbols corresponds to: (red
circle) this work (blue triangles) LDA calculations of Ref. \cite{Luo2013}, 
(black stars) experimental data from Ref. \cite{Lee2010} and from Ref. \cite{Luo2013} (green diamonds).}
\label{raman-layers}
\end{figure}

Figure \ref{raman-layers} shows the frequency of the $A_1$ and $E'$ modes as
a function of layer number. We compare LDA calculations (circles comes from Ref. \cite{Molina-Sanchez2011}
and triangles from Ref.\cite{Luo2013}) with the experimental data 
of Refs. \cite{Lee2010} and \cite{Luo2013}. 
The out-of-plane mode $A_{1}'$ increases in frequency with increasing number of 
layers. The explanation lies in the interlayer interaction, caused by
weak van-der-Waals bonding of the sulphur atoms of neighbouring layers. 
The stacking of layers can thus be seen as the addition of a spring between the sulfur atoms
of neighboring layers. Within the picture of connected harmonic
oscillators, this results in an increase of the frequency. The LDA calculations
reproduces well this trend, as can be seen in Fig.~\ref{raman-layers}. The small
disagreement between theory and experiment can be attributed to the inaccuracy of 
the interlayer interaction given by LDA. 

The in-plane mode $E'$ displays the opposite trend, decreasing in frequency  by about 4 cm$^{-1}$
from single-layer to bulk (Fig.~\ref{raman-layers}, lower panel). This is - at first sight - unexpected, because the additional
``spring'' between the sulfur atoms should lead to an increased restoring force and thus to a
frequency increase as in the case of the $A_{1}'$ mode. Several attempts have been made in the
past to explain this anomalous behaviour, ascribing it to long-range Coulomb 
interactions\cite{Lee2010}. In our previous previous publication\cite{Molina-Sanchez2011},
we have investigated how the dielectric screening in the bulk environment reduces
the long-range (Coulomb) part of the force constants. However, the long-range part plays
only a minor role. We have verified this by performing an {\it ab-initio} phonon calculation
of the $E'$ mode of single-layer MoS$_2$ sandwiched between graphene-layers. If the distance
between the sulfur atoms and the graphene layer is higher than 6{\AA}, there is no ``chemical''
interaction between the different layers and the graphene just enhances the dielectric
screening of the MoS$_2$ layer. Since the $E'$ remains unaffected, we conclude that
the long-range Coulomb effect can be discarded as a possible effect for the anomalous
frequency trend. 
 
The solution to the problem has been given by Luo et al.\cite{Luo2013} and is related to a weakening
of the nearest neighbor Mo--S force-constant in the bulk environment. To be precise, one
has to compare the Real space force constants $C^{Mo,S}_{x,x}(\bm{0})$
for the force and displacement parallel to the layer. (See Eq.~(\ref{realforceconst}) for the definition
of the force constant). This is the dominant force constant determining the frequency of the $E'$ mode
as becomes immediately clear from the displacement pattern in Fig.~\ref{sl-modes}.
Force constants between more distant atoms play a very small role.
There are two reasons why $C^{Mo,S}_{x,x}(\bm{0})$ is larger for the single-layer than for bulk.
One reason is that in the single-layer, the Mo--S bond is slightly shortened. However, even without
this change in bond-length, the Mo--S force constant is slightly larger in the single-layer than in bulk.
This can be obtained in an {\it ab-initio} calculation of the single-layer using the interatomic distances from bulk.
The results of our calculations are shown in Table~\ref{fclayer} and are very similar to the values
of Fig.~3 in Ref.~\cite{Luo2013}. The frequency of the $E'$ mode is proportional to the force-constant: $\omega_{E'} \propto \sqrt{C^{Mo,S}_{x,x}(\bm{0})}$.
Even though the differences seem tiny, they explain the calculated and observed frequency differences between
single-layer and bulk in table~\ref{symmetry}. The finding that the Mo--S force constant is smaller in bulk than in single-layer,
even at equal bond-length, is related to a (slight) redistribution of the charge density when a neighboring layer is present.

\begin{table}
\begin{center}
\caption{{\it Ab-initio} Force constant $C^{Mo,S}_{x,x}(\bm{0})$ in (Ha/Bohr$^2$) in single layer (SL) and bulk MoS$_2$.}
\begin{tabular}{|c|c|c|}
\hline
    \multirow{2}{*}{bulk} &  \multicolumn{2}{c|}{SL} \\
          &  bulk geom. &     opt. geom. \\
\hline
    -0.1102   & -0.1119  &  -0.1127  \\
\hline
\end{tabular}
\end{center}
\label{fclayer}
\end{table}

The fact that the $A_1'$ and the $E'$ modes move in opposite directions with increasing number of layers, makes the distance between the two 
corresponding peaks in the Raman spectra a reliable measure for the layer number\cite{Lee2010,Luo2013}. But this is not the only way
to detect the layer number in Raman spectroscopy. The low frequency Raman active modes display an even stronger dependence as explained
in the following.


The shearing mode (C), denoted in bulk as $E^2_{2g}$, is the rigid-layer displacement in-plane. 
This mode is Raman active in bulk, as indicated in Table \ref{symmetry}. 
The layer-breathing mode (LBM) corresponds to vertical rigid-layer vibrations, 
in the case of bulk, where is has $B^2_{2g}$ symmetry, it is a silent mode. 
However, in the bi-layer case it has $A_{1g}$ symmetry and is (weakly) visible.
Several groups have investigated the low frequency behaviour of few-layer MoS$_2$ \cite{Plechinger2012,Zhang2013,Zhao2013}.
The frequency trends as a function of the layer number can be explained via a simple analytical
model that was first developed to explain the corresponding modes in few-layer graphene\cite{Tan2012,Michel2012}.
In this model, $N$ layers with a mass per unit area $\mu$ are connected via harmonic springs.
One distinguishes between force constants (per unit area) for displacement perpendicular $\alpha_{\perp}$ and parallel $\alpha_{\parallel}$ 
to the layer, respectively.
Mathematically, the model is equivalent to a linear chain of $N$ atoms.
Assuming a time dependence Assuming a time dependence of $u_n(t) = u_n^0 \exp[i\omega t]$ for all the $N$
atoms, Newton's equation of motion lead to the secular equation
\begin{equation}
\omega^2 \left(
\begin{array}{c}
u_1 \\ \vdots \\ \vdots  \\ \vdots \\ \vdots \\ u_n
\end{array} \right)
= \frac{\alpha}{\mu}
\left(
\begin{array}{cccccc}
1      & -1     & 0      & \cdots &      0 & 0      \\
-1     & 2      & -1     & \ddots &        & 0      \\
0      & -1     & \ddots & \ddots & \ddots & \vdots \\
\vdots & \ddots & \ddots & \ddots & -1     & 0      \\
0      &        & \ddots & -1     & 2      & -1     \\
0      & 0      & \ddots & 0      & -1     & 1      \\
\end{array} \right)
\left(
\begin{array}{c}
u_1 \\ \vdots \\ \vdots  \\ \vdots \\ \vdots \\ u_n
\end{array} \right),
\label{secular-chain}
\end{equation}
where $\alpha=\alpha_{\perp}$ for the layer-breathing modes and $\alpha=\alpha_{\parallel}$ for the shear modes.
The frequency of the $i$th phonon mode $(i=1,\ldots,N)$ is
\begin{equation}
\omega_i = \sqrt{
\frac{2\alpha}{\mu}\left( 1 - \cos\left[\frac{(i-1)\pi}{N}\right] \right)}.
\label{layerfreq}
\end{equation}
For $i=1$ one obtains the acoustic (translational) mode, $i=2$ yields the lowest non-vanishing frequency
of the mode with a nodeless envelope function, and $i=N$ yields the highest frequency mode where neighboring
layers are vibrating with a phase shift of (almost) $\pi$.

\begin{figure*}
\begin{center}
\includegraphics[width=15 cm]{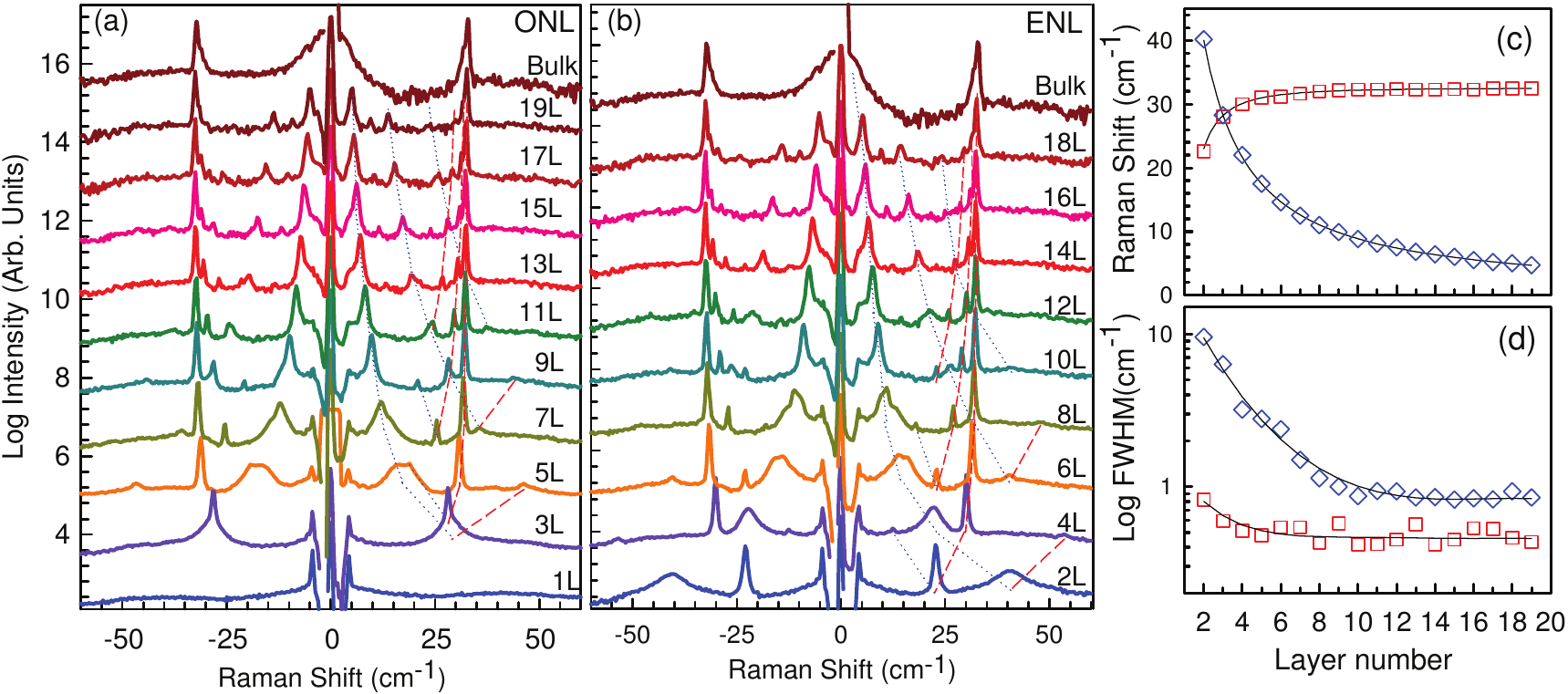}
\end{center}
\caption{Stokes and anti-Stokes Raman spectra for (a) odd and (b) even number of 
layers (ONL and ENL, respectively). The spectrum of bulk is included in (a) and (b).
(c) Raman frequency and (d) FWHM,  as a function of the number of layers 
of the breathing mode (LBM) and shear mode (C) (Reprinted
with permission from Ref. \cite{Zhang2013}. Copyright (2013) by the American
Physical Society).}
\label{low-modes}
\end{figure*}

Fig. \ref{low-modes} shows the Raman spectra published in Ref. \cite{Zhang2013}. 
The number of layers ranges from 1 to $19$ in the 
case of odd number of layers (ONL) and from 2 to $18$ in the case 
of even number of layers (ENL). Evidently, the single-layer \MoS ~Raman spectra has no 
signature of low-frequency modes. The peaks at $4.55$ cm$^{-1}$ are due to the 
Brillouin scattering of the LA mode from the silicon substrate. One can see that
some peaks stiffen (dashed lines) for increasing thickness and others 
are softened (dotted lines). Fig. \ref{low-modes}(c) shows the shear (C) and 
breathing (LBM) mode as a function of the number of layers. 
We can see also in Fig.~\ref{low-modes}(d) that the  
full width at half maximum (FWHM) behaves in a different way for the C or the LBM modes. 
In the case of the LBM mode (blue dots in in Fig.~\ref{low-modes} (a,b)),
the branch with the largest weight is the one with $i=2$. According
to Eq.~\ref{layerfreq}, the frequency of this branch as a function of layer number $N$ goes like
\begin{equation}
\omega^{\mbox{LBM}}(N)=\omega^{\mbox{LBM}}(2)\sqrt{1-\cos(\pi/N)},
\end{equation}
where $\omega^{\mbox{LBM}}(2)=\sqrt{2\alpha_{\perp}/\mu}$. This is the functional 
form of the blue diamonds in Fig.~\ref{low-modes} (c). 
For $i=N$, the LBM increases in frequency according to
\begin{equation}
\omega^{\mbox{LBM}}(N)=\omega^{\mbox{LBM}}(2)\sqrt{1+\cos(\pi/N)}
\end{equation}
and approaches, for $N \rightarrow \infty$, the value of the $B^2_{2g}$ bulk mode
at 55.7 cm$^{-1}$. However, since the bulk mode is not Raman active, the intensity of this
mode quickly decreases with increasing $N$ and the mode is already almost invisible for $N=4$.
For intermediate values of $2 < i < N/2$, side branches of the LBM appear and are clearly
visible in the Raman spectra of Fig.~\ref{low-modes}.

The same analysis can be done for the shear (C) mode. In this case, it is the $i=N$ branch
that has the dominant weight and converges towards the bulk $E^2_{2g}$ shear mode with increasing $N$:
\begin{equation}
\omega^{\mbox{C}}(N)=\omega^{\mbox{C}}(2)\sqrt{1+\cos(\pi/N)}.
\end{equation}
This is the functional form of of the red squares in Fig.~\ref{low-modes} (c).
The frequency ratio of bulk and double layer shear modes is $\omega_{bulk}/\omega(2)=\sqrt{2}$ which is
verified by the experimental data\cite{Plechinger2012,Zhang2013}.
Some side-branches for $N \le i < N$ are visible in the spectra as well, however with lower intensity
than the $i=N$ branch.

Due to the strong layer dependence of the frequencies, the shear and compression mode are a very
sensitive tool for the determination of layer-thickness by Raman spectroscopy \cite{Plechinger2012,Zhang2013}. 
The monoatomic chain model is able to explain the main physics of these modes. Small deviations from
the analytic result have been observed \cite{Plechinger2012} and might be due to
anharmonic effects\cite{Boukhicha2013}.
The disadvantage of the use of these modes lies in the detection of frequencies close to those of the Brillouin scattering. 

\section{Electronic structure}
\label{band}

This section is devoted to the main properties of the electronic structure 
of \MoS. Based on
first principles calculations, the characteristics of the band structures of single-layer, double-layer and bulk \MoS
are discussed. In particular, we analyze the electronic band gaps 
and show \abinitio 
how the band structures depend on the unit cell parameters and the structural 
optimization. We explain the reasons for the differences in the results 
obtained through the different computational approaches.

Historically, TMDs are used in the field
of tribology as lubricants. The attention given to TMDs
decades ago has lead to several theoretical studies of 
the band structure of \MoS~ in single-layer and 
bulk forms\cite{Mattheiss1973,Mattheiss1973a}. These studies were 
complemented more recently with angle-resolved
photoelectron spectroscopy measurements for bulk \MoS~ 
accompanied by \abinitio~ calculations \cite{Boker2001}. 

The current interest in \MoS~ \cite{RadisavljevicB.2011, Peelaers2012}, 
the availability of high-quality single-layer flakes \cite{Coleman2011}, 
and the improvement of experimental results have prompted new theoretical studies in the past 5 years.
Regarding the electronic structure, the most efficient approach with
respect to computational cost and accuracy is the use of
DFT-LDA/GGA.
Due to the potential application of single-layer \MoS~ in transistors, most calculations are focused on the band gap. By using LDA,  
single-layer MoS$_2$ is determined as a direct semiconductor with a band gap of
1.78 eV at the $\bf{K}$ point of the 
Brillouin zone \cite{Lebegue2009}. 
The transition from a direct to an indirect gap semiconductor with increasing 
number of layers is also confirmed
\cite{Scalise2012, Scalise2014}. 
The extensive use of standard DFT in MoS$_2$ (and the comparison with
more advanced methods) has demonstrated 
that this approach is adequate for investigating trends. 
However, an inherent problem of DFT-LDA/GGA is the underestimation of the band gap which is related to the lack of the derivative discontinuity 
in (semi)local exchange-correlation potentials\cite{Jones1989}.

In a first attempt to improve the electronic band gaps and band dispersions 
at low computational cost, the modified Becke-Johnson  (MBJ) potential \cite{becke:jcp:124} 
combined with LDA correlation was applied. The MBJLDA approach was developed 
by F. Tran and P. Blaha \cite{tran:prl:102} and implemented in VASP \cite{Kim2010}.
The MBJ potential is a local approximation to an atomic 
exact exchange potential plus a screening term (screening parameter $c$) and is given by

\begin{equation}
\label{eq:MBJ}
 V_{x,\sigma}^{\rm MBJ}(\mathbf{r}) = cV_{x,\sigma}^{\rm BR}(\mathbf{r}) + (3c-2)\frac{1}{\pi}\sqrt{\frac{5}{12}}\sqrt{\frac{2\tau_{\sigma}(\mathbf{r})}{\rho_{\sigma}(\mathbf{r})}}.
\end{equation}

In this expression, $\rho_{\sigma}$ denotes the electron 
density, $\tau_{\sigma}$ the kinetic energy density, and $V_{x,\sigma}^{\rm BR}(\mathbf{r})$ the Becke-Roussel (BR) potential \cite{becke:pra:39}.
In the parameter-free MBJLDA calculation employed in this 
study, the $c$ parameter is chosen to depend linearly 
on the square root of the average of $|\nabla \rho|/\rho$ 
over the unit cell volume and is self-consistently determined. 

Alternatively, 
the screened hybrid functional HSE \cite{heyd:jcp:118,paier:jcp:122,krukau:jcp:125}
and the improved HSEsol functional \cite{schimka:jcp:134} were tested.
The success of HSE in predicting band gaps with accuracy comparable to that of schemes based on many body perturbation theory
($GW$ methods) but significantly reduced computational cost is multiply demonstrated in the work of G.~Scuseria
(see the review in Ref. \cite{janesko:pccp:11}) 
as well as in independent studies including a variety of materials 
and properties \cite{paier:jcp:124,paier:jcp:125:erratum,hummer:prb:75,paier:prb:79,kim:prb:80}.
In general, hybrid functionals are constructed by using the DFT correlation 
energy $E_{\mathrm{c}}$ and adding 
an exchange energy $E_{\mathrm{x}}$ that consists of 25\% Hartree-Fock (HF) exchange and 75\% DFT
exchange.
Furthermore, in the concept of the screened HSE functional \cite{heyd:jcp:118}
the expensive integrals of the slowly decaying long-ranged part of the
HF exchange are avoided 
by further separating the $E_{\mathrm{x}}$ into a short- (SR) and long-ranged
(LR) term, where the latter is replaced by its DFT counterpart. An additional parameter $\mu$ defines the
range-separation \cite{note:HSE03}.
The HSEsol functional is analogous to the HSE functional, 
but is based on the PBEsol functional for all DFT parts according to
\begin{eqnarray}
\label{eq:HSEsol}
 E_{\mathrm{xc}}^{\mathrm{HSEsol}} & = & E_{\mathrm{c}}^{\mathrm{PBEsol}} + E_{\mathrm{x}}^{\mathrm{PBEsol}} \nonumber \\
                                   & - & \frac{1}{4} E_{\mathrm{x}}^{\mathrm{SR,PBEsol}}(\mu)  \nonumber    \\
                                   & + & \frac{1}{4}E_{\mathrm{x}}^{\mathrm{SR,HF}}(\mu).
\end{eqnarray}

Concerning the electronic properties, the highest level of accuracy has been achieved by $GW$ calculations.
In this work the band structures were studied using the single-shot ($G_0W_0$) and the self-consistent $GW$ (sc$GW$) 
approximation. In both approaches, the dynamically (frequency dependent) 
screened Coulomb interaction $W$ and the self energy $\Sigma(r,r',\omega)$ were
calculated using the DFT-LDA wave functions.
In the sc$GW$ case, 
both, the quasiparticle (QP) energies (one electron 
energies) and the one electron orbitals (wave functions) are 
updated in $G$ and $W$. 
Note that the attractive electron-hole 
interactions were not 
accounted for by vertex corrections in $W$.
Thus the calculated band gaps are expected to be slightly 
overestimated compared to experiment \cite{shishkin:prl:99}.
In the particular case
of \MoS, the $GW$ method has been used in many flavours, yielding
a significant spread of values for the band gap, to be discussed
below.

\subsection{Characterization of the band structure of single-layer and bulk MoS$_2$}

\begin{figure*}
\begin{center}
\includegraphics[width=14 cm]{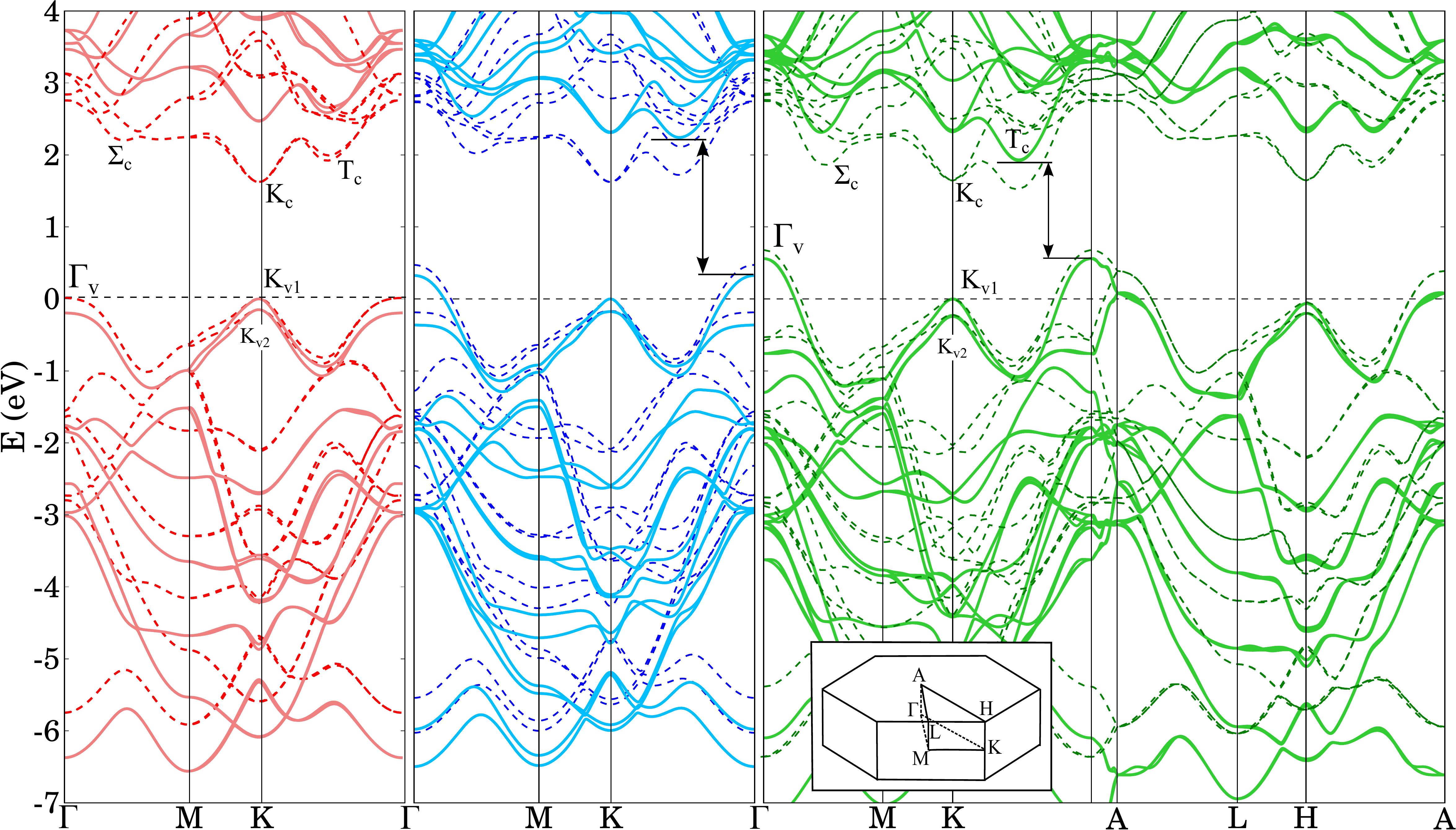}
\end{center}
\caption{Band structure of \MoS~ single-layer (left), double-layer (center), and bulk (right) in
the LDA (thin dashed lines) and in the $G_0W_0$ approximation (thick continuous lines). The lattice parameters 
have been obtained from the structure optimization using the optB86b-VdW functional.}
\label{bands}
\end{figure*}

First of all, we discuss the main features of the electronic structure of 
\MoS. Figure \ref{bands} shows the band structure for 
single-layer (1L), double-layer (2L) and bulk MoS$_2$. The vdW-DF optimized lattice constant 
(Section \ref{structure}) is used in all calculations.
The suitability of using either the experimental or the theoretical 
lattice parameter in the calculations is still controversial.
For the calculations presented in this review, we have chosen 
the latter, which guarantees a strain-free structure. 
Thereby, we will further be able to investigate strain 
effects on the electronic structure.

Figure \ref{bands} shows the relevant features of the bulk \MoS~ band structure: 
\begin{itemize}
 \item two distinct valence band edges (VBEs) located at $\bf{K}$ and $\Gamma$,
 \item three conduction band extrema (CBEs) at $\bf{T}$ (half way between $\Gamma$
and $\bf{K}$), $\bf{K}$, and $\Sigma$ (half way between $\Gamma$ and $\bf{M}$),
 \item the valence band maximum (VBM) located at $\Gamma$ and the conduction
band minimum (CBM) at $\bf{T}$,
\item a fundamental electronic band gap of indirect nature that is defined by the energy difference $T_c-\Gamma_v$,
 \item the splitting of the VBM at $\bf{K}$ into states $K_{v1}$ and $K_{v2}$ due to interlayer interaction,
 \item two-fold spin degeneracy of all states due to inversion symmetry, and
 \item nearly parabolic band dispersions at $\Gamma$, $\bf{T}$, and $\bf{K}$.
\end{itemize}
On the $G_0W_0$ level of accuracy the CBM in bulk \MoS ($T_c$ point) is $\sim$0.4 eV lower in energy than the CBE at $K_c$ and
the VBM at $\Gamma$ is favored by $\sim$0.5 eV energy difference with respect to
the band edge at $\bf{K}$. 
In bulk \MoS, the valence band splitting at $\bf{K}$ amounts to roughly 240 meV. 

In contrast to bulk and multi-layer \MoS, the main attribute of the single-layer \MoS~ band structure is the direct fundamental band gap
located at $\bf{K}$ since both, the VBM and the CBM, are at $\bf{K}$.
This direct band gap at $\bf{K}$ has been clearly demonstrated by several preceding works on 
single-layer \MoS \cite{Komsa2012,Ramasubramaniam2012,Yun2012,Cheiwchanchamnangij2012,Shi2013,Molina-Sanchez2013} and explains
the significant increase of photoluminescence yield in single-layer 
\MoS \cite{Mak2010}. 
The other important feature in the band-structure of 1L-\MoS~ is
the splitting of the valence band maximum $K_{v,1}-K_{v,2}$ which amounts 
to $\sim$150 meV. Since inter-layer interactions are absent, this splitting 
must have a origin different from the splitting in bulk.
Indeed, due to missing inversion symmetry, the spin-degeneracy of the bands
is lifted, resulting in a particularly large
spin-orbit splitting at $\bf{K}$\cite{zhu:prb:84,Cheiwchanchamnangij2012}. 
This splitting explains the doublet structure of the strong 
peak in the absorption spectrum of 1L-\MoS~\cite{Mak2010}.

The CBM in 1L-\MoS~ is also located at $\bf{K}$ but the splitting due to SOC is 
one order of magnitude weaker than the splitting of the VBM. Its absolute
value is strongly affected by the exchange correlation functional used in the calculations~\cite{Kuc2015} as will be discussed later.
Both, the valence and conduction bands exhibit nearly parabolic dispersion at this point, which explains the small 
effective charge carrier masses and indicates promising conductivity. 
However, compared to bulk \MoS~ the valence band at $\Gamma$ is considerably flattened. 
This flattening results in a much higher effective hole mass in 1L-\MoS, which
was also observed in Angle-Resolved Photoemission Spectroscopy (ARPES) experiments \cite{Jin2013}.
A second local conduction band minimum close in energy is observed at $\bf{T}$. 
The relative energy positions of the states $K_c$ and $T_c$ and the location of
the VBM (either $\bf{K}$ or $\Gamma$)
determine, whether the material is a direct or an indirect semiconductor. 
We observed that in 1L-\MoS~ the $K_c-T_c$ energy difference is very sensitive to (i) the structural optimization, (ii) the applied in-plane strain, and 
(iii) the $GW$ accuracy (around 0.05 eV). We discuss these issues in more detail later in this section. 

We now describe the changes stemming from quasiparticle corrections
in the band-structures of bulk, single-, and double-layer \MoS.
The most notable change is the sizable increase of the band gap on the level
of the $GW$ method. Also the valence band width increases slightly.
Note that for the single-layer, the VBM at $\Gamma$ is shifted downwards
as compared to the VBM at $\bf{K}$. The conduction band is even more profoundly 
affected. The upshift of the CBM at $K_c$ is larger than the one of the
secondary CBM $T_c$.
In the single-layer, this results in the lowering of the $T_c$ energy relative to $K_c$ and thus in the reduction 
of the $K_c-T_c$ energy difference by 60\% compared to LDA turning the material almost indirect on the $G_0W_0$ level.
In bulk \MoS~ the CBM $T_c$ is lower in energy than $K_c$ 
(due to inter-layer interaction) already on the DFT level. 
On the $G_0W_0$ level, this trend is amplified even more.
In both, double-layer and bulk \MoS, the $GW$ corrections

Besides that, one has also to consider that the $GW$ correction to the band structures slightly depends on the 
number of layers in multi-layer systems. 
We find that the band gap correction at $\bf{K}$ decreases in double-layer and bulk 
with respect to the case of single-layer \MoS. The larger number of layers is associated with
an increase of the dielectric screening, which results in a smaller correction \cite{Wirtz2006,Komsa2012}.

In order to better understand, the origin of the differences between single-, double-, and bulk \MoS ~band structures, we have summarized the orbital 
composition of the highest valence and lowest conduction bands at
points of interest in the Brillouin zone in Table \ref{orbital-table}. 
In single-layer \MoS, the S-$p$ and Mo-$d$ orbitals dominate the composition of the wave functions, with a 
minor contribution of the $s$ orbitals. The conduction band edge at $\bf{K}$ is mainly a Mo-$d_{z^2}$ (86 \%) and 
the remaining part shared between the S-$p_{xy}$ and Mo-$s$ orbital. The valence states at $K_{v1}$ and $K_{v2}$ 
are predominantly Mo-$d_{xy}$ (80 \%) and 20 \% of S-$p_{xy}$ without any
contribution from $s$ orbitals. 
The wave function at the local minimum at $T_c$ has a more complex 
composition, typical for points of low symmetry, as summarized in Table \ref{orbital-table}.
These $GW$ findings qualitatively reproduce previous DFT-PBE results (e.g.,
Fig. 4 in Ref. \cite{Scalise2014}, Fig. 5 in Ref. \cite{Guzman2014})
as well as the tight-binding (TB) model of Liu and co-workers \cite{Liu2013, Liu2015}. The latter, however, suggest also a significant contribution
of the Mo-$d_{x^2-y^2}$ states to the valence band edge at $K_{v}$ which 
we contribute to a deficiency of the TB model using only three Mo-$d$ bands.
The composition of the $\Gamma_c$ and $K_c$ states in bulk \MoS~ is very
similar to the single-layer values. 
The bulk $K_{v1}$ and $K_{v2}$ states however, are now predominantly composed of Mo-$d_{x^2-y^2}$
and the valence band states at $\Gamma_v$ change the weight of
the orbital $p_z$ of sulphur atoms. The latter increase is related to the bonding between sulphur atoms of different layers, 
which produces the interlayer coupling \cite{Cappelluti2013}.

\begin{table*}
\begin{center}
 \caption{Orbital composition of the wave functions at the points $\bf{K}$,
$\bf{T}$ and $\Gamma$, in the case of
single-layer and bulk \MoS.}
 \begin{tabular}{l|ccc|cccccc}
 \hline
 \hline
 \multicolumn{10}{l}{Single-layer} \\
 \hline
  Atom      &\multicolumn{3}{c}{Sulphur} &               \multicolumn{6}{c}{Molybdenum}        	          \\
 Orbital    &  $s$  &   $p_{xy}$ & $p_z$ & $s$  & $d_{x^2-y^2}$ & $d_{yz}$ & $d_{z^2}$ &  $d_{xz}$ &  $d_{xy}$ \\ 
\hline
 $\Gamma_c$ &  -    &    0.54    & -     & -    &      -        & 0.46     &     -     &      -    &     -     \\
 $\Gamma_v$ &  -    &     -      & 0.23  & 0.02 &      -        & -        &    0.75   &      -    &     -     \\
 $K_{c}$    &  -    &    0.09    & -     & 0.05 &      -        & -        &    0.86   &      -    &     -     \\
 $K_{v1}$   &  -    &    0.20    & -     & -    &      -        & -        &     -     &      -    &    0.80   \\
 $K_{v2}$   &  -    &    0.20    & -     & -    &      -        & -        &     -     &      -    &    0.80   \\
 $T_c$      & 0.03  &    0.22    & 0.06  & -    &     0.54      & -        &    0.12   &      -    &    0.01    \\
\hline 
\multicolumn{10}{l}{Bulk} \\
 \hline
  Atom      &\multicolumn{3}{c}{Sulphur} &               \multicolumn{6}{c}{Molybdenum}        	          \\
 Orbital    &  $s$  &   $p_{xy}$ & $p_z$ & $s$  & $d_{x^2-y^2}$ & $d_{yz}$ & $d_{z^2}$ &  $d_{xz}$ &  $d_{xy}$ \\ 
\hline
 $\Gamma_c$ &  -    &    0.53    & -     & -    &      -        &  0.47    &     -     &      -    &     -     \\
 $\Gamma_v$ &  0.07 &     -      & 0.30  & 0.03 &      -        & -        &    0.60   &      -    &     -     \\
 $K_{c}$    &  -    &    0.09    & 0.05  &  -   &      -        & -        &    0.86   &      -    &     -     \\
 $K_{v1}$   &  -    &    0.21    & -     & -    &     0.79      & -        &     -     &      -    &        \\
 $K_{v2}$   &  -    &    0.18    & -     & -    &     0.82      & -        &     -     &      -    &        \\
 $T_c$      &  0.07 &    0.18    & 0.06  & -    &     0.52      & -        &    0.14   &      -    &    0.03     \\
\hline
\hline
 \end{tabular}
 \label{orbital-table}
\end{center}
\end{table*}

\begin{figure}
\begin{center}
\includegraphics[width=7 cm]{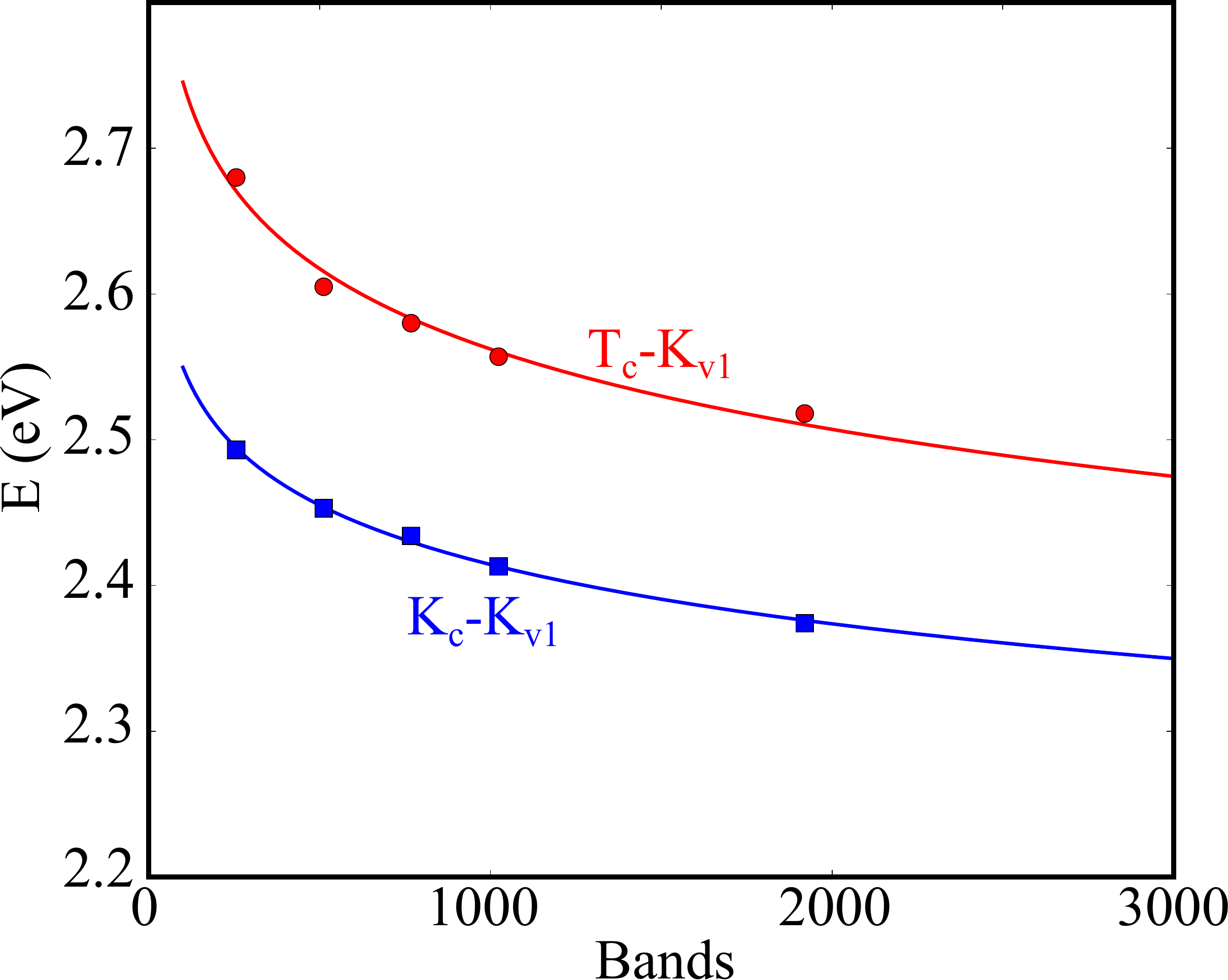}
\end{center}
\caption{Convergence of the direct and indirect quasiparticle energy gap of 1L-\MoS~ calculated within the $G_0W_0$ approximation with respect to the total number of
bands (occupied + virtual) taken into account for the calculation of the screening ($W$). The solid lines represent logarithmic fits and serve as guides to the eye.}
\label{fig:convtest}
\end{figure}

A correct description of the electronic properties requires 
(i) the inclusion of Molybdenum semi-cores states ($4s$ and $4p$ orbitals) in the basis set,
(ii) a plane wave cutoff of 350 eV, 
(iii) at least a $12\times12\times3$ ($12\times12\times1$) $\Gamma$-centered $\kvec$ mesh for bulk (1L) \MoS, and 
(iv) the explicit inclusion of the spin-orbit interaction \cite{Sangalli2012}. We interpolate the band structure to a finer grid using the WANNIER90 
code \cite{mostofi:cpc:178} and the VASP2-WANNIER90 
interface \cite{vaspwannier}. 
With respect to $GW$ calculations, it is important to mention that 
(i) solely including valence electrons leads to an erroneous wave-vector dependence of the $GW$ correction \cite{Molina-Sanchez2013},
(ii) the convergence with respect to virtual states when calculating $W$ is particularly slow for 1L-\MoS \cite{Qiu2013}, and
(iii) the default value for the number of quasiparticle energies that are calculated and updated in the sc$GW$ VASP calculation must be substantially increased. 
While taking into account $\approx$ 500 virtual states is sufficient to converge the quasiparticle gaps of bulk \MoS~ within 20 meV, more than 1000 bands are required for 1L-\MoS~
gaps to be stable within 40 meV. The logarithmic scaling of the direct and indirect gap in 1L-\MoS~ with respect to the number of bands included in the calculation
is shown in Fig. \ref{fig:convtest}. 
Concerning the number of quasiparticle energies that have been updated in the sc$GW$ calculations (NBANDSGW parameter in VASP), it is emphasized that more than 200 
are required to converge the quasiparticle gaps. In particular the conduction
band extremum at point $\bf{T}$ strongly
depends on this parameter.

\subsection{Dependence on the crystal structure}

The analysis of the preceding paragraphs underlines the importance of accurately calculating
the energy difference between the conduction band minima at $K_c$ and $T_c$. This is a challenge for the different theoretical approaches mentioned before, because these quantities also sensitively depend on the details of the crystal
structure. In order to discuss this, we focus on single-layer and 
bulk \MoS~ but the
conclusions can be extended to multi-layer \MoS.

One source of controversy between the results of several calculations performed for 1L-\MoS ~\cite{Komsa2012,Ramasubramaniam2012,Cheiwchanchamnangij2012,Shi2013} 
could be the underlying crystal structure.
In particular, the lattice constant, interlayer distance 
(relevant for multi-layer and bulk \MoS),
and atomic positions defining the interatomic distance (Mo-S bond 
length and S-Mo-S bond angle) may significantly affect the energy 
gaps and band dispersion. The dependence of the \MoS~ band 
structure on the details of the crystal structure has not been addressed so far and will be elucidated in the following.

\begin{figure}
\begin{center}
 \includegraphics[width= 7.6 cm]{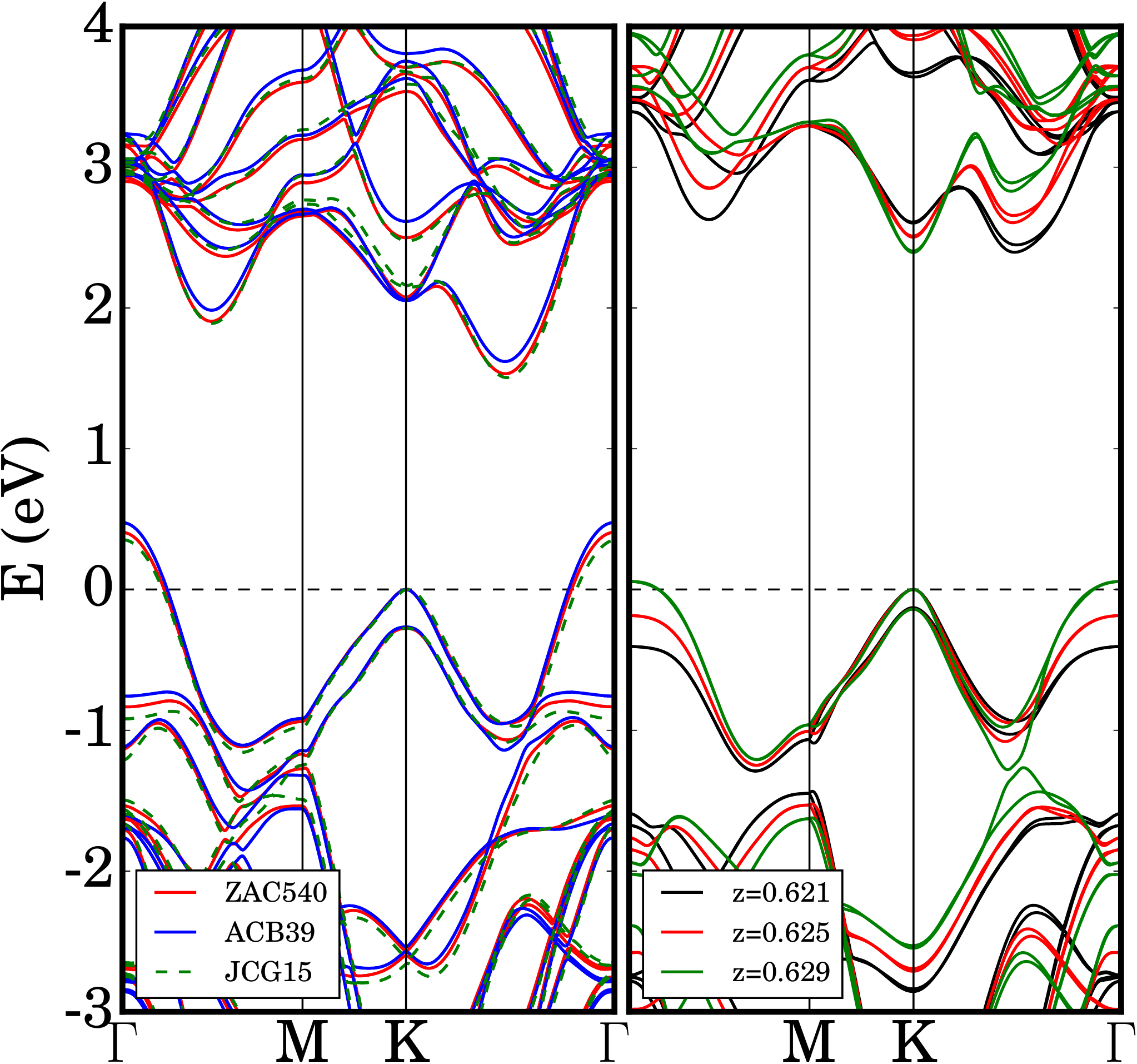}
 \caption{\label{fig:structtest}
(Color online) Band structures of bulk (a) and 
single-layer \MoS~ ~(b) 
calculated within the $G_0W_0$ approximation explicitly 
including SOC for different experimental crystal structure parameters.
The valence band extremum at $\bf{K}$ is aligned at zero 
energy. The abbreviations ACB39
($3.161$ \AA, $c=12.295$ \AA, $z=0.6275$),
JCG15 ($3.14$ \AA, $c=12.327$ \AA, $z=0.621$), ZAC540 ($3.16$ \AA, $c=12.29$ \AA, $z=0.629$) refer to the
crystal structure details published in 
Ref. \cite{schoenfeld:acb:39}, 
Ref. \cite{alhilli:jcg:15}, and 
Ref. \cite{bronsema:zaac:540}, respectively.}
\end{center}
\end{figure}

Most calculations reported so far, have used the experimental room temperature lattice constant of bulk \MoS, \cite{dickinson:jacs:45} 
\ie, $a=3.16$ \AA, and 10-15 \AA ~vacuum along the 
$c$ axis for the single-layer (1L) \MoS~ slab structure. However, less information is given about the choice of the 
origin of the unit cell (atomic positions) and the $z$ parameter. Unfortunately, according to Bronsema \etal ~\cite{bronsema:zaac:540} an inconsistency
exists in literature regarding the choice of the atomic positions and the corresponding $z$ parameter. In fact, the latter
determines the interatomic distances and thus the S-Mo-S layer thickness.
For this reason, the band
structure of bulk and 1L-\MoS~ has been calculated by LDA+SOC and $G_0W_0$+SOC for some of the crystal structures summarized in 
Tab. \ref{tab:lattice}. The $G_0W_0$+SOC results are depicted in Fig. \ref{fig:structtest}.

For bulk \MoS, small variations of the $z$ parameter (compare red and blue
lines in Fig. \ref{fig:structtest}) and the lattice constants $a$ and $c$ 
(compare the red and green lines in Fig. \ref{fig:structtest})
have only little influence on the band-structure, since there is anyway a strong inter-layer interaction that leads to a splitting of the valence
and conduction bands. The situation is quite different for the single-layer:
as illustrated in Fig. \ref{fig:structtest}(b).
The  significant role of the 
internal $z$ parameter that is defining the interatomic 
distances (Mo-S bond length and S-Mo-S bond angle) is revealed. With 
increasing $z$ from 0.621 to 0.629, the Mo-S bond length is 
reduced from 2.42 \AA ~to 2.35 \AA, respectively. This favors 
hybridization between the Mo-$d$ and S-$p$ states that comprise 
the highest occupied and lowest unoccupied bands and therefore 
the band dispersion (band width) increases. As 
a consequence, the CBE at $\bf{T}$  and the VBE at $\Gamma$ are pushed 
to higher energies and the CBE at $\bf{K}$ becomes the CBM giving rise 
to a direct fundamental energy gap. It is strongly emphasized that 
an improper choice of atomic positions and corresponding $z$ parameter 
can fortuitously yield a direct fundamental band gap 
in 1L-\MoS~ and may partly explain the inconsistency among  
$G_0W_0$ band structures\cite{Komsa2012,Ramasubramaniam2012,Cheiwchanchamnangij2012,Shi2013} reported so far.

\begin{figure}
\begin{center}
 \includegraphics[width= 7.6 cm]{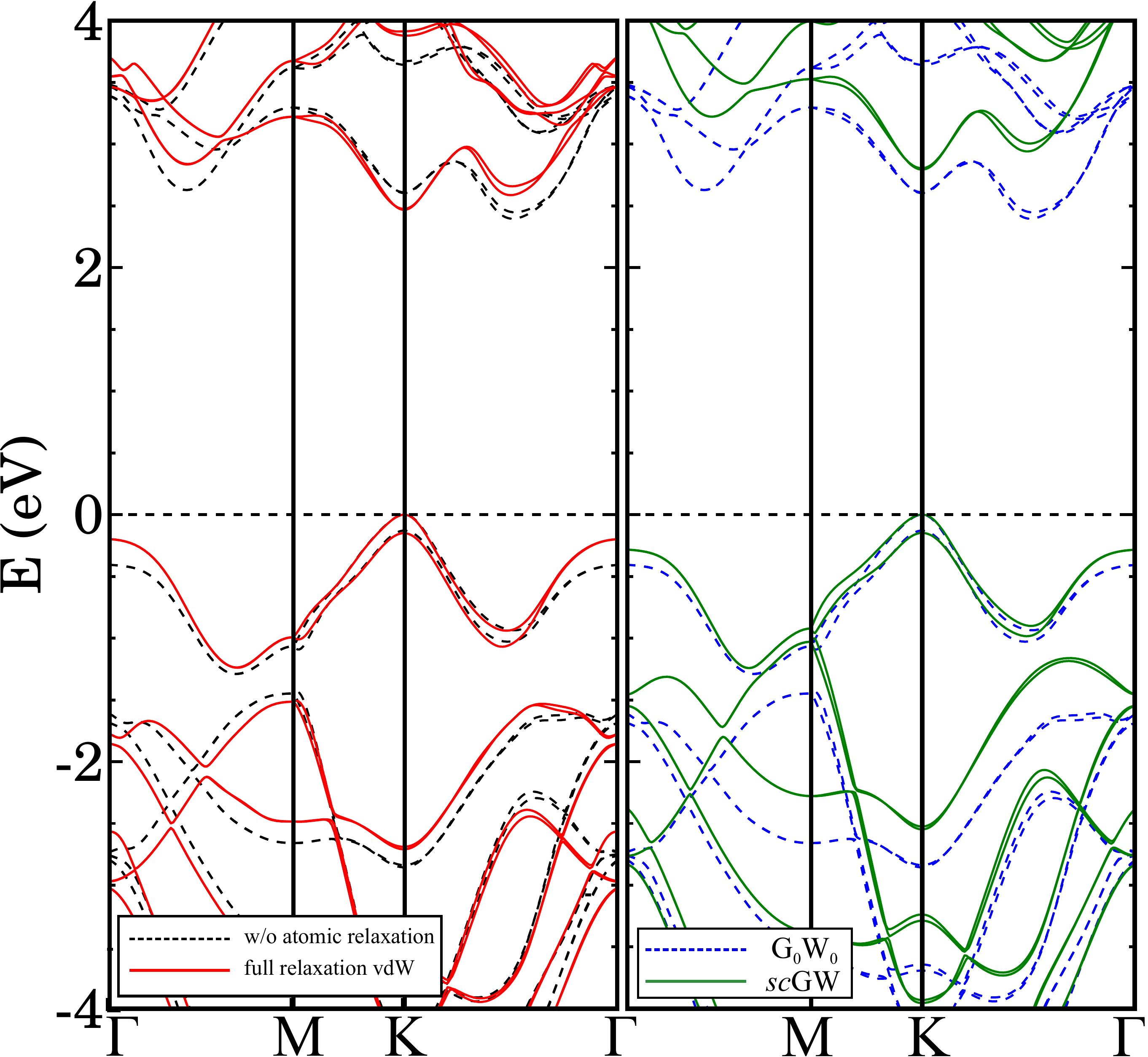}
 \caption{\label{fig:relaxtest}
(Color online) Band structure of 1L-\MoS~ calculated with the experimental bulk lattice constant of $a$=3.1602 \AA. 
In (a) the $G_0W_0$ band structure obtained by omitting the atomic force minimization in the single-layer is
compared to the corresponding results with atomic position relaxation on LDA level. 
In (b) the $G_0W_0$+SOC and the sc$GW$+SOC approach are compared.
The VBM at $\bf{K}$ is set at zero energy. The Fermi level is indicated by the dashed horizontal line.
}
\end{center}
\end{figure}

Another source of discrepancies between single-layer \MoS~ calculations
might be related to the relaxation of the atomic positions.
In Fig. \ref{fig:relaxtest}(a), the $G_0W_0$ band structures of 1L-\MoS~ calculated using the 
experimental bulk unit cell lattice constant ($a$=3.16 \AA) without and with 
LDA-relaxed atomic positions in the single-layer are compared.
Due to the overbinding in DFT-LDA, the Mo-S bond length gets 
reduced with relaxation, which again strengthens the Mo-$d$--S-$p$ hybridization
resulting in an increase of the band dispersion along $\overline{\Gamma K}$.
This results in a raise of the VBM at $\Gamma$ accompanied by
an increase of the conduction band $\bf{T}$ valley energy and the
stabilization of the CBM at $\bf{K}$.
Consequently, $G_0W_0$+SOC yields the correct direct gap band structure for 1L-\MoS,  
if the atomic forces are minimized on the LDA level.
As can be seen from Fig. \ref{fig:relaxtest}(a), the direct gap
at $\bf{K}$ is reduced in the position relaxed case by $\sim$90 meV. 
While the CBM energy at $\bf{T}$ is not affected, further reduction of the direct gap
at $\bf{K}$ by $\sim$40 meV is obtained by using the lattice constant of the
optB86b-VdW  fully relaxed bulk unit cell ($a$=3.164 \AA) and LDA-relaxed atomic positions in the single-layer [not shown in Fig. \ref{fig:relaxtest}(a)]. 

A final test on the level of $G_0W_0$ for the influence of the structural details on the band structure of 1L-\MoS~ was performed with
a fully optB86b-VdW optimized single-layer structure, \ie, the in-plane lattice constant $a$ = 3.162 \AA ~and LDA-relaxed atomic positions. 
Since the VdW interactions are not relevant in the single-layer, the obtained structure is very close to the experimental bulk one. Thus 
the $G_0W_0$+SOC band structure resembles that one calculated without any atomic position relaxation [indirect gap, not shown in Fig. \ref{fig:relaxtest}(a)].
From this analysis we conclude, that the location of the valence and conduction
band extrema at $\Gamma$ and $\bf{K}$
are very sensitive to the relaxation of the atomic positions and
if the atomic positions in the single-layer are relaxed within DFT-LDA, the CBM
at $\bf{K}$ is stabilized with respect the CBE at $\bf{T}$.

The finding of an direct gap 1L-\MoS~ on the $G_0W_0$ level 
seemed to be controversial to the results obtained by Shi \etal ~\cite{Shi2013},
who performed an analogous comparison of $G_0W_0$ and $scGW$ calculations for 1L-\MoS~
as presented in Fig. \ref{fig:relaxtest}(b) and concluded that single-shot $G_0W_0$ 
is insufficient in describing 1L-\MoS. As shown here, omitting the relaxation of the atoms in 
the single-layer structure constructed from the experimental bulk structure leads indeed to
an incorrect description of 1L-\MoS~ on the $G_0W_0$ level (indirect band gap). 
The $scGW$ calculation cures this problem, but further increases the direct gap
at $\bf{K}$.
The tendency of $scGW$ to overestimate semiconductor band gaps is known \cite{shishkin:prl:99}
and thus one must assume that the $scGW$ direct gap of 1L-\MoS~ is too large. 

\subsection{Performance of different methodologies}

\begin{figure*}
\begin{center}
\includegraphics[width=13 cm]{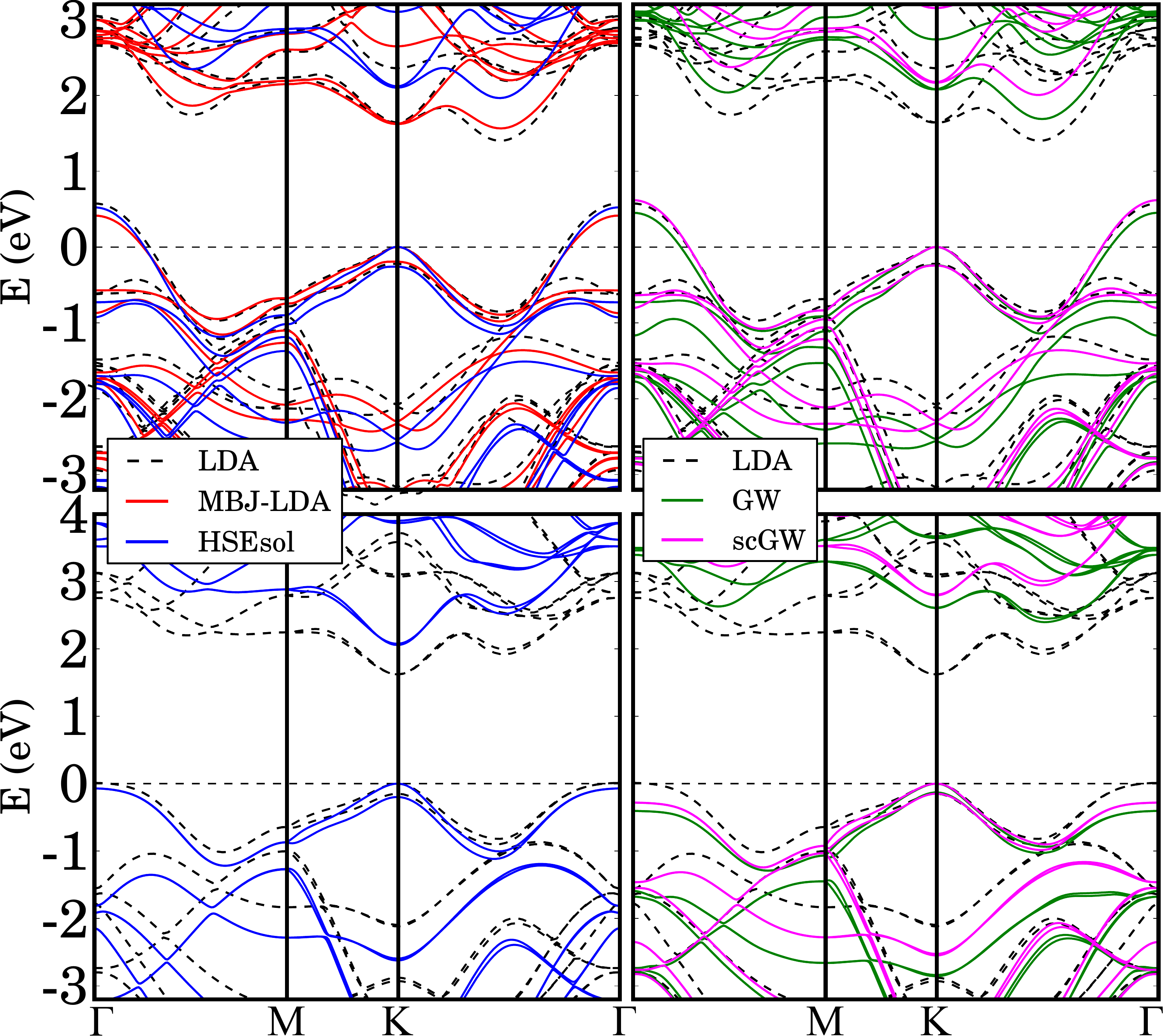}
\caption{(Color online) Band structure calculations of bulk \MoS~ [(a) and (b)] 
and single-layer \MoS~ [(c) and (d)] performed on different 
levels of theory. The valence band extremum at $\bf{K}$ is 
aligned at zero energy.}
\label{fig:methodtest}
\end{center}
\end{figure*}

After the discussion of the relation between structural and electronic properties,
we focus on how the results depend on the XC functionals.
The results of this analysis for bulk, double-, and single-layer \MoS~ are summarized in Table \ref{tab:master} and Fig. \ref{fig:methodtest}.
Note, the fully optimized structures using the van der Waals 
functional as previously described were used in these calculations.

\begin{table*}
\begin{center}
\caption{\label{tab:master}Direct band gaps and interband transitions in 
\MoS ~(in eV) as well as the energy difference between the two lowest conduction band
extrema $K_c-T_c$ calculated on different levels
of theory explicitly including SOC in comparison to available literature 
data listed in brackets.
}
\begin{tabular}{lccccccc}
\hline
\hline
   (eV)    &  $K_c-K_{v1}$  &  $K_c-K_{v2}$ & $K_c-\Gamma_v$  & 
$T_c-K_{v1}$   &  $T_c-\Gamma_v$ & $K_c-T_c$ &  $\Gamma_c-\Gamma_v$\\
\hline
Bulk &  &  &  & & & &\\
\hline
   LDA     &  1.64   &   1.86  &  1.07  &  1.40  &  0.83  & -0.24 & 2.08 \\
   LDA \cite{Peelaers2012} & (1.80) & & & & (0.81) & & \\ 
   PBEsol  &  1.65   &   1.87  &  1.10  &  1.42  &  0.87  & -0.23 & 2.11 \\
   PBE \cite{Yun2012}&   & & & & (0.87) & & \\
   PBE \cite{Peelaers2012} & (1.58) & & & & (0.86) & & \\
   MBJ     &  1.62   &   1.82  &  1.21  &  1.56  &  1.15  & -0.06 & 2.27 \\
   HSEsol  &  2.10   &   2.36  &  1.58  &  1.96  &  1.44  & -0.14 & 2.84 \\
   HSE \cite{Peelaers2012} & (2.16) & & & & (1.48) & & \\
   $G_0W_0$&  2.08   &   2.32  &  1.63  &  1.69  &  1.24  & -0.39 & 2.53 \\
   $G_0W_0$ \cite{Jiang2012} & (2.07)   &  &  &  & (1.23) &  & \\
   $G_0W_0$ \cite{Komsa2012} & (2.00)   &  &  &  & (1.30) &  & \\
   sc$GW$  &  2.17   &   2.41  &  1.59  &  2.02  &  1.44  & -0.16 & 2.88 \\
   sc$GW$ \cite{Cheiwchanchamnangij2012}  & (2.099)  & (2.337)  &  &  & (1.287) & &  \\
   EXPT. \cite{Mak2010}  &  (1.78)   &         &        &       & (1.29)  & &  \\
   EXPT. \cite{Frey1998}  &  (1.95)   &         &        &       & (1.20)  & &  \\
\hline
Single layer &  &  &  & & & &\\
\hline
   LDA     &  1.62   &   1.77  &  1.61  &  1.92  &  1.91  & \phantom{-}0.30 & 2.74 \\
   PBEsol  &  1.65   &   1.79  &  1.69  &  1.89  &  1.93  & \phantom{-}0.24 & 2.78 \\
   PBE \cite{Yun2012}& (1.75)  & & & &  & & \\
   PBE \cite{Ramasubramaniam2012}  & (1.60) &  & & & & & \\
   HSEsol  &  2.09   &   2.28  &  2.23  &  2.45  &  2.59  & \phantom{-}0.36 & 3.63 \\
   HSE     &  2.06   &   2.25  &  2.13  &  2.51  &  2.58  & \phantom{-}0.45 & 3.60\\
   HSE \cite{Ramasubramaniam2012}  & (2.05) &  & & & & & \\
   $G_0W_0$&  2.45   &   2.60  &  2.61  &  2.59  &  2.74  & \phantom{-}0.14 & 3.60 \\
   $G_0W_0$\cite{Ramasubramaniam2012}  & (2.82) &  & & & & & \\
   $G_0W_0$ \cite{Komsa2012} & (2.97)   &  &  &  & (3.26) & & \\
   sc$GW$  &  2.72   &   2.87  &  2.90  &  2.98  &  3.16  & \phantom{-}0.26 & 4.29 \\
   sc$GW_0$ \cite{Shi2013} (w/o SOC)  &  (2.78) &   &   &   &   & & \\
   sc$GW$ \cite{Cheiwchanchamnangij2012}  & (2.759)  & (2.905)  &  &  & &  & \\
   EXPT. \cite{Mak2010} &  (1.88)  & (2.05) & (1.6) &   &  & & \\
\hline
\hline
\end{tabular}
\end{center}
\end{table*}

On all levels of theory the band structure of 
bulk \MoS~ shown in Fig.~\ref{fig:methodtest}(a) and (b) 
corresponds to an indirect semiconductor with 
the VBM located at $\Gamma$ and the CBM at $\bf{T}$. 

When comparing the numbers given in Table \ref{tab:master}, LDA and PBEsol underestimate the $T_c-\Gamma_v$ transition (indirect gap of $\sim$ 0.85 eV) are found to be
in agreement with previous calculations \cite{Yun2012,Ramasubramaniam2012}.
Compared to LDA, the inclusion of local exchange as provided by the MBJ potential mainly affects the $T_c$ and $\Gamma_v$ energies resulting
in a larger indirect $T_c-\Gamma_v$ gap of 1.15 eV. However, the band dispersion
along $\bf{K}$ towards $\Gamma$ is reduced resulting in almost energetically balanced
CBEs at $\bf{K}$ and $\bf{T}$, \ie, $|K_c-T_c|$ is only 60 meV. Therefore the difference between indirect $T_c-\Gamma_v$ and the direct $K_c-K_{v1}$ gap is less pronounced than in LDA.
Further improvement is achieved by taking into account non-local exchange using the HSEsol functional that increases
the indirect as well as the direct $K_c-K_{v1}$ gap. The $|K_c-T_c|$ difference in HSEsol is roughly 140 meV, significantly less compared to LDA ($\sim$ 240 meV). 
Taking into account the dielectric screening in the $G_0W_0$ approach strongly enhances the $|K_c-T_c|$ difference to $\sim$ 390 meV. 
This results in an indirect for bulk \MoS~ of 1.24 eV, in good agreement with experiment (1.2-1.3 eV) as well as previous 
calculations. \cite{Jiang2012, Komsa2012} 
The outstanding agreement of both, the indirect (1.24 eV) and direct (2.08eV) bulk gaps, with the values obtained by an \textit{all-electron} $GW$ code \cite{Jiang2012}
verify the accuracy of the present PAW results.

A higher level of accuracy is reached by performing sc$GW$ calculations. Compared to the $G_0W_0$ band structure, 
the $|K_c-T_c|$ difference is again significantly reduced (to $\sim$ 160 meV), since $T_c$ is pushed up in energy almost back to the HSEsol position.
The fundamental indirect sc$GW$ gap is 1.39 eV and slightly overestimated compared to experiment, which is due neglecting the attractive electron-hole interactions 
via Vertex correction in $GW$. \cite{shishkin:prl:99}
In the $\bf{K}$--$\Gamma$ region, the sc$GW$ band structure resembles the HSEsol,
whereas we observe remarkable differences in the $\Gamma$--$\bf{M}$--$\bf{K}$ range. The $G_0W_0$ and sc$GW$ results are consistent to previous calculations \cite{Cheiwchanchamnangij2012} given in Tab. \ref{tab:master} within the uncertainties  
originating from computational aspects. 

Single-layer \MoS~ is described as a semiconductor with a direct gap at $\bf{K}$
on all levels of theory beyond standard DFT-LDA, provided that the crystal structure is fully relaxed as stressed in Sec. \ref{structure}. 
Standard DFT (LDA and PBEsol) severely underestimates the direct gap of the single-layer structure. 
Besides that, LDA wrongly sets the VBM at $\Gamma$ at slightly higher energy
than the VBE at $\bf{K}$. 
It is important to emphasize that the underestimated $K_{v1}-\Gamma_v$ energy difference 
is observed if the single-layer slab is constructed from the optB86b-VdW relaxed bulk structure (in-plane lattice constant $a$=3.164 \AA),
but not in case of the experimental bulk lattice constant $a$=3.16 \AA.
Therefrom we conclude that the relative positions of the $K_{v1}$ and $\Gamma_v$ energies are strongly dependent on the in-plane lattice
constant. Hence is imperative the investigation of strain effects on the 1L-\MoS~ band structure presented later in this section.

Employing the HSEsol functional to 1L-\MoS~ shifts the conduction bands almost uniformly upwards in energy compared to
DFT-LDA resulting in a rather constant $|K_c-T_c|$, \ie,  $\sim$360 meV and $\sim$300 meV, in HSEsol and LDA, respectively. 
Band dispersions in the valence band region are increased within the HSEsol
description and the VBM splitting at $\bf{K}$ is enhanced to 
$\sim$ 200 meV compared to the LDA value of $\sim$ 150 meV. The VBM at $K_{v1}$ is stabilized by roughly 200 meV compared to $\Gamma_v$ in HSEsol calculations.
Concerning $GW$ approaches, the subtle changes of the band dispersions between
$\Gamma$ and $\bf{K}$ result in a
significant change of the $K_c-T_c$ energy difference: $G_0W_0$ stabilizes the
CBM at $\bf{K}$ by 130 meV, whereas sc$GW$ enhances this energy difference by a factor
of two. Note that the energy differences strongly depend on the total number of bands (NBANDS) included in the $GW$ calculations (see Fig. \ref{fig:convtest}) and the
convergence with NBANDS itself is influenced by the amount of vacuum included in the single layer \MoS~ cell (20 \AA ~in the present case). This means that using 
a larger amount of vacuum requires an increase of the NBANDS parameter as well. For this reason, the comparison between the present results and previously
reported values, as summarized in Tab. \ref{tab:master}, is difficult. The values listed in Tab. \ref{tab:master} refer to $GW$ calculations with NBANDS=512.
Increasing NBANDS from 512 to 1920 reduces the direct gap at $\bf{K}$ by roughly 80 meV, the indirect gap by 60 meV, but the $K_c-T_c$ energy difference increases by
40 meV.

Analogous to bulk \MoS, including non-local exchange by HSEsol increases the $K_c-K_{v1}$ gap (2.09 eV) considerably.
The calculated $K_c-K_{v1}$ quasiparticle $G_0W_0$ gap amounts to 2.45 eV, which is smaller by 0.3-0.5 eV to reported values. \cite{Ramasubramaniam2012,Komsa2012,Liang2013}
This difference is attributed to structural and computational details: A $G_0W_0$ calculation performed with the experimental crystal structure and a 
reduced $\kvec$ mesh of 8$\times$8$\times$2 yields 2.86 eV. 
Liang \etal reported a direct band gap of 1L-\MoS~ of 2.75 eV, \cite{Liang2013} which was obtained by $G_0W_0$ calculations taking into account 
a Coulomb interaction truncation to avoid spurious interlayer interaction between the periodically repeated monolayers, but using the generalized
plasmon-pole model (GPP) for the dynamical screening and omitting SOC. 
The issues of the Coulomb interaction truncation, $k$-point sampling, and vacuum
layer thickness were also addressed by H\"user \etal \cite{Huser2013}, who argued that
the band gap values converged with respect to $k$-point sampling and slab distance are rougly 0.4 eV too small compared to the free standing monolayer (including Coulomb truncation).
Once again this reflects the difficulty to achieve accurate results and explains the plethora of band gap data in literature.

Compared to the single-shot $G_0W_0$ result for the direct $K_c-K_{v1}$ gap, 
the sc$GW$ further increases the gap to 2.72 eV in agreement with previously reported values as listed in Table \ref{tab:master}. 
The slow convergence of the 1L-\MoS~ $GW$ band gaps with the NBANDS parameter was only recently stressed \cite{Qiu2013} and might explain
the larger values previously reported.

At this point it is important to recall that quasiparticle gaps are 
single-particle gaps. Their overestimation by roughly 
0.5 eV compared to some experiment is explained by the missing 
electron-hole interactions (excitonic effects), which are strong 
in 2D materials due to confinement and lead to the formation 
of bound electron-hole pairs. These bound excitons reduce the direct 
band gap by their binding energy and define the optical gap, which 
is experimentally accessed by optical measurements such as 
photoluminescence or photoconductivity. Excitonic effects are 
addressed in the next section. 

To conclude, it is evident based on the above discussion that
the different 
levels of accuracy and/or complexity applied in the $GW$ 
methods substantially alter the results. The 
non-local exchange and dynamical screening are inevitable 
for an accurate description of the electronic properties of \MoS. 
Based on our calculations of single-layer \MoS, a reasonable estimate for the direct band gap 
is 2.4$\pm$0.2eV and the spin-orbit splitting of the valence band edge at
$\bf{K}$ is 150-160 meV. 
The energy difference between the two valence band extrema in
1L-\MoS~ is much smaller than in bulk \MoS~ and
very sensitive to the in-plane lattice constant.
As put forward by Kuc and Heine\cite{Kuc2015}, the estimations of the 
weak spin-orbit splitting of the conduction band edge
is strongly dependent on the XC functional used in the DFT calculation
 and a better description by methods beyond ground state DFT is required.
From our $GW$ calculations, we deduce 
a conduction band splitting of 10 meV, which however falls within the 
estimated uncertainty range.

A final remark concerns the starting point of the $scGW$ calculations.
One should keep in mind that the $scGW$ result can be influenced
by the wave functions (orbitals) used for calculating $G$ and $W$
as pointed out in Ref. \cite{Caruso2012}. Thus, in the complexity of 
the $GW$ method one can go a step forward 
by applying a full self-consistent procedure in the 
Green's function $G$ and the screened interaction $W$, consisting 
in the iterative solution of the Dyson equation. However, the 
extraordinarily cumbersome calculations required for this procedure
restrict the application of this approach to small systems, such
as binary molecules as N$_2$ or CO \cite{Caruso2012}. Up to now 
this scheme has not been applied to single-layer \MoS ~and 
we think that its implementation for layered materials is still 
far.

\subsection{Strain effects in single-layer \MoS}
\label{subsec:strain}

The ideal scenario of free-standing 2D layers as considered in most theoretical
simulations is hardly fulfilled in reality. 
In the course of experiments or device fabrication with 2D materials, 
it is important to consider strain resulting e.g. from the
lattice constant mismatch between the substrate and the 2D layer.
Equally important in this context is the interaction of the
2D material with the substrate as shown in Ref. \cite{Jin2013}.
Therein, the ARPES scans of exfoliated single-layer
\MoS~ compared to those of chemical vapor deposition grown 
single-layer \MoS~ on silicon  
revealed that the presence of substrate alone is sufficient to modify the 
\MoS~ band structure. In particular, the \MoS-substrate interactions 
are responsible for the pronounced flattening 
of the VBM at $\Gamma$ of \MoS~ on silicon.

In addition, recent experiments 
have demonstrated that application of tensile strain
changes the gap from direct to indirect \cite{Conley2013}.
In particular, the \MoS~ flake deposited on a flexible substrate 
which is subsequently deformed in a controllable manner, experiences
uniaxial tensile strain up to 2.2 \%.
The photoluminescence spectra of these samples
show a clear transformation of the band character, 
and an associated reduction of the integrated intensity of the optical
signal. 

The sensitivity of TMDs band 
structure on the lattice constant opens the possibility to 
modify the band gap and thus the optical properties in 
a controlled way by external strain. This issue 
has been theoretically addressed 
either through LDA/GGA calculations \cite{Ellis2011, Scalise2012, Scalise2014, Dong2014}, or 
the $GW$ method \cite{Shi2013}.
The effect of hydrostatic pressure on the vibrational, electronic, and
optical properties of bulk, multi-, and single layer 
\MoS~ was investigated by Nayak \etal \cite{Nayak2014,Nayak2015} by combining various
experiments (high resolution transmission electron microscopy, electrical resistance measurements,
laser Raman spectroscopy, synchrotron X-ray diffraction experiments under high-pressure) with
DFT calculations. Interestingly, while the direct bulk band gap decreases with increasing pressure, the
direct band gap of 1L-\MoS~ increases by 11.7\% up to $\sim$ 12 GPa before it is reduced. Thus the pressure
induced electronic transition from the semiconducting to a semimetallic state occurs at much larger pressures
in the latter. \cite{Nayak2015}

\begin{figure*}
\begin{center}
 \includegraphics[width=15 cm]{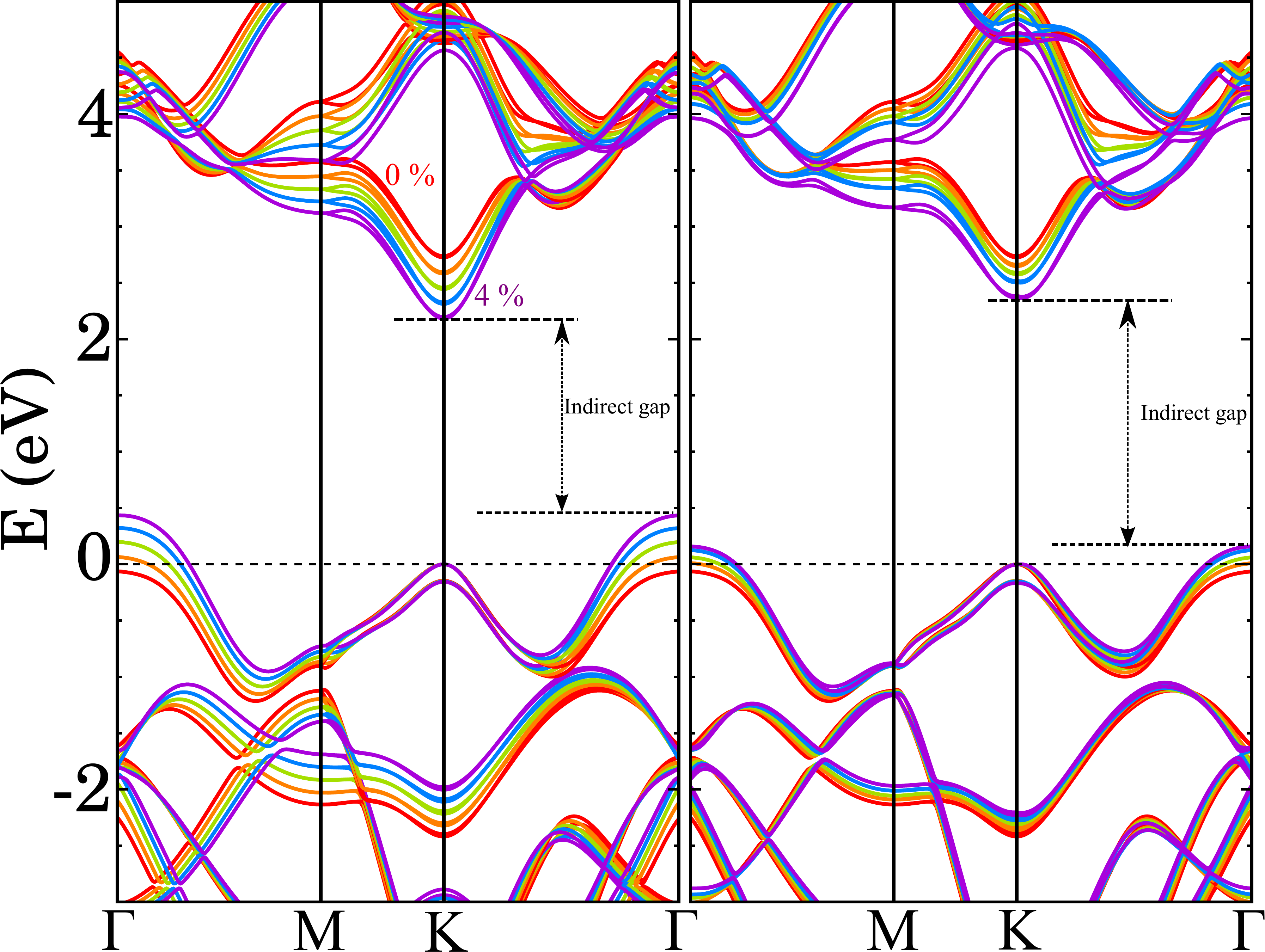}
 \caption{\label{fig:strain}
(Color online) The band structure of 1L-\MoS~ as a function 
of biaxial (left) and uniaxial (right) tensile strain calculated with $G_0W_0$+SOC is depicted.
The VBM at $\bf{K}$ is aligned at zero energy. The unstrained material refers 
to the experimental lattice constant $a$=3.1602 \AA.}
\end{center}
\end{figure*}

Being aware of the importance of substrate interactions, we investigated the 
strain effects on the electronic properties 
of 1L-\MoS ~within the $GW$ approach
and the model of free-standing 2D layers. Biaxial tensile 
strain has been realized by increasing the in-plane lattice 
constant of 1L-\MoS. The band structures of the strained 
materials were calculated  with 
relaxed atomic positions and are shown in Fig. \ref{fig:strain}. 
The direct $K_c-K_{v1}$ gap and interband transitions as 
a function of strain deduced from these band structures are
collected in Tab. \ref{tab:biaxial}.

\begin{table*}
\begin{center}
\caption{The $K_c-K_{v1}$ and $K_c-\Gamma_v$ gap as well as the $K_c-T_c$ energy difference in units of (eV) as a 
function of biaxial tensile strain.} 
\begin{tabular}{lcccccc}          
\hline
\hline
  strain (\%)    &  \multicolumn{3}{c}{$G_0W_0$+SOC} & \multicolumn{3}{c}{sc$GW$+SOC}\\
                 & $K_c-K_{v1}$  & $K_c-\Gamma_v$ (eV)& $K_c-T_c$ & $K_c-K_{v1}$ (eV) & $K_c-\Gamma_v$ (eV) & $K_c-T_c$\\ 
\cline{2-4} \cline{5-7}
  0  &  2.51  &  2.71 & 0.08  & 2.76  & 2.85 & 0.39 \\   
  1  &  2.37  &  2.44 & 0.24  & 2.61  & 2.58 & 0.55 \\
  2  &  2.23  &  2.18 & 0.41  & 2.47  & 2.31 & 0.72 \\
  3  &  2.11  &  1.93 & 0.55  & 2.34  & 2.06 & 0.86 \\
  4  &  1.98  &  1.69 & 0.69  & 2.21  & 1.82 & 0.99 \\
\hline
\hline
\end{tabular}
\label{tab:biaxial}
\end{center}
\end{table*}

\begin{figure}
\begin{center}
 \includegraphics[width=7.6 cm]{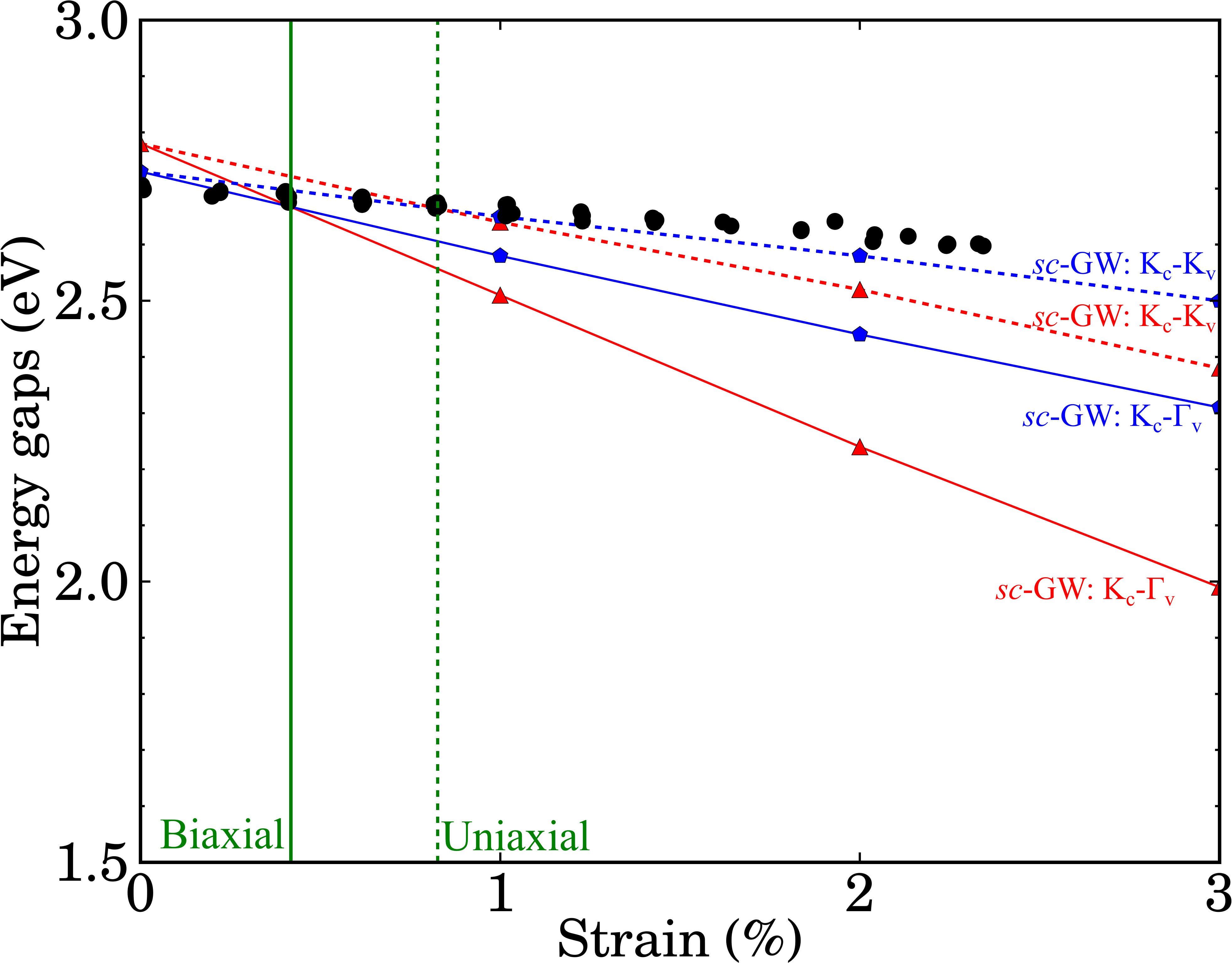}
 \caption{\label{fig:strain_gaps}
(Color online) The $K_c-K_{v1}$ gap and the $K_c-\Gamma_v$ gap as a function of uniaxial [100] (dashed lines) and biaxial (full lines)
tensile strain calculated with $G_0W_0$+SOC.
The obtained transition strain is marked by the vertical full (dashed) lines in green for biaxial and 
uniaxial strain, respectively.
The unstrained material refers to the experimental lattice constant $a$=3.1602 \AA. The experimental points from Ref. \cite{Conley2013} 
are represented by black dots. Note, the experimental data have been constantly shifted upwards in energy, since the electron-hole
interaction is not included in the calculations.}
\end{center}
\end{figure}

With increasing in-plane lattice constant (biaxial tensile strain) 
the bond distances within the $xy$ plane of the Mo-S-Mo sheets are changed, but also in
the perpendicular direction through the relaxation along $z$.
According to Tab. \ref{orbital-table}, the valence states at $\bf{K}$ (VBM) are mainly composed of
S-$p_{xy}$ and Mo-$d_{xy}$ orbitals, whereas the valence states at $\Gamma$ have
predominantly Mo $d_{z^2}$ character. Concerning the CBM at $\bf{K}$, the states are mainly Mo $d_{z^2}$ 
orbitals and the conduction band states around $\bf{T}$ have predominantly Mo $d_{x^2-y^2}$ character.
By changing the S-Mo-S bond lengths and angles due to tensile strain, 
the overlap of the Mo $d_{z^2}$ with the S $p_{xy}$ is reduced, whereas
the coupling between the Mo $d_{xy}$ and S $p_{xy}$ is increased.\cite{Guzman2014}
As a consequence, the $\Gamma_v$  energy raises with respect to $K_v$ and the $K_c$ energy 
decreases compared to $T_c$.
Concomitantly, the $K_c-K_{v1}$ gap decreases, but 
not as fast as the $K_c-\Gamma_v$ gap, which results in a 
transition to an indirect 1L-\MoS~ as illustrated in Fig. \ref{fig:strain_gaps}. Also 
evident in Fig. \ref{fig:strain_gaps} is the 
linear dependence of the band gaps on biaxial tensile strain. 
$G_0W_0$+SOC suggests a transition strain of $\sim$1.6\% (\ie, 3.21 \AA), which is in good agreement with
the value obtained by recent $GW_0$ calculations ($\sim$1.5\%) \cite{Shi2013}.
Thus iterating the QP energies and one-electron 
wave functions in $G$ only, seems to change 
the linear decrease of the band gaps only marginally.
The sc$GW$ approach though, increases the direct $K_c-K_{v1}$ gap
rather constantly (by $\sim$ 0.2 eV) and significantly 
affects the band dispersions. 
The former results in a rigid shift of the direct gap dependence with strain,
whereas the latter changes the $K_c-T_c$ and
$K_{v1}-\Gamma_v$ energy differences (compare data
summarized in Tab. \ref{tab:biaxial}).
Consequently, a much lower 
strain of $\sim$0.7\% (\ie, 3.18 \AA) 
for the direct-to-indirect transition in 1L-\MoS~ is
obtained. 

At this point, the present $GW$ results are also compared to conclusions drawn from previous DFT-PBE calculations.
Scalise \etal \cite{Scalise2012} proposed a stain-induced semiconductor to metal transition in 1L-\MoS~ on the basis of DFT-PBE calculations
omitting SOC. When the slope of the linear fit to their band gap data as a
function of biaxial tensile strain (Fig. 2(b) in Ref. \cite{Scalise2012})
is compared to that deduced from the present $GW$ results, the obtained difference is roughly 30\%., \ie, the DFT-PBE slope is 30\% smaller. However, 
in both cases the closure of the indirect band gap with increasing tensile strain is estimated at (10$\pm$1)\% tensile strain suggesting that the
trend of the electronic properties as a function of strain is well reproduced by standard-DFT.

To compare with recent photoluminescence data \cite{Conley2013}, we also 
calculated the band structure of 1L-\MoS~ as a function of uniaxial tensile
strain along the [100] direction with the $G_0W_0$ approach.
These results are summarized in Tab. \ref{tab:uniaxial}. 
Since the electron-hole interaction 
is not included in the calculations at this stage, the experimental data have been rigidly shifted
in energy for better comparison. Thereby it is assumed that the band gap renormalization due to
excitonic effects does not change for small strains as applied in this case.

\begin{table*}
\begin{center}
\caption{The $K_c-K_{v1}$ and $K_c-\Gamma_v$ gap as well as the $K_c-T_c$ energy difference in units of (eV) as a 
function of uniaxial [100] tensile strain.} 
\begin{tabular}{lcccccc}          
\hline
\hline
  strain (\%)    &  \multicolumn{3}{c}{$G_0W_0$+SOC} & \multicolumn{3}{c}{sc$GW$+SOC}\\
                 & $K_c-K_{v1}$  & $K_c-\Gamma_v$ (eV)& $K_c-T_c$ & $K_c-K_{v1}$ (eV) & $K_c-\Gamma_v$ (eV) & $K_c-T_c$\\ 
\cline{2-4} \cline{5-7}
  0  &  2.51  &  2.71 & 0.08  & 2.85  & 2.76 & 0.39 \\
  1  &  2.43  &  2.56 & 0.19  & 2.58  & 2.56 & 0.53 \\
  2  &  2.37  &  2.44 & 0.27  & 2.47  & 2.49 & 0.60 \\
  3  &  2.29  &  2.31 & 0.36  & 2.33  & 2.42 & 0.69 \\
  4  &  2.21  &  2.17 & 0.46  & 2.20  & 2.35 & 0.77 \\
\hline
\hline
\end{tabular}
\label{tab:uniaxial}
\end{center}
\end{table*}

The effect of the uniaxial strain on the band structure as shown 
in Fig. \ref{fig:strain} is similar to 
that of biaxial strain. Again
the $K_c$ energy decreases, whereas the $\Gamma_v$ energy increases
with increasing uniaxial tensile strain giving rise to 
the direct-indirect gap transition, when $\Gamma_v$ becomes the VBM.
Compared to biaxial strain, the slope of the linear dependence
of the direct and indirect energy gaps is significantly smaller in the 
uniaxial case, \ie, roughly by a factor of two. 
As a consequence, the transition between direct and indirect
single-layer \MoS~ occurs at a larger strain of 3.3\% (equivalent to $a$=3.26\AA), which is roughly
twice as high as in the biaxial case. 
Since the strain induced transition to an indirect gap significantly reduces the photoluminescence
yield, understanding of the strain effects on the 
opto-electronic properties of single-layer \MoS~ is particularly relevant
for technological applications.

\subsection{Tight-binding modelling of single-layer MoS$_2$}

Tight-binding calculations can give further insight on the electronic properties of 
single-layer \MoS. The tight-binding method expands the wave functions in terms of an 
atomic orbital basis, thus giving a simple and
intuitive physical picture of the electronic bands. The atomic orbital
weight of each band state is directly accessible and changes
in the band structure can be attributed to the
change of a single tight-binding parameter.

\begin{figure}
 \includegraphics[width=7.6 cm]{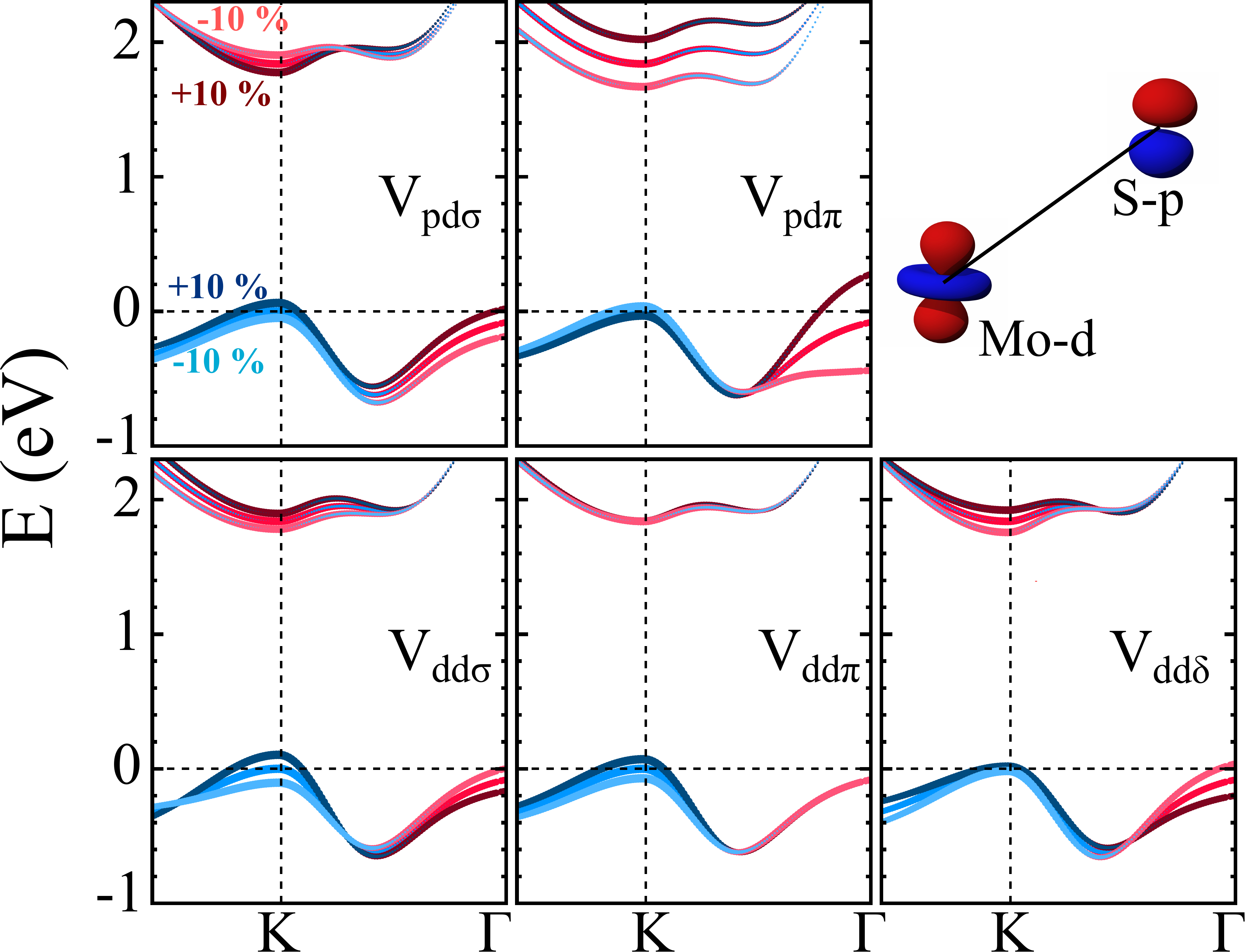}
 \caption{\label{tb-bands} Band structures for different values of the 
tight-binding hopping parameters (indicated in each graph). For each graph, we vary
the indicated parameter $\pm 10$ \%. The red circles are associated with the $d_0$ orbitals
and the blue circles with the $d_2$ orbitals. Their size is related to the orbital weight.}
\end{figure}

Fig. \ref{fig:structtest} shows the change
of the conduction and valence bands with variation of the atomic 
positions. However, it does not tell which orbitals are responsible for such variations. The 
parametrization given in 
Ref. \cite{Cappelluti2013} was used, but for clarity, we only consider $d$-orbitals 
for Mo and $p$-orbitals for S. Interatomic interactions
up to the second nearest neighbors were taken into account. Figure \ref{tb-bands} 
illustrates the band structure altering the hopping parameter $V$ 
while keeping fixed the remaining parameters. The subindices denote the 
kind of orbitals ($p$ or $d$) and the symmetry of the bond, 
$\sigma$,$\pi$, and $delta$ (see details in \cite{Slater1954,Enderlein1997}). 
The size of the red circles indicates the weight of the $d_0$ orbital 
and that of the blue circles the weight of the $d_2$ 
orbitals. We have increased each parameter $\pm 10$\%.

Focusing on nearest neighbor interaction between Mo-S, the 
parameters $V_{pd\sigma}$ and $V_{pd\pi}$ carry the effect of a 
vertical displacement of the S atoms. In the \textit{ab initio}
calculations shown in Fig. \ref{fig:structtest} we have seen the total 
effect of the displacement on the band structure, but with the 
tight-binding formalism we can assign interactions to bands. The 
change in $V_{pd\sigma}$ modifies
the relative positions of the 
two local minima of the conduction band, $K_c$
and $T_c$, producing almost a rigid shift of the valence band. In 
the case of the hopping parameter $V_{pd\pi}$, the effect is rather
different, giving a rigid shift of the conduction band at $K_c$ 
and $T_c$. For the valence band, the dispersion at $\Gamma_v$ is
strongly shifted to higher energies for $+10$\% and to lower
energies for $-10$\% $V_{pd\pi}$. The combination of these two
weights is equivalent to displacing vertically 
the S atoms. In the case of the hopping
parameters related with the $d$ orbitals of Mo atoms, the results
are more complex. This can be seen for the cases of $V_{dd\sigma}$
and $V_{dd\delta}$ in Fig. \ref{tb-bands}. 
Only for the  $V_{dd\pi}$ case a minor influence on the band structure is obtained. 
The latter three parameters are also important in the change of the lattice parameter. The analysis in terms
of the tight-binding model establishes the rather complex nature of covalent bonding formed
by $p-d$ orbitals in TMDs. Note that a proper description of the conduction band
would require a tight-binding model with more parameters to capture the interaction
with farer neighbors and to enlarge the basis of atomic orbitals \cite{Zahid2013}. 

\subsection{Effective charge carrier masses}

Electron and hole carrier mobility is inversely proportional to their effective masses. The strain effects on effective mass
are an important issue from the technological point of view. The effective electron and hole masses 
of bulk \MoS~ as a function of strain were studied on the DFT-HSE level by Peelaers and
Van de Walle \cite{Peelaers2012} and on DFT-LDA level by Scalise \etal.\cite{Scalise2014}
For the sake of completeness, the values for the bulk effective masses given in Ref. \cite{Peelaers2012}
are briefly summarized here. As pointed out in Ref. \cite{Cheiwchanchamnangij2012} and evident from Tab. III of
Ref. \cite{Peelaers2012}, the valleys as well as edges at $\bf{K}$ exhibit rather isotropic effective masses 
concerning both longitudinal and transversal directions. This justifies averaging over both
directions as suggested in Ref. \cite{Ramasubramaniam2012}. However, the longitudinal and transversal
masses are quite different at the $\bf{T}$ point 
(the same as $\Lambda$ in Ref. \cite{Peelaers2012} and $\Sigma_{min}$ in Ref. \cite{Cheiwchanchamnangij2012})
and the $\Gamma$ point, which are the CBM and VBM in bulk \MoS, respectively. 
The value given in Ref. \cite{Peelaers2012} for the transversal hole mass (\ie, $\Gamma$[$\Sigma$] in Ref. \cite{Peelaers2012}) 
is 0.62$m_0$, where  $m_0$  is the electron rest mass. This number is in good agreement with the present result of 0.64$m_0$, evaluated
by a parabolic fit of the Wannier interpolated $G_0W_0$+SOC band structure, calculated with the experimental structure data 
according to $E = \frac{\hbar^2 k^2}{2 m_0 m_{t}}$ in the $\Gamma M$ direction (equal to $\Gamma K$). 
For the optB86b-VdW relaxed bulk, we obtain 0.69$m_0$ and 0.70$m_0$ on the $G_0W_0$+SOC and sc$GW$+SOC level, respectively, indicating that the
effective masses are well described on the $G_0W_0$+SOC level.
The longitudinal hole mass given in Ref. \cite{Peelaers2012} is 0.80$m_0$ reflecting the anisotropy between transveral and longitudinal hole masses. 
The corresponding estimate from the $G_0W_0$+SOC band structure of the optB86b-VdW relaxed bulk is 1.05$m_0$ and 1.03$m_0$ for the experimental bulk structure.
This 20\% overestimation  can be explained by the neglect of the spin-orbit interaction in the DFT-HSE calculations of Ref. \cite{Peelaers2012}.

The value given in Ref. \cite{Peelaers2012} for the transversal electron mass (\ie, $\Lambda_{min}$[$\Lambda$] in Ref. \cite{Peelaers2012})
is 0.53$m_0$ that is close to the present calculations, which yield 0.58$m_0$ at
$\bf{T}$ (through fitting of $E(\kvec)$ along $TK$ or $T\Gamma$). 
When evaluating the effective masses for both, electron and hole, at the $\bf{K}$ point, their longitudinal and transversal component are strongly isotropic. 
On the $G_0W_0$+SOC level, an average hole mass $\overline{m_h}$ of 0.40$m_0$ and an average electron mass $\overline{m_e}$ of 0.63$m_0$ is obtained. 
Performing $G_0W_0$+SOC (sc$GW$+SOC) calculations with the optB86b-VdW optimized bulk structure yields $\overline{m_h}$ = 0.39$m_0$ ($\overline{m_h}$ = 0.40$m_0$)
and $\overline{m_e}$ = 0.52$m_0$ ($\overline{m_e}$ = 0.52$m_0$), for the average hole and electron masses, respectively. 
For comparison, the corresponding values reported in Ref. \cite{Peelaers2012} are 0.45$m_0$ and 0.46$m_0$ for $\overline{m_h}$ and $\overline{m_e}$, 
respectively. Despite the different theoretical approaches, \ie, DFT with the HSE functional omitting spin-orbit coupling in Ref. \cite{Peelaers2012}
compared to $GW$ calculations including SOC in the present work, the overall agreement between the results is good.

In the following, we focus on single-layer \MoS~ and present the effective
electron and hole masses at the $\bf{K}$ point for the longitudinal and transversal directions
in Tab. \ref{tab:effmass}. For comparison with available literature data, the average of the longitudinal and transversal component of the electron ($\overline{m_e}$)
as well as hole mass ($\overline{m_h}$) are included too. Note that both, $G_0W_0$ and sc$GW$ calculations were performed with the optB86b-VdW optimized
in-plane lattice constant and LDA-relaxed atomic positions. The values corresponding to the single-layer constructed from the experimental bulk
structure are given for comparison in brackets in Tab. \ref{tab:effmass}.
As emphasized by Shi \etal ~\cite{Shi2013}, the effective masses are sensitive to strain (\ie, the in-plane lattice constant), spin-orbit interaction, the $GW$ accuracy
as well as the convergence criteria of the $GW$ calculations, in particular the $\kvec$ point sampling. 
It is difficult to disentangle these dependencies and the values summarized in Tab. \ref{tab:effmass} exhibit some spread.
However they are consistent within the order of magnitude.
The average values obtained in the present work for the effective electron mass $\overline{m_e}$ range from 0.35-0.40$m_0$, whereas average
effective hole masses $\overline{m_h}$ between 0.43 and 0.49$m_0$ have been estimated. 

In contrast to the electron masses, the anisotropy between longitudinal and transversal 
hole masses is strongly pronounced, thus yielding a larger uncertainty
range for the average value. This may be explained by the significance of spin-orbit interactions: 
SOC affects the valence band dispersion at $\bf{K}$ more strongly than the CBM \cite{Shi2013}. A common observation is, 
that the electron mass is slightly smaller than the effective hole mass. As shown in the previous section, 
tensile strain shifts the valence band maximum to the $\Gamma$ point.  At this point the 
band dispersion is significantly smaller than at the $\bf{K}$ point resulting
in much larger effective hole masses (2-3$m_0$ estimated at 2\% biaxial tensile strain) and 
therefore decreased hole mobilities.

\begin{table*}
\begin{center}
\caption{Values of  1L-\MoS~ effective charge carrier masses in units of the electron rest mass $m_0$ evaluated
from parabolic fits of the valence and conduction band edges. The subscripts refer to the
longitudinal ($l$) and transversal ($t$) directions that are further specified by the points given
in parenthesis. The first one denotes the location of the
band extremum, whereas the second defines the direction from that point.
As suggested in Ref.\cite{Ramasubramaniam2012}, the average effective masses ($\overline{m_e}$ and $\overline{m_h}$)
determined from the longitudinal and transversal directions are also included. The values obtained from
band structures calculated with the experimental in-plane lattice constant are given in brackets.} 
\begin{tabular}{lcccccc}          
\hline
\hline
  Single-layer    &  \multicolumn{3}{c}{electron} & \multicolumn{3}{c}{hole}\\
                 & $m_{l}$ ($K-\Gamma$)  & $m_{t}$ ($K-M$) & $\overline{m_e}$ & $m_{l}$ ($K-\Gamma$) & $m_{t}$ ($K-M$) & $\overline{m_h}$\\ 
\cline{2-4} \cline{5-7}
  $G_0W_0$+SOC  &  0.33  &  0.36 & 0.35  & 0.40  & 0.50 & 0.45 \\
            &  [0.36]  &  [0.36] & [0.36]  & [0.35]  & [0.50] & [0.43] \\
  $G_0W_0$+SOC \cite{Ramasubramaniam2012} &    &   & [0.60]  &   &   & [0.54] \\
  HSE \cite{Peelaers2012} &  [0.37]  & [0.38] & [0.38]  & [0.44] & [0.48]  & [0.46] \\
  $G_1W_0$+SOC \cite{Qiu2013} &    &   & 0.37  &   &   & 0.21 \\
  sc$GW$+SOC &  0.39  &  0.42 & 0.40  & 0.42  & 0.56 & 0.49 \\
            &  [0.33]  &  [0.36] & [0.35]  & [0.38]  & [0.50] & [0.44] \\
  sc$GW$+SOC \cite{Cheiwchanchamnangij2012}   &  [0.34]  & [0.35]  & [0.35]  & [0.46]  & [0.43] & [0.44] \\        
  sc$GW_0$ \cite{Shi2013}  &    &   & 0.36  &   &   & 0.39 \\
  sc$GW_0$ \cite{Shi2013}  &    &   & [0.32]  &   &   & [0.37] \\ 
\hline
\hline
\end{tabular}
\label{tab:effmass}
\end{center}
\end{table*}

\section{Optical properties}
\label{exciton}

In semiconductors like \MoS, electron-hole pairs interact via Coulomb attraction, forming
excitons. Excitonic effects determine the optical properties of
\MoS\cite{Neville1976,Coehoorn1987,Coehoorn1987a,Splendiani2010,Mak2010}. For example,
experiments like photoluminescence \cite{Korn2011,Shi2013a} and second 
harmonic generation \cite{Kumar2013,Malard2013} are strongly influenced by excitonic
effects. The most common excitonic effects are a red-shift in the optical
gap (with respect to the quasiparticle gap) and, in some cases, 
a radical change in the optical spectra shape with respect to the 
independent particle spectra. This is in particularly the case when bound excitons (absorption peaks below
the onset of the continuum) are formed. It has been shown before for hexagonal boron nitride (a prototypical
wide-band gap layered material), that the anisotropic dielectric constant and the layered, quasi 2D
confinement of excitons, leads to very strongly bound
excitons\cite{Arnaud2006,Wirtz2006,Wirtz_comment}.
Also in \MoS, there is a series of strongly bound excitons \cite{Komsa2012,Komsa2013,Molina-Sanchez2013,Qiu2013}(albeit with comparatively lower binding energies).
Their binding energy depends on the number of layers (and the inter-layer distance in the case of a 
periodic supercell calculation for single-layers). These excitons determine the shape
of the optical absorption spectra as will be explained in this section.

We analyze the optical properties of \MoS~multi-layers in the framework of the 
Bethe-Salpeter equation. In addition, 
we compare \MoS~single-layer optical
properties with other transition metal dichalcogenides. Finally, we discuss 
results obtained applying empirical model Hamiltonians~\cite{Berghauser2014}.

\subsection{Bethe-Salpeter equation formalism}

The Bethe-Salpeter equation (BSE) formalism gives an accurate description of 
the electron-hole interaction \cite{Onida2002}. BSE is based on many-body 
perturbation theory \cite{Rohlfing2000}. Starting from the 
eigenvalues and wave functions of the system, obtained by \textit{ab initio} methods, BSE gives 
the dielectric function and the excitonic binding energy without introducing any additional parameter  
\cite{Strinati1982,Strinati1984,Rohlfing2000,Onida2002,Marini2009}. BSE can be written as:

\begin{equation}
(\varepsilon_{c{\bf k}}-\varepsilon_{v{\bf k}}) A^X_{vc{\bf k}} +
\sum_{{\bf k'}v'c'} \left\langle \xi_{vc{\bf k}}
|\mathbb{K}_{eh}|\xi_{v'c'\bf k'}\right\rangle A^X_{v'c'{\bf k'}} =
\Upomega^X A^X_{vc{\bf k}}.
\label{bethesalpeter}
\end{equation}

The electronic excitations are expressed in the basis of electron-hole pairs, $\xi_{vc{\bf k}}$. We assume vertical 
excitations at $\bf{k}$, from a valence-band state with 
quasiparticle energy $\varepsilon_{v{\bf k}}$, to a conduction-band state with energy $\varepsilon_{c{\bf k}}$.
$A^X_{vc{\bf k}}$ denote the expansion coefficients of the excitons and $\Upomega^X$
is the exciton energy. The 
interaction kernel $\mathbb{K}_{eh}$ describes the screened Coulomb and the exchange interaction between electrons and holes, which includes local field effects \cite{Onida2002,Marini2009}. In absence of
electron-hole interaction $\mathbb{K}_{eh}=0$. In this review, we consider only interband
transitions. This is consistent with the experimental data, where the 
energy of optical excitations is
always above the band gap value. Another important physical aspect is 
the omission of phonon-mediated transitions. They 
are important in indirect semiconductors, especially 
in the study of photoluminescence \cite{Marini2008}. We 
focus mainly on the optical absorption spectra, where the weight of direct 
transitions is much higher than
indirect transitions.

The exciton wave function, expressed in the basis of the electron-hole pairs

\begin{equation}
\Phi^X(\bm{r}_e,\bm{r}_h) = \sum_{{\bf k}vc}A^X_{vc{\bf k}}\phi_{v{\bf k}}(\bm{r}_h)\phi_{c{\bf k}}(\bm{r}_e)
\end{equation}

is a function of six coordinates, where $\bm{r}_e$ and $\bm{r}_h$ are the spatial coordinates of electron and hole. $\phi(\bm{r})$ are the Kohm-Sham orbitals. We can define useful magnitudes from the exciton 
wave function. The weight of a transition $v\rightarrow c$ is
defined as the sum over all $\bf k$

\begin{equation}
 w^X_{vc} = \sum_{\bf k}A^X_{vc{\bf k}}.
\end{equation}

Analogously, we define the weight of each $\bf k$, by summing over all transitions:

\begin{equation}
 w^X_{\bf k} = \sum_{vc}A^X_{vc{\bf k}}.
\end{equation}

The amplitude of electron-hole
pairs that compose each exciton, as a function of the transition energy is

\begin{equation}
g^X(\omega) = \sum_{vc {\bf k}} \left|\left\langle \xi_{vc{\bf k}}|\Phi^X \right\rangle\right|^2 \delta(\omega-\omega_{vc \bf k}).
\end{equation}

Finally, the optical absorption spectrum is the imaginary part of the dielectric function, 
$\varepsilon(\hbar\omega)$, written as

\begin{equation}
 \varepsilon_2(\hbar\omega) \propto \sum_{X}\left|\sum_{{\bf k}vc} A^X_{vc{\bf k}} \frac{\left\langle \phi_{c{\bf k}}
|p_i|\phi_{v\bf k}\right\rangle }{\epsilon_{c \bf}-\epsilon_{v \bf}}  \right|^2\delta(\Upomega^X-\hbar\omega-\Gamma),
\end{equation}
where $\left\langle \phi_{c{\bf k}}|p_i|\phi_{v\bf k}\right\rangle$ are the dipole matrix elements of the transitions
$v \quad c$.  The vector $\bm{p}$  represents the light polarization. We restrict to light linearly polarized
in the basal plane. The polarization perpendicular to the basal plane of MoS$_2$ has a negligible 
contribution to absorption for energies close to the band gap.

Realistic results are only possible by performing an adequate 
convergence. In BSE calculations, we have to check carefully the number
of valence and conduction bands, as well the $\bf k$-point mesh. Coarse $\bf k$-grids tend
to overestimate the exciton binding energy. The building blocks of the BSE kernel,
$\mathbb{K}_{eh}$,  are the screened Coulomb
and the exchange interaction. Therefore, the dielectric function which enters in the Coulomb interaction has also to be 
converged with the number of bands and the $\bf{k}$-point grid (see Refs. \cite{Marini2009,Deslippe2012} for details). 

The supercell geometry is also an important factor in BSE calculations. The long-range Coulomb 
interaction between replicas decreases slowly with distance. Consequently, GW and BSE 
corrections converge also slowly with the separation between replicas \cite{Wirtz2006,Hueser2013}. In general,
these corrections have opposite sign and partially cancel each other, and the total correction 
of the band gap is close to the experimental value. However, exact determination of the 
exciton binding energy requires overcoming this problem. An 
efficient technique is truncating the Coulomb potential (or Coulomb cut-off), simulating an 
infinite distance between replicas \cite{Rozzi2006}. In single-layer \MoS, both GW band gap 
and exciton binding energy increases notably altough not their difference. A drawback
of the Coulomb cut-off technique is the
slower convergence with respect to the $\bf k$-point grid \cite{Huser2013}. 

\begin{figure}
\begin{center}
\includegraphics[width=7.6 cm]{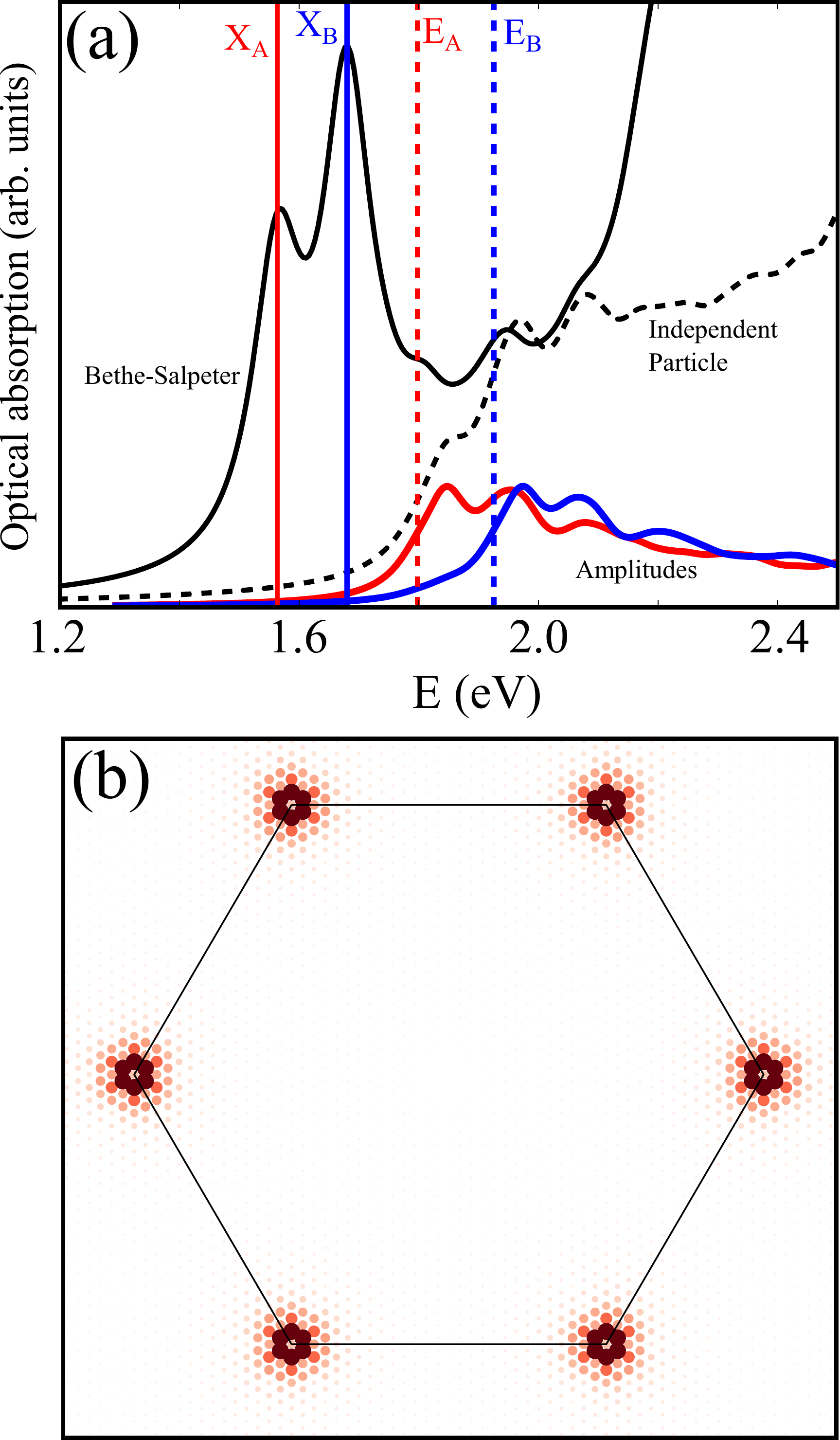}
\end{center}
\caption{(a) Optical absorption with/without 
(solid/dashed lines) electron-hole interaction, together with the amplitude $g(\omega)$ of excitons $X_A$ 
and $X_B$. Absorption thresholds are indicated by vertical solid/dashed lines. (b) Weigth 
$w_{\bf k}^{X}$ of the exciton $X_A$.} 
\label{exciton-kmap}
\end{figure}

Figure \ref{exciton-kmap}(a) shows a typical BSE calculation for single-layer \MoS. Absoprtion is 
depicted with and without excitonic effects (solid and dashed black curves). Electron-hole
interaction down-shifts the absorption threshold and increases the absorption 
coefficient. Amplitude functions $g^{X_{A}}(\omega)$ and $g^{X_{B}}(\omega)$ show
the typical profile of an exciton built from a transition between nearly
parabolic bands. Contribution decays for increasing energy, with the maximum close to the 
band edges. Panel (b) of Figure \ref{exciton-kmap} represents the weight $w_{\bf k}^{X_A}$ of
the exciton $A$ (for exciton $B$ we obtain a similar result). The contribution is localized
at $\bf{K}$ and the $\bf k$-grid must be fine enough to describe accurately 
excitons $A$ and $B$ \cite{Molina-Sanchez2013,Qiu2013}. The distribution of the weight $w_{\bf k}^{X_A}$
reflects the importance of a proper convergence of the $\bf k$-grid to obtain realistic calculations.

In the following, we discuss the calculations of Ref. \cite{Molina-Sanchez2013}. The $\bf k$-grids
are $51\times 51\times 1$ (for single- and double-layer), and 
$21\times 21\times 3$ for bulk. We have included the bands in
the energy window from -3 to 3 eV.

\begin{figure}
\begin{center}
\includegraphics[width=7.6 cm]{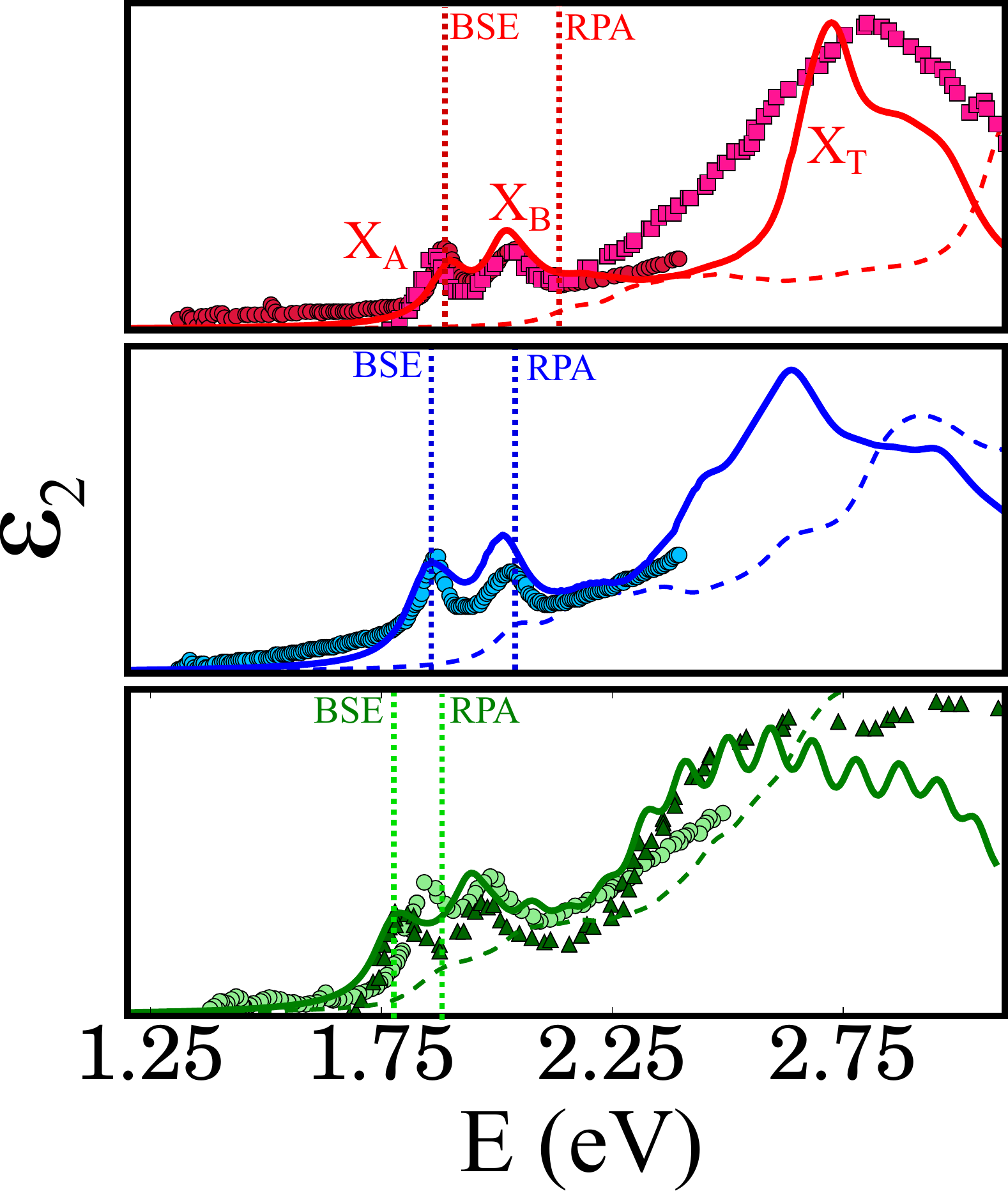}
\end{center}
\caption{Optical spectra for single-, double-layer and bulk \MoS, BSE (solid lines) and
RPA (dashed lines). The experimental data has been collected from several publications, red 
squares\cite{Malard2013}, red and blue circles\cite{Mak2010}, green circles\cite{Splendiani2010}
and green triangles \cite{Stacy1985}.}
\label{bse-spectra}
\end{figure}

Figure \ref{bse-spectra} shows the optical absorption for single-layer, double-layer and
bulk \MoS~(solid lines). For comparison, we have included the independent-particle absorption spectra, without the electron-hole interaction, obtained in 
the random-phase approximation (RPA, dashed lines). The starting point of BSE is the GW eigenvalues
and the LDA wave functions. We have drawn the experimental optical absorption (dots). We have rigidly shifted the 
theoretical spectra to match with the experimental points. The discrepancy is around 0.2 eV, within the error margin
of GW and BSE calculations. Nonetheless, BSE describes well the main trends of the spectra. We remark
the agreement in the absorption threshold, where BSE spectra reproduce accurately the spin-orbit splitting. The 
theoretical absorption at high energies also matches with the experiments. These high-energy peaks come from transitions located
around the $\Gamma$ point.

We can also make a comparative analysis of single-layer, double-layer and bulk \MoS. The spectra have the same line-shape
at the absorption threshold, two peaks that correspond to excitons $X_A$ and $X_B$, followed by a plateau. The differences
arise from the exciton binding energy, which decreases with the number of layers. The reason of such decreasing 
is related to a larger dielectric screening in double-layer and bulk. The high-energy exciton
exhibits a sharp peak in the case of single-layer MoS$_2$, and it becomes difficult to distinguish in double-layer and bulk. Experimental observation
confirms the latter result, in which we observe a broad absorption for bulk MoS$_2$, in contrast to the relatively narrow peak of 
single-layer \MoS. Putting together the theoretical and experimental data
we can deduce that the interlayer interaction affects mainly exciton $X_A$ and $X_B$. This conclusion agrees with the study
of MoS$_2$/WS$_2$ heterostructures of Ref. \cite{Komsa2013}: the optical 
spectra of MoS$_2$/WS$_2$ is the addition of the spectra of independent single-layers
of MoS$_2$ and WS$_2$, and not the combination of spectra modified by a strong interlayer coupling. Inspecting the
exciton wave functions we can get a better insight into the interlayer coupling.

\begin{figure*}
\begin{center}
\includegraphics[width=15 cm]{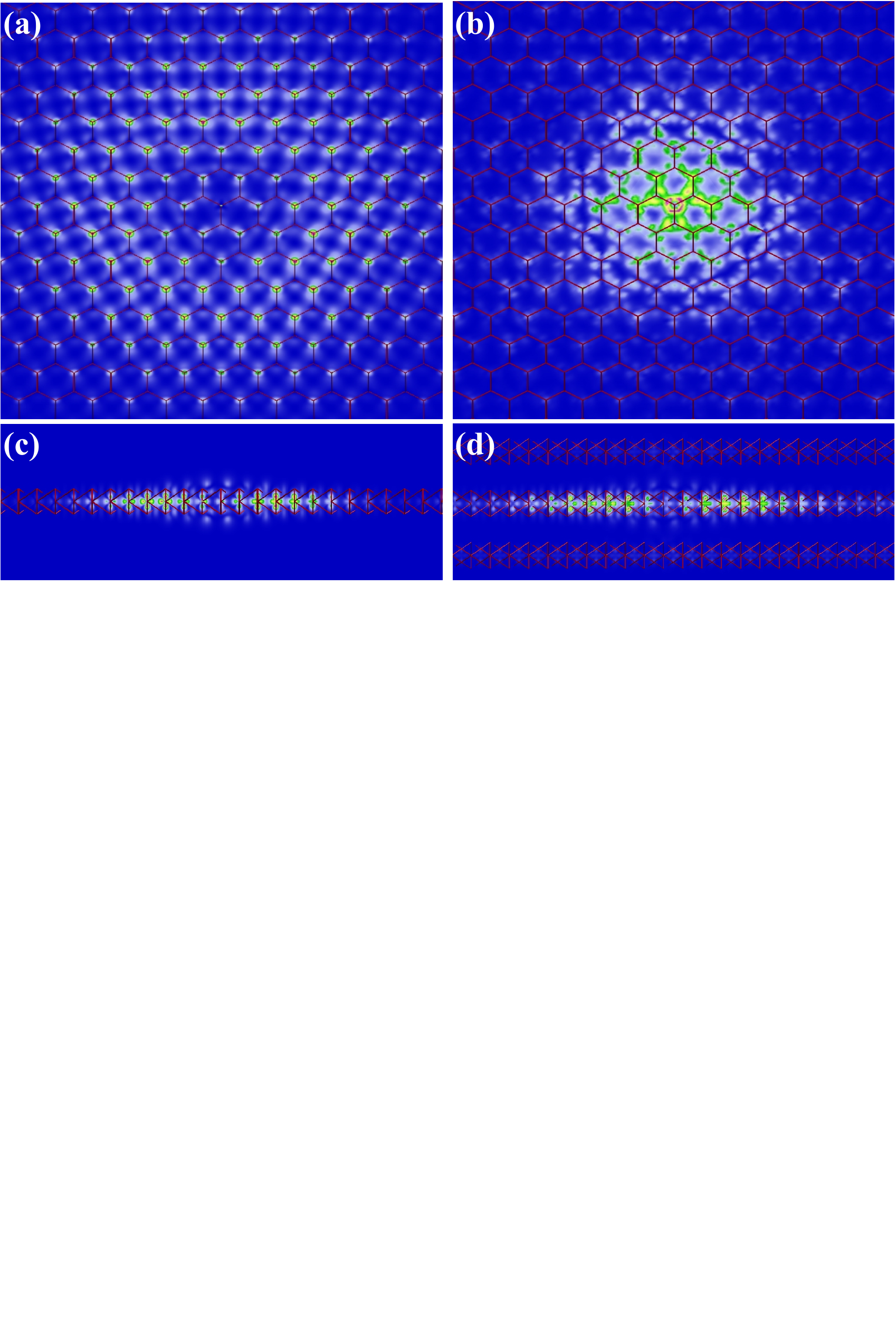}
\end{center}
\caption{Top view of excitons (a) $X_A$ and (b) $X_H$ in 
single-layer \MoS. Lateral view of the exciton $X_A$ 
for (c) single-layer and (d) bulk \MoS.}
\label{exciton-wf-2D}
\end{figure*}

The intensity of the excitonic peaks is directly related to the spatial localization of the 
wave function. Figure \ref{exciton-wf-2D} shows the exciton wave functions for the excitons
(a) $X_A$ and (b) $X_H$ for the case of single-layer \MoS. We have placed the hole at the 
Mo atom. This makes sense as the valence band states at $\bf{K}$ are composed primarily by
Mo $d$-orbitals (see Table \ref{orbital-table}). The exciton $X_A$ is extended more 
than 50 \AA, in consonance with the localization at the momentum space (see 
Fig. \ref{exciton-kmap}). Exciton $X_B$ has an identical wave function (not shown here). On the 
contrary, the high-energy exciton is localized in a few unit cells, in less that 30 \AA. The exciton $X_H$ 
is built from a contribution of a larger set of $\bf k$ points than in the case of $X_A$. We will discuss
the features of this exciton below.

By looking at the exciton wave function in a plane parallel to the vertical axis we 
can learn more details about the interlayer coupling. Figures \ref{exciton-wf-2D}(c) and (d) show the 
lateral view of the $X_A$ exciton in single-layer and bulk, respectively. The exciton density lies
mainly on the Mo atoms for the single-layer and only a tiny fraction is outside the layer. Bulk
\MoS~ does not exhibit big differences with respect to the single-layer. Therefore, excitons $X_A$ and $X_B$ remain 
in one layer, without coupling between layers, and optical transitions take place at the same layer. In other 
layered materials like hexagonal boron nitride, we
find that excitons can spread over several layers \cite{Galambosi2011}. 

\begin{figure}
\begin{center}
\includegraphics[width=7.6 cm]{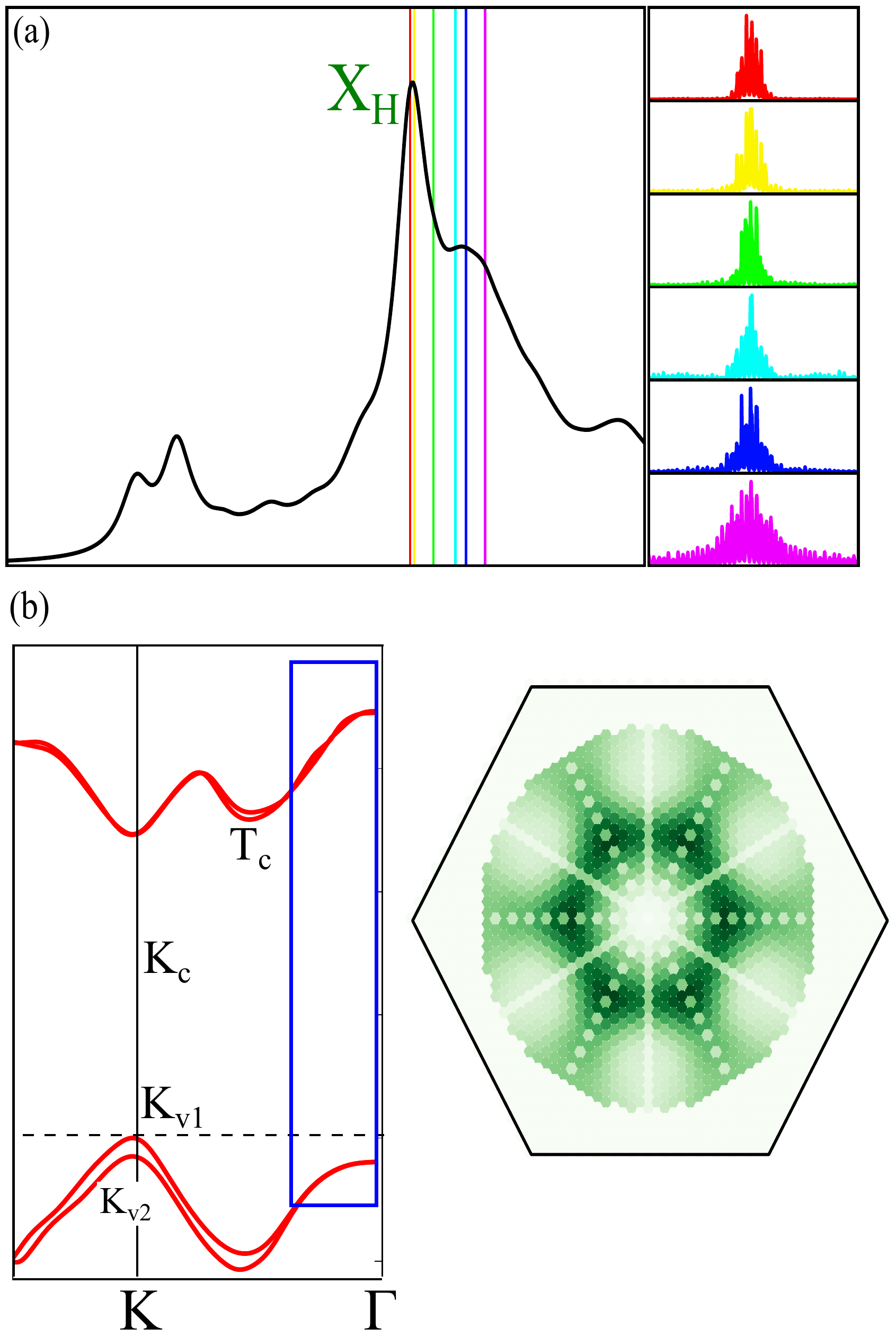}
\end{center}
\caption{(a): Bethe-Salpeter spectra of \MoS~single-layer together
with the side view of the exciton wave functions (marked with a vertical 
line). (b) Band structure of single-layer \MoS~close to $\Gamma$ and
wave function of $X_H$ exciton represented in $\bf k$-space.} 
\label{van-hove}
\end{figure}

Another distinctive optical property of single-layer \MoS~is the exciton $X_H$, visualized by a sharp peak 
at high energy (2.75 eV). Is this peak built from a single exciton or do we see the collective
effect of many excitons? Figure~\ref{van-hove} (a) shows the excitonic peak $X_H$, together with
each contribution (vertical lines). On the right side we have drawn the profile of  
the corresponding exciton wave function. Figure~\ref{van-hove} (b) depicts the weigth
$w_{\bf k}^{X_H}$ of the first vertical line (in red). The characteristics of this exciton 
are radically different from the case of
$X_A$ and $X_B$ excitons. First, the exciton is localized around $\Gamma$ in the
$\bf k$ space, forming a kind of hexagonal wheel. Second, defining the binding
energy is ambiguous. We know that for bound excitons, like $X_A$ and $X_B$, the binding energy is defined as the difference between
exciton energy and the band gap energy. The transition energies of the $X_H$ exciton 
fall within the continuous of states, making difficult to
define such binding energy. Third, the sharp peak is the collective contribution 
of several excitons with similar energy and wave function, as we can see from
the wave functions of Fig.~\ref{van-hove} (b). The parallel transitions lead to a singularity in the density of states, and often
the term \textit{Van-Hove} exciton is used to denominate such
peak \cite{VanHove1953,cardona,Riefer2011}.

Recently, several experiments have found fingerprints of the \textit{Van-Hove} 
exciton, e.g., two-photon spectroscopy \cite{Malard2013},
photocurrent spectroscopy \cite{Klots2014}, and light scattering 
spectra \cite{Mertens2014}. Some properties of the \textit{Van Hove}
exciton, under discussion nowadays, are the large electric field required
to dissociate the exciton, and the spontaneous decay of \textit{Van Hove}
excitons into a free electron-hole pair.

Additionally, single-layer \MoS~ exhibits a Rydberg-like 
exciton series in the optical spectra \cite{Qiu2013}. To capture these
excited states the convergence of the $\bf k$-grid 
requires up to $72\times 72\times 1$. In comparison with the Rydberg series
for a 2D hydrogen model, the excitation spectrum of single-layer \MoS~ 
is completely different (see in supplementary material
of Ref. \cite{Qiu2013}). The reason for this difference is the spatial
variation of the dielectric function in \MoS. The Rydberg-like series and its
particular behaviour has been also observed in 
single-layer WS$_2$ \cite{Chernikov2014}.

\subsection{Optical spectra of other transition metal dichalcogenides}

The rest of semiconductor transition metal dichalcogenides, MoSe$_2$, WS$_2$, and WSe$_2$, shares the interesting
optical properties of \MoS. They have similar lattice parameters, band gap and spin-orbit splitting
\cite{Antoci1972,Chernikov2014,Kumar2014,Ramirez-Torres2014,Yuan2014}. However, band gaps
are different enough to generate
band mismatch in heterostructures. We can combine TMDs
in vertical heterostructures for quantum well growth \cite{Britnell2013,Komsa2013,Chiu2014}, with the 
purpose of selective
confinement of photogenerated excitons.

\begin{table}
\begin{center}
\begin{tabular}{lccc}
\hline
\hline
                      &   WS$_2$   &    WSe$_2$     &    MoSe$_2$    \\
\hline
$a$ (Ang.)            &   3.127      &   3.250       &   3.253       \\
$E_g$ (eV)            &   1.739      &   1.458       &   1.516       \\         
$\Delta_{so}$ (meV)   &   406.8      &   457.2       &   185.6       \\        
\hline
\hline
\end{tabular}
\end{center}
\caption{Lattice parameters, band gap at $\bf{K}$ ($E_g$), and spin-orbit splitting at
$\bf{K}$ ($\Delta_{so}$) of TMDs,
obtained with DFT-LDA.}
\end{table}

\begin{figure}
\begin{center}
\includegraphics[width=7.0 cm]{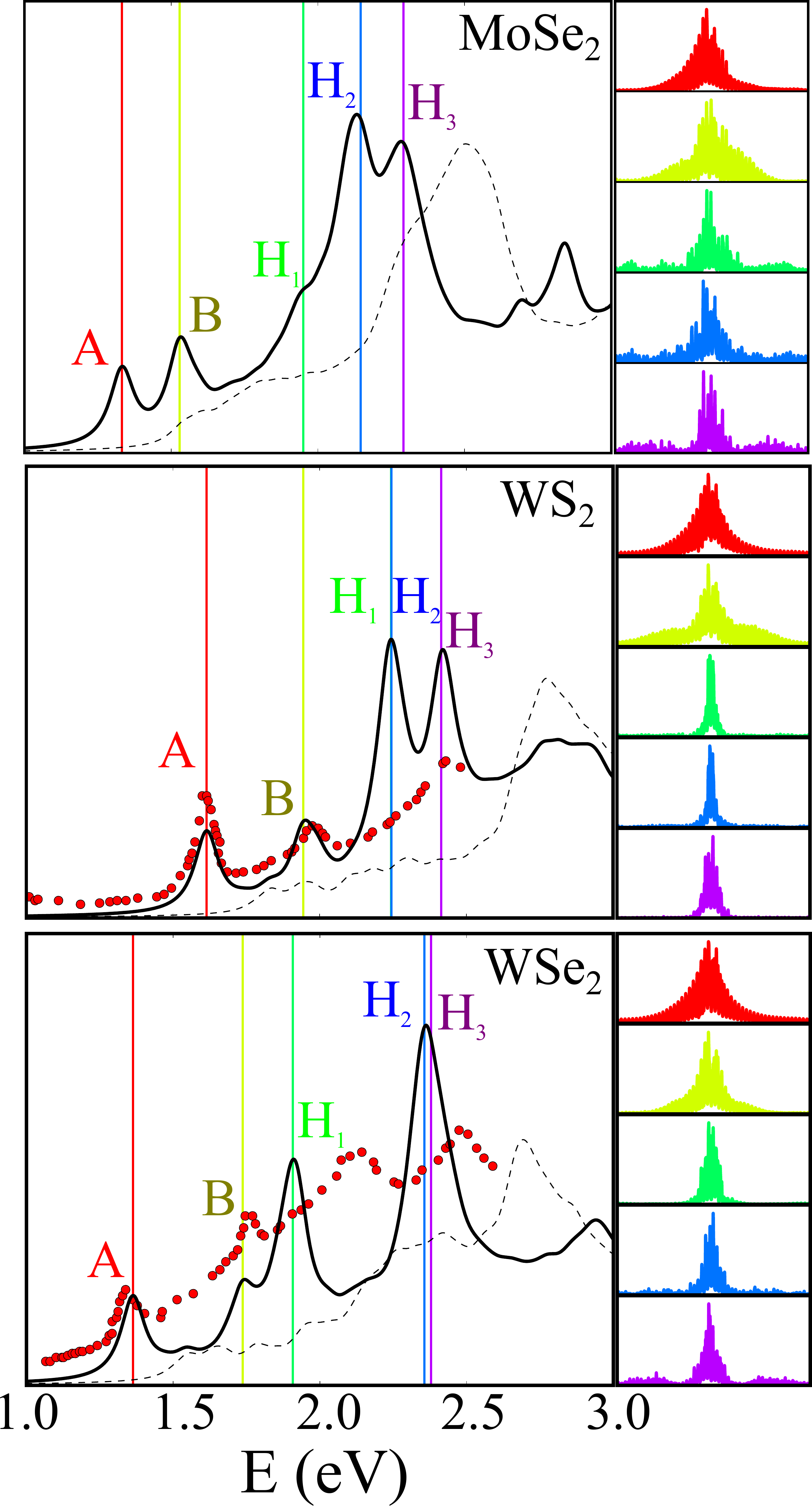}
\end{center}
\caption{From top to bottom, optical spectra of single-layer MoSe$_2$, WS$_2$, WSe$_2$, with and without excitonic
effects (solid and dashed black line). The exciton wave functions on the right are associated with the vertical lines marked
in the spectra. We have placed the hole 0.5 Bohr on top of Mo (W). Experimental data are extracted from Ref. \cite{Kozawa2014}.}
\label{tmds-bse}
\end{figure}

Figure~\ref{tmds-bse} shows the BSE spectra of single-layer MoSe$_2$, WS$_2$, and WSe$_2$ (solid lines) and the 
RPA spectra (dashed lines). We have included experimental data for WS$_2$, and WSe$_2$. On the right side we present a 
lateral view of the exciton wave functions. Starting points
are the LDA calculations, including spin-orbit. We have used
a $51\times 51\times 1$ $\bf k$-grid and we have included four valence and conduction bands. The static dielectric 
function is obtained with 60 bands. More accurate spectra require using 
self-consistent GW 
quasiparticle eigenvalues, as shown in Section~\ref{band}. For introductory purposes, using LDA as starting
point allows to describe the main physics of the optical properties of TMDs.

All the spectra exhibit two well differentiated excitons, $A$ and $B$, which come from transitions centered at the
$\bf{K}$ point, analogously to single-layer \MoS. The spin-orbit splitting determines the separation 
between the peaks $A$ and $B$. The theoretical splitting agrees very well with the experiments. Exciton $A$ is 
uniquely composed of the top of the valence band and the bottom of
the conduction band. 
Accordingly, exciton $B$ is mostly composed by the second
valence band ($K_{v2}$) and the conduction band with opposite spin.
In comparison with the $A$ exciton of single-layer \MoS, we observe a
similar spreading of the wave functions for the other TMDs. Evidently, the spin-orbit splitting is much higher
in compounds which include tungsten. In valley-physics, this has important consequences. The tuning of the excitation
energy is crucial to obtain an excitonic population with certain polarization. Tuning in
WS$_2$ and WSe$_2$ will in principle be easier due to the energy separation and the generation of a valley
polarization will be more efficient. We will comment on this again in subsection~\ref{spin}, devoted to valley physics.

The high-energy excitons (from $H_1$ to $H_3$) show a strong spatial confinement, analogously to \MoS. However, they split
in several and well differentiated peaks. Spin-orbit also splits the bands around $\Gamma$, and this results in 
the splitting of the excitons. The experimental spectra of WSe$_2$ agree with the calculation in the relative separation
between the $H_1$ and $H_2$ peaks. This latter compound presents the strongest spin-orbit coupling effects, either for
the bound excitons $A$ and $B$, or the Van-Hove exctions $H$s. In summary, selenium-based TMDs have an $H$-peak separation close to 0.5 eV 
and sulphur-based TMDs have a difference below 0.2 eV.

\subsection{Empirical models for excitons in \MoS}

Analytical approaches are very useful in the study of excitons in single-layer
\MoS~and they can be easily expanded to explore excitonic complexes such as
trions and biexcitons \cite{Mak2013}. These approaches are also suitable to obtain
the Rydberg excitonic series \cite{Chernikov2014}.

The tight-binding ansatz of the 
electronic wave function takes the orbital composition of 
the valence and conduction band states obtained from DFT results. Using
the density matrix formalism, one obtains the analytical solution of the
band structure close to the points $\bf{K}$ and $\bf{K}'$ \cite{Berghauser2014}. 

$$ \epsilon_{\bf k,\xi}^{\lambda_s} = \pm \frac{1}{2}\sqrt{(\Delta\varepsilon_{\xi}^{\lambda_s})^2 + 4|t^{\lambda_s}|^2 f(\bf k) },$$

where $\xi$ is the valley and $\lambda=v_{\uparrow},v_{\downarrow}, c_{\uparrow},c_{\downarrow}$ denotes band and spin. Momentum dependence is given by

$$ f({\bf k}) = 3 + 2\cos(k_y) +4\cos(k_y/2)\cos(\sqrt{3}k_x/2) .$$

\begin{figure}
\begin{center}
\includegraphics[width=7.6 cm]{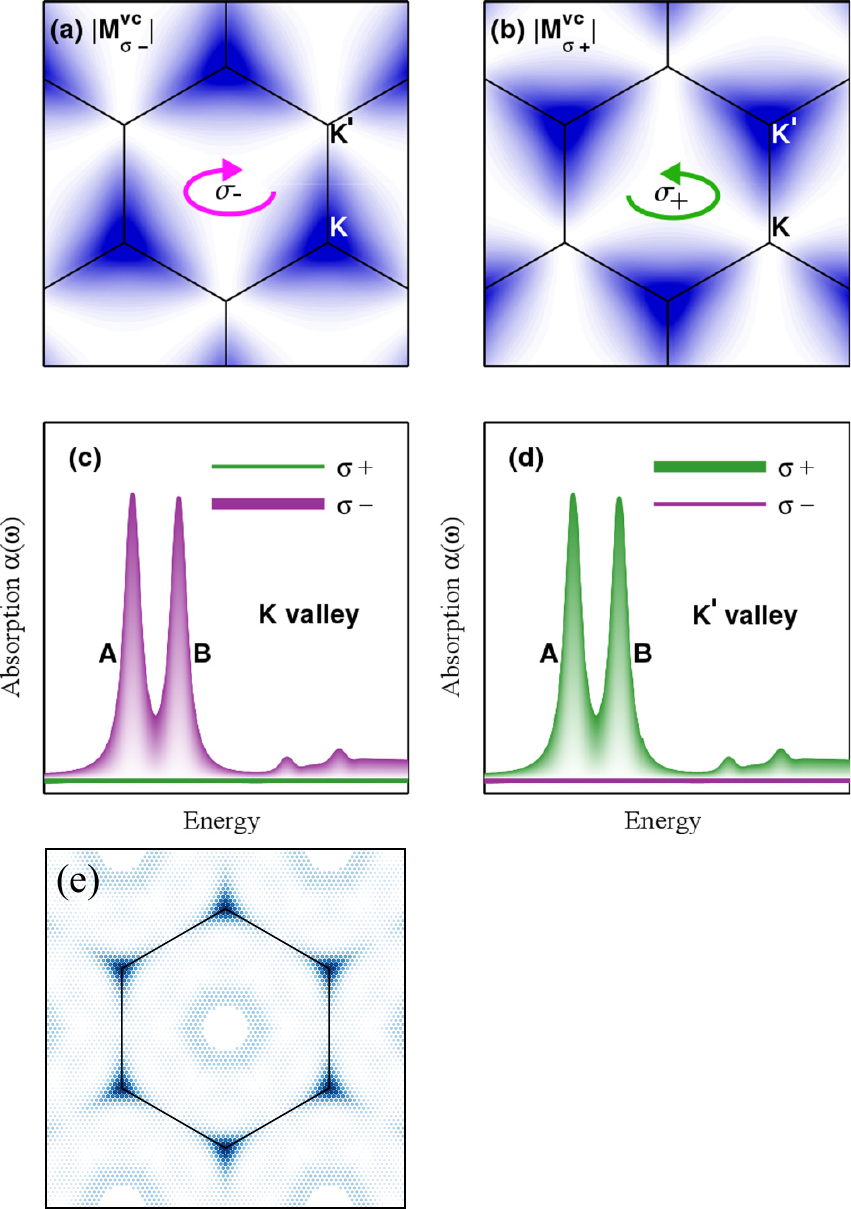}
\end{center}
\caption{(a) and (b): Optical matrix elements for negative ($\sigma-$) and positive
($\sigma+$) circular polarization. The corresponding spectra are represented in (c) and
(d) (reprinted with permission from Ref. \cite{Berghauser2014}. 
Copyright (2014) by the American Physical Society). (e) Optical matrix elements calculated
with {\tt Yambo} \cite{Marini2009} for linear light polarization.}
\label{matrix-elements}
\end{figure}

The Taylor expansion simplifies the eigenvalue momentum dependence to a 
parabolic band structure

$$ \epsilon_{\bf k,\xi}^{\lambda_s} \approx \pm \left( \frac{\Delta\varepsilon_{\xi}^{\lambda_s}}{2} + \frac{3|t^{\lambda_s}|^2}{4\Delta\varepsilon_{\xi}^{\lambda_s}} \right), $$

and the solution of the model Hamiltonian gives the 
eigenvectors, from which one can obtain the carrier-light matrix 
elements. Transitions at $\bf{K}$ and $\bf{K}'$ are between Mo-$d$ orbitals
of the valence band and S-$p$ orbitals of the conduction band. These transitions
can be optically excited by circularly polarized light. The right-handed
circularly polarized light will excite states at $\bf{K}$ and the left-handed light
at $\bf{K}'$, allowing a valley selection. This is the
cornerstone of \textit{valley physics}, which will be presented in subsection \ref{spin}. Figure \ref{matrix-elements} shows 
the results of Ref. \cite{Berghauser2014}
for the matrix elements and the absorption spectra for negative (a,c) and positive (b,d)
light polarization. Fig. \ref{matrix-elements}(e) shows our own calculations 
of the optical matrix elements for linear polarization. In the latter case we
excite the valleys $\bf{K}$ and $\bf{K}'$ with the same probability. 
Agreement between DFT and the model
Hamiltonian is excellent and justifies the use of the analytical approach.

Excitonic effects are included
introducing the Coulomb interaction into the 
model Hamiltonian  \cite{Berghauser2014}. In the framework of the 
semiconductor Bloch equation, from microscopic polarization, one can
obtain an analytical expression of the absorption coefficient (for a 
complete derivation, see Ref. \cite{Koch2004}). In addition, the effects of
the substrate on the exciton binding energy can be quantified by a proper choice of
the dielectric constant.

The application of the model shows
a binding energy of the $X_A$ exciton 
for free-standing single-layer \MoS~of 860 meV. The binding energy decreases to 
455 meV on top of a silicon oxide substrate. Among limitations of this analytical 
approch are the calculation of the relative intensity of the $X_A$ and $X_B$ peaks, which can be 
attributed to higher-order effects beyond the Hartree-Fock approximation. The prediciton
of the high-energy excitons will also require a much more complicated reformulation of the
model.

\subsection{Experimental determination of the band gap and exciton binding energy in MoSe$_2$}

The experimental determination of electronic band gap and exciton binding energy requires at least two techniques 
for determining univoquely these magnitudes. We have seen that the electronic band gap is related 
to the single-particle excitation. In addition, the binding energy is the difference between the electronic and the
optical band gap. In Ref. \cite{Ugeda2014}, Ugeda \textit{et. al.} have used a combination of experiments and theory to give an accurate value for the binding energy of single-layer MoSe$_2$. The measurements by scanning tunneling spectroscopy (STS) have measured the electronic gap, and photoluminescence (PL) has defined
the optical gap. The reported values are $E_g=2.18\pm 0.04$ eV (electronic band gap) and $E_{opt}=1.63\pm 0.01$ eV, what gives a 
binding energy of $0.55\pm 0.04$. Experimental findings are well supported
by GW and BSE calculations, taking into account the incidence of the substrate (bilayer graphene). As mentioned, the substrate
increases the dielectric constant and reduces the electronic and optical band gaps. However, a full GW and BSE calculation of
the system MoSe$_2$ plus substrate would be computationally very complex. Alternatively, authors have made apart the 
calculation of the substrate dielectric screening. The MoSe$_2$ contribution is 
obtained in the random phase approximation, including local 
fields. Afterwards, the MoSe$_2$ and substrate contributions
are merged in the Bethe-Salpeter equation. The binding energy reduces from
a value of 0.65 eV to 0.52 eV, with an uncertainty of 0.10 eV, in
fair agreement with experimental values. Therefore, a wise treatment of substrate influence on electronic and optical properties appears as an important aspect in calculations aiming of having predictive character.

\subsection{Spin-orbit interaction and valley physics in MoS$_2$}
\label{spin}

The lack of inversion symmetry and the strong spin-orbit interaction 
in single-layer \MoS~lead to what is called valley-physics. The valley index denotes 
the momentum of the valence band state, $\mathbf{K}$ or $\mathbf{K}'$, and the spin (up or down). As shown
above, excitons $A$ and $B$ can be generated exclusively from the valley 
$\bf{K}$ or $\mathbf{K}'$ by selecting the appropiate light polarization. Among the potential
uses and research related with this new concept are information transport by means of a new carrier, defined in terms of the
valley index, or the generation of a valley-Hall effect \cite{Cao2012,Mak2012,Zeng2012,Mak2014}.

\begin{figure}
\begin{center}
\includegraphics[width=7.6 cm]{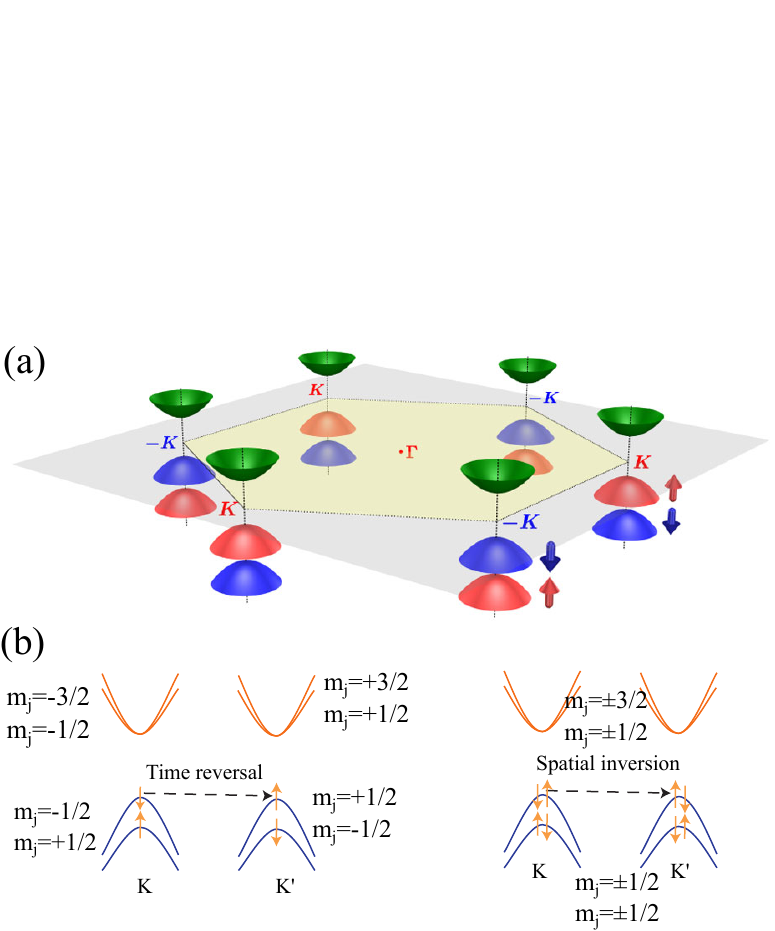}
\end{center}
\caption{(a) Illustration of the valence and conduction bands at the $mathbf{K}$ points 
of the Brillouin zone, including the spin-orbit interaction
(reprinted with permission of Ref. \cite{Xiao2012}, copyright (2014) by the American Physical Society).
(b)  (reprinted by permission from Macmillan Publishers Ltd: Nature
Nanotechnology, Ref. \cite{Mak2012}, copyright 2012.}
\label{spin-flip}
\end{figure}

In the case of single-layer \MoS~, spin-orbit coupling splits the valence band 
maximum at $\bf{K}$ and $\bf{K}'$ 
by $\sim 150$ meV. Moreover, the point group symmetry $D_{3h}$ does not have inversion symmetry. Under
these conditions, Kramer's degeneracy states that $E_{\uparrow}(\bf{k})=E_{\downarrow}(-\bf{k})$. In the case
of the valence band states at $\bf{K}$, we find the 
relation $E_{\uparrow}(\bf{K})=E_{\downarrow}(\bf{K}')$. The spin of the valence band states $K_{v1}$ and $K_{v2}$ flips
under change of inequivalent $\bf{K}$ points. In other words, valence band edge valley and spin are coupled. Figure~\ref{spin-flip}(a)
illustrates the spin composition of the valence band edges at $\bf{K}$ and $\bf{K}'$. The
spin-orbit splitting is negligible (a few $meV$) for conduction states and we 
can consider a two-fold
degenerate state. Following Ref. \cite{Xiao2012}, we can assign 
the following total angular momenta, $m_j$, in the $z$-direction
to the conduction and valence band states. The conduction band state at $\bf{K}$ is energy degenerate with $m_j=-3/2$ and
$m_j=-1/2$. In the case of valence band states, $K_{v1}$ and $K_{v2}$ have $m_j=-1/2$ and $m_j=1/2$ respectively,
split in energy by $\sim 150$ meV. The values for the total angular momentum
of the valence and conduction states at $\bf{K}'$ are obtained by multiplying 
the ones at $\bf{K}$ by $-1$.

Exposing the crystal to circularly polarized light of momentum $+1$ will promote a valence electron with momentum
$m_j=+1/2$ to the conduction state of $m_j=+3/2$, or from the valence state $m_j=-1/2$ to the conduction state $m_j=+1/2$, both transitions taking place in the valley $\bf{K}'$. These transitions are 
consequently forbidden at valley $\bf{K}$. By tuning the excitation energy one can precisely select
 which couple of valence-conduction band states are participating and from which valley. Thus, light with the energy of the band gap and momentum $+1$ will only
promote valence electrons with $m_j=+1/2$. In this way, one can create a stable population of electron-hole pairs with defined 
valley index. In other words, a ``valley polarization'' is generated. 
This has been demonstrated 
theoretically and experimentally in Refs. \cite{Cao2012,Mak2012,Xiao2012}. The persistence of this valley polarization is related to the conservation of spin. Hence, the flipping of valley index implies a spin flip, which is only possible by scattering with magnetic impurities or by means of relaxation via intra- and inter-valley scattering \cite{Xiao2012,Mai2014}.

Nowadays, the selective generation of valley polarizations is presented as 
a way to develop the so-called valleytronics and
the stability of the valley polarization is being investigated. Techniques 
such as ultra-fast spectroscopy make it possible 
to trace the time evolution of valley polarization.  Although theory predicts a long lifetime for the valley polarization, recent studies observe a non-negligible decoherence \cite{Jones2013}, which points towards other reasons 
than magnetic impurities as the origin of the spin-flip. Experimental studies have been undertaken to identify the reasons of decoherence of valley polarizations \cite{Wang2013, Shi2013a, Cui2014, Lagarde2014, Mai2014}. The analysis of the time-dependent optical response of single-layer \MoS~supported in several substrates or suspended in vacuum can give a hint \cite{Shi2013a} of how the environment can alter the electronic structure and hence the valley polarization. Other causes of decoherence can be the interplay between intra-valley and inter-valley scattering. Here the detuning of the valley polarization can take place via intermediate transitions through the $\Gamma$ point \cite{Mai2014}. The proliferation of experimental results using time-dependent spectroscopy is increasing our knowledge of the electronic and optical properties of \MoS~and other two-dimensional materials.

\section{Conclusions}

This review has summarized the theoretical and computational
description of the fundamental 
spectroscopic properties of \MoS ~in its single-layer, few-layer, and bulk form.
We have focussed on \MoS ~but many of the findings are similarly valid for 
the other semiconducting transition-metal dichalcogenides (TMDs) of 2H polytype.
We have summarized the numerous published investigations that report partly 
controversial findings. We have analysed the possible reasons for the
discrepancies by performing comprehensive density functional theory 
calculations on different levels of approximation. 
For the geometry, we have investigated the inclusion of Van der Waals 
interactions. For the quasiparticle gap, we have compared the results
obtained by DFT with different hybrid functionals with the result obtained
by many-body perturbation theory. Excitonic effects in the optical properties
are described with the Bethe-Salpeter approach.
Thereby, our review provides a general idea on the most important computational 
issues that arise when studying TMDs. We hope that it gives 
stimulation to the scientific community to achieve accurate and 
converged results.

The structural and vibrational properties are well described by density
functional theory approaches including Van der Waals interactions.
Excellent results are obtained for the lattice parameters and the calculated 
phonon frequencies that agree well with different experimental data from 
Raman spectroscopy and neutron scattering.
In this context, the open question about the anomalous Davydov splitting 
has been explained in terms of many neighbors interaction between Mo-S atoms
of different layers. The anomalous trend of the $E_{2g}$ mode as a function
of layer number is a consequence of the renormalization of the atomic 
distances due to the free surface\cite{Luo2013}.

There is a variety of results in the literature for the band structures and 
band gaps based on different levels of computational approach.
The spread of results is connected to the inherent problem of local and 
semilocal exchange correlation functionals
to yield accurate band gaps and can be overcome by including nonlocal exchange 
(hybrid functionals) or using the $GW$ approximation. The latter is
commonly applied in a one-shot manner ($G_0W_0$) on top of DFT wave functions.
Different schemes of self-consistency lead to a large spread in the 
quasiparticle gaps. 
Reliable band structures and relative positions between 
the two lowest conduction band extrema $K_c$ and $T_c$, but slightly 
overestimated band gaps, are obtained by self-consistent quasiparticle $GW$ 
calculations provided that one starts from
a fully optimized crystal structure. 
Particularly important in the case of single-layer \MoS ~is the 
convergence of the calculations with respect to the $\kvec$-point grid, the number of virtual states,
and the vacuum layer in the slab approach.

As for most other layerd materials, excitonic effects are very pronounced
in the optical properties of \MoS. 
The absorption spectra of mono-, bilayer, and bulk \MoS~ display three 
pronounced peaks.
The excitons are calculated with the Bethe-Salpeter equation based on the
full spinorial wave functions in order to include the effects of spin-orbit
interaction. 
In all single-layer, double-layer and bulk \MoS, there is 
a pronounced splitting between A and B excitons which can be traced back
to the splitting of the valence band maximum at the high-symmetry point $\mathbf{K}$. For single-layer \MoS, the splitting is due to the
strong spin-orbit splitting of the valence band maximum at $\mathbf{K}$, for
for double-layer and bulk \MoS, the splitting is due to the
interlayer interaction. Interestingly, the brightest exciton of 
single-layer \MoS ~is not found at the absorption threshold, 
but at higher energies (around 3 eV). This exciton, also called ``Van Hove exciton'' stems from a large joint density of states due to parallel conduction
and valence bands around $\Gamma$. It plays an important role in the resonant
Raman spectroscopy of various semiconducing transition-metal dichalcogenides.

\section{Acknowledgements}

A. M.-S.\ and L.W. acknowledge support by the National Research Fund, Luxembourg
(Projects C14/MS/773152/FAST-2DMAT and  INTER/ANR/13/20/NANOTMD). Calculations
were performed on the Vienna Scientific Cluster (VSC) and on the IDRIS
supercomputing center, Orsay (Proj. No. 091827). We acknowledge stimulating
discussions with D. Sangalli and A. Marini. We thank E. Kalesaki, S. Reichardt and H.P.C. Miranda 
for critically reading the
manuscript.

and E. Kalesaki for critically reading the manuscript.


\begin{thebibliography}{100}

\bibitem{Novoselov2005}
K.~S. Novoselov, D.~Jiang, F.~Schedin, T.~J. Booth, V.~V. Khotkevich, S.~V.
  Morozov, and A.~K. Geim.
\newblock Proceedings of the National Academy of Sciences of the United States
  of America \textbf{102}, 30, 10451 (2005).

\bibitem{Wilson1969}
J.~Wilson and A.~Yoffe.
\newblock Advances in Physics \textbf{18}, 73, 193 (1969).

\bibitem{Jiang2015}
S.~Jiang, M.~Q. Arguilla, N.~D. Cultrara, and J.~E. Goldberger.
\newblock Accounts of Chemical Research \textbf{48}, 1, 144 (2015).
\newblock PMID: 25490074.

\bibitem{Novoselov2004}
K.~S. Novoselov, A.~K. Geim, S.~V. Morozov, D.~Jiang, Y.~Zhang, S.~V. Dubonos,
  I.~V. Grigorieva, and A.~A. Firsov.
\newblock Science \textbf{306}, 5696, 666 (2004).

\bibitem{Geim2009}
A.~K. Geim.
\newblock Science \textbf{324}, 5934, 1530 (2009).

\bibitem{Katsnelson2006}
M.~I. Katsnelson, K.~S. Novoselov, and A.~K. Geim.
\newblock Nat Phys \textbf{2}, 9, 620 (2006).

\bibitem{Katsnelson2012}
M.~I. Katsnelson.
\newblock \textit{Graphene: Carbon in Two Dimensions}.
\newblock Cambridge University Press (2012).

\bibitem{Lembke2015}
D.~Lembke, S.~Bertolazzi, and A.~Kis.
\newblock Accounts of Chemical Research \textbf{48}, 1, 100 (2015).
\newblock PMID: 25555202.

\bibitem{RadisavljevicB.2011}
RadisavljevicB., RadenovicA., BrivioJ., GiacomettiV., and KisA.
\newblock Nat Nano \textbf{6}, 3, 147 (2011).

\bibitem{Mak2010}
K.~F. Mak, C.~Lee, J.~Hone, J.~Shan, and T.~F. Heinz.
\newblock Phys. Rev. Lett. \textbf{105}, 136805 (2010).

\bibitem{Splendiani2010}
A.~Splendiani, L.~Sun, Y.~Zhang, T.~Li, J.~Kim, C.-Y. Chim, G.~Galli, and
  F.~Wang.
\newblock Nano Letters \textbf{10}, 4, 1271 (2010).
\newblock PMID: 20229981.

\bibitem{Lembke2012}
D.~Lembke and A.~Kis.
\newblock ACS Nano \textbf{6}, 11, 10070 (2012).

\bibitem{Baugher2013}
B.~W.~H. Baugher, H.~O.~H. Churchill, Y.~Yang, and P.~Jarillo-Herrero.
\newblock Nano Letters \textbf{13}, 9, 4212 (2013).

\bibitem{Radisavljevic2011}
B.~Radisavljevic, M.~B. Whitwick, and A.~Kis.
\newblock ACS Nano \textbf{5}, 12, 9934 (2011).

\bibitem{Zhang2012}
Y.~Zhang, J.~Ye, Y.~Matsuhashi, and Y.~Iwasa.
\newblock Nano Letters \textbf{12}, 3, 1136 (2012).

\bibitem{Lopez-Sanchez2013}
O.~Lopez-Sanchez, D.~Lembke, M.~Kayci, A.~Radenovic, and A.~Kis.
\newblock Nat Nano \textbf{8}, 7, 497 (2013).

\bibitem{Zhang2015}
Y.~Zhang, H.~Li, L.~Wang, H.~Wang, X.~Xie, S.-L. Zhang, R.~Liu, and Z.-J. Qiu.
\newblock Sci. Rep. \textbf{5}, 7938, 07938 (2015).

\bibitem{Fontana2013}
M.~Fontana, T.~Deppe, A.~K. Boyd, M.~Rinzan, A.~Y. Liu, M.~Paranjape, and
  P.~Barbara.
\newblock Sci. Rep. \textbf{3},  (2013).

\bibitem{Furchi2014}
M.~M. Furchi, A.~Pospischil, F.~Libisch, J.~Burgdörfer, and T.~Mueller.
\newblock Nano Letters \textbf{14}, 8, 4785 (2014).
\newblock PMID: 25057817.

\bibitem{Pospischil2015}
A.~Pospischil, M.~M. Furchi, and T.~Mueller.
\newblock Nature Nanotechnology \textbf{9}, 4, 257–261 (2014).

\bibitem{Gutierrez2013}
H.~R. Gutiérrez, N.~Perea-López, A.~L. Elías, A.~Berkdemir, B.~Wang, R.~Lv,
  F.~López-Urías, V.~H. Crespi, H.~Terrones, and M.~Terrones.
\newblock Nano Letters \textbf{13}, 8, 3447 (2013).

\bibitem{Li2012c}
H.~Li, Z.~Yin, Q.~He, H.~Li, X.~Huang, G.~Lu, D.~W.~H. Fam, A.~I.~Y. Tok,
  Q.~Zhang, and H.~Zhang.
\newblock Small \textbf{8}, 1, 63 (2012).

\bibitem{Castellanos-Gomez2012b}
A.~Castellanos-Gomez, M.~Barkelid, A.~Goossens, V.~E. Calado, H.~S. van~der
  Zant, and G.~A. Steele.
\newblock Nano letters \textbf{12}, 6, 3187 (2012).

\bibitem{Coleman2011}
J.~N. Coleman, M.~Lotya, A.~O’Neill, S.~D. Bergin, P.~J. King, U.~Khan,
  K.~Young, A.~Gaucher, S.~De, R.~J. Smith, I.~V. Shvets, S.~K. Arora,
  G.~Stanton, H.-Y. Kim, K.~Lee, G.~T. Kim, G.~S. Duesberg, T.~Hallam, J.~J.
  Boland, J.~J. Wang, J.~F. Donegan, J.~C. Grunlan, G.~Moriarty, A.~Shmeliov,
  R.~J. Nicholls, J.~M. Perkins, E.~M. Grieveson, K.~Theuwissen, D.~W. McComb,
  P.~D. Nellist, and V.~Nicolosi.
\newblock Science \textbf{331}, 6017, 568 (2011).

\bibitem{Korn2011}
T.~Korn, S.~Heydrich, M.~Hirmer, J.~Schmutzler, and C.~Schuller.
\newblock Applied Physics Letters \textbf{99}, 10, 102109 (2011).

\bibitem{Sundaram2013}
R.~S. Sundaram, M.~Engel, A.~Lombardo, R.~Krupke, A.~C. Ferrari, P.~Avouris,
  and M.~Steiner.
\newblock Nano Letters \textbf{13}, 4, 1416 (2013).

\bibitem{Scalise2012}
E.~Scalise, M.~Houssa, G.~Pourtois, V.~Afanas’ev, and A.~Stesmans.
\newblock Nano Research \textbf{5}, 1, 43 (2012).

\bibitem{Scalise2014}
E.~Scalise, M.~Houssa, G.~Pourtois, V.~Afanas'ev, and A.~Stesmans.
\newblock Physica E: Low-dimensional Systems and Nanostructures \textbf{56}, 0,
  416 (2014).

\bibitem{He2013}
K.~He, C.~Poole, K.~F. Mak, and J.~Shan.
\newblock Nano Letters \textbf{13}, 6, 2931 (2013).

\bibitem{Conley2013}
H.~J. Conley, B.~Wang, J.~I. Ziegler, R.~F. Haglund, S.~T. Pantelides, and
  K.~I. Bolotin.
\newblock Nano Letters \textbf{13}, 8, 3626 (2013).

\bibitem{Dong2014}
L.~Dong, R.~Namburu, T.~O’Regan, M.~Dubey, and A.~Dongare.
\newblock Journal of Materials Science \textbf{49}, 19, 6762 (2014).

\bibitem{Guzman2014}
D.~M. Guzman and A.~Strachan.
\newblock Journal of Applied Physics \textbf{115}, 24, 243701 (2014).

\bibitem{Nayak2014}
A.~P. Nayak, S.~Bhattacharyya, J.~Zhu, J.~Liu, X.~Wu, T.~Pandey, C.~Jin, A.~K.
  Singh, D.~Akinwande, and J.-F. Lin.
\newblock Nature Communications \textbf{5}, 3731 (2014).

\bibitem{Nayak2015}
A.~P. Nayak, T.~Pandey, D.~Voiry, J.~Liu, S.~T. Moran, A.~Sharma, C.~Tan, C.-H.
  Chen, L.-J. Li, M.~Chhowalla, J.-F. Lin, A.~K. Singh, and D.~Akinwande.
\newblock Nano Letters \textbf{15}, 1, 346 (2015).
\newblock PMID: 25486455.

\bibitem{Xiao2012}
D.~Xiao, G.-B. Liu, W.~Feng, X.~Xu, and W.~Yao.
\newblock Phys. Rev. Lett. \textbf{108}, 196802 (2012).

\bibitem{Mak2012}
K.~F. Mak, K.~He, J.~Shan, and T.~F. Heinz.
\newblock Nat Nano \textbf{7}, 8, 494 (2012).

\bibitem{Zeng2012}
H.~Zeng, J.~Dai, W.~Yao, D.~Xiao, and X.~Cui.
\newblock Nat Nano \textbf{7}, 8, 490 (2012).

\bibitem{Cao2012}
T.~Cao, G.~Wang, W.~Han, H.~Ye, C.~Zhu, J.~Shi, Q.~Niu, P.~Tan, E.~Wang,
  B.~Liu, and J.~Feng.
\newblock Nat Commun \textbf{3}, 887 (2012).

\bibitem{Kumar2014}
N.~Kumar, J.~He, D.~He, Y.~Wang, and H.~Zhao.
\newblock Nanoscale pages~-- (2014).

\bibitem{Mak2013}
K.~F. Mak, K.~He, C.~Lee, G.~H. Lee, J.~Hone, T.~F. Heinz, and J.~Shan.
\newblock Nat Mater \textbf{12}, 3, 207 (2013).

\bibitem{Plechinger2015}
G.~Plechinger, P.~Nagler, J.~Kraus, N.~Paradiso, C.~Strunk, C.~Schüller, and
  T.~Korn.
\newblock physica status solidi (RRL) – Rapid Research Letters \textbf{9}, 8,
  457 (2015).

\bibitem{Kumar2013}
N.~Kumar, S.~Najmaei, Q.~Cui, F.~Ceballos, P.~M. Ajayan, J.~Lou, and H.~Zhao.
\newblock Phys. Rev. B \textbf{87}, 161403 (2013).

\bibitem{Malard2013}
L.~M. Malard, T.~V. Alencar, A.~P.~M. Barboza, K.~F. Mak, and A.~M. de~Paula.
\newblock Phys. Rev. B \textbf{87}, 201401 (2013).

\bibitem{Baugher2015}
B.~W.~H. Baugher, H.~O.~H. Churchill, Y.~Yang, and P.~Jarillo-Herrero.
\newblock Nature Nanotechnology \textbf{9}, 4, 262–267 (2014).

\bibitem{Ross2015}
J.~S. Ross, P.~Klement, A.~M. Jones, N.~J. Ghimire, J.~Yan, D.~G. Mandrus,
  T.~Taniguchi, K.~Watanabe, K.~Kitamura, W.~Yao, D.~H. Cobden, and X.~Xu.
\newblock Nature Nanotechnology \textbf{9}, 4, 268–272 (2014).

\bibitem{geim2013}
A.~K. Geim and I.~V. Grigorieva.
\newblock Nature \textbf{499}, 7459, 419 (2013).

\bibitem{Huang2014}
C.~Huang, S.~Wu, A.~M. Sanchez, J.~J.~P. Peters, R.~Beanland, J.~S. Ross,
  P.~Rivera, W.~Yao, D.~H. Cobden, and X.~Xu.
\newblock Nat. Mater. \textbf{13}, 1096 (2014).

\bibitem{Gong2014}
Y.~Gong, J.~Lin, X.~Wang, G.~Shi, S.~Lei, Z.~Lin, X.~Zou, G.~Ye, R.~Vajtai,
  B.~I. Yakobson, H.~Terrones, M.~Terrones, B.~Tay, J.~Lou, S.~T. Pantelides,
  Z.~Liu, W.~Zhou, and P.~M. Ajayan.
\newblock Nat. Mater. \textbf{13}, 1135 (2014).

\bibitem{Britnell2013}
L.~Britnell, R.~M. Ribeiro, A.~Eckmann, R.~Jalil, B.~D. Belle, A.~Mishchenko,
  Y.-J. Kim, R.~V. Gorbachev, T.~Georgiou, S.~V. Morozov, A.~N. Grigorenko,
  A.~K. Geim, C.~Casiraghi, A.~H.~C. Neto, and K.~S. Novoselov.
\newblock Science \textbf{340}, 6138, 1311 (2013).

\bibitem{he:prb:89}
J.~He, K.~Hummer, and C.~Franchini.
\newblock Phys. Rev. B \textbf{89}, 075409 (2014).

\bibitem{debbichi:prb:89}
L.~Debbichi, O.~Eriksson, and S.~Leb\`egue.
\newblock Phys. Rev. B \textbf{89}, 205311 (2014).

\bibitem{liu:natcommun:5}
K.~Liu, L.~Zhang, T.~Cao, C.~Jin, D.~Qiu, Q.~Zhou, A.~Zettl, P.~Yang, S.~G.
  Louie, and F.~Wang.
\newblock Nat. Commun. \textbf{5}, 4966 (2014).

\bibitem{Verble1970}
J.~L. Verble and T.~J. Wieting.
\newblock Phys. Rev. Lett. \textbf{25}, 362 (1970).

\bibitem{Wieting1971}
T.~J. Wieting and J.~L. Verble.
\newblock Phys. Rev. B \textbf{3}, 4286 (1971).

\bibitem{Wakabayashi1975}
N.~Wakabayashi, H.~G. Smith, and R.~M. Nicklow.
\newblock Phys. Rev. B \textbf{12}, 659 (1975).

\bibitem{Bertrand1991}
P.~A. Bertrand.
\newblock Phys. Rev. B \textbf{44}, 5745 (1991).

\bibitem{Lee2010}
C.~Lee, H.~Yan, L.~E. Brus, T.~F. Heinz, J.~Hone, and S.~Ryu.
\newblock Acs Nano \textbf{4}, 5, 2695 (2010).

\bibitem{Windom2011}
B.~Windom, W.~Sawyer, and D.~Hahn.
\newblock Tribology Letters \textbf{42}, 3, 301 (2011).

\bibitem{Li2012}
H.~Li, Q.~Zhang, C.~C.~R. Yap, B.~K. Tay, T.~H.~T. Edwin, A.~Olivier, and
  D.~Baillargeat.
\newblock Advanced Functional Materials \textbf{22}, 7, 1385 (2012).

\bibitem{Plechinger2012}
G.~Plechinger, S.~Heydrich, J.~Eroms, D.~Weiss, C.~Schuller, and T.~Korn.
\newblock Applied Physics Letters \textbf{101}, 10, 101906 (2012).

\bibitem{Boukhicha2013}
M.~Boukhicha, M.~Calandra, M.-A. Measson, O.~Lancry, and A.~Shukla.
\newblock Phys. Rev. B \textbf{87}, 195316 (2013).

\bibitem{Luo2013}
X.~Luo, Y.~Zhao, J.~Zhang, Q.~Xiong, and S.~Y. Quek.
\newblock Phys. Rev. B \textbf{88}, 075320 (2013).

\bibitem{Zhao2013}
Y.~Zhao, X.~Luo, H.~Li, J.~Zhang, P.~T. Araujo, C.~K. Gan, J.~Wu, H.~Zhang,
  S.~Y. Quek, M.~S. Dresselhaus, and Q.~Xiong.
\newblock Nano Letters \textbf{13}, 3, 1007 (2013).

\bibitem{Zhang2013}
X.~Zhang, W.~P. Han, J.~B. Wu, S.~Milana, Y.~Lu, Q.~Q. Li, A.~C. Ferrari, and
  P.~H. Tan.
\newblock Physical Review B \textbf{87}, 11, 115413 (2013).

\bibitem{ZhangX2015}
X.~Zhang, X.-F. Qiao, W.~Shi, J.-B. Wu, D.-S. Jiang, and P.-H. Tan.
\newblock Chem. Soc. Rev. \textbf{44}, 2757 (2015).

\bibitem{Ataca2011}
C.~Ataca, H.~Sahin, E.~Akturk, and S.~Ciraci.
\newblock Journal of Physical Chemistry C \textbf{115}, 10, 3934 (2011).

\bibitem{Molina-Sanchez2011}
A.~Molina-S\'anchez and L.~Wirtz.
\newblock Phys. Rev. B \textbf{84}, 155413 (2011).

\bibitem{Terrones2014}
H.~Terrones, E.~D. Corro, S.~Feng, J.~M. Poumirol, D.~Rhodes, D.~Smirnov, N.~R.
  Pradhan, Z.~Lin, M.~A.~T. Nguyen, A.~L. Elias, T.~E. Mallouk, L.~Balicas,
  M.~A. Pimenta, and M.~Terrones.
\newblock Sci. Rep. \textbf{4}, 4215 (2014).

\bibitem{Rice2013}
C.~Rice, R.~J. Young, R.~Zan, U.~Bangert, D.~Wolverson, T.~Georgiou, R.~Jalil,
  and K.~S. Novoselov.
\newblock Physical Review B \textbf{87}, 8, 081307 (2013).

\bibitem{dickinson:jacs:45}
R.~Dickinson and L.~Pauling.
\newblock J. Am. Chem. Soc. \textbf{45}, 1466 (1923).

\bibitem{schoenfeld:acb:39}
B.~Sch{\"o}nfeld, J.~J. Huang, and S.~C. Moss.
\newblock Acta Cryst. B \textbf{39}, 404 (1983).

\bibitem{alhilli:jcg:15}
A.~A. Al-Hilli and B.~L. Evans.
\newblock J. Cryst. Growth \textbf{15}, 93 (1972).

\bibitem{petkov:prb:65}
V.~Petkov, S.~J.~L. Billinge, P.~Larson, S.~D. Mahanti, T.~Vogt, K.~K. Rangan,
  and M.~G. Kanatzidis.
\newblock Phys. Rev. B \textbf{65}, 092105 (2002).

\bibitem{he:submitted}
J.~He, K.~Hummer, and C.~Franchini.
\newblock submitted  (2013).

\bibitem{Wirtz2004}
L.~Wirtz and A.~Rubio.
\newblock Solid State Communications \textbf{131}, 3–4, 141  (2004).

\bibitem{Kern1999}
G.~Kern, G.~Kresse, and J.~Hafner.
\newblock Phys. Rev. B \textbf{59}, 8551 (1999).

\bibitem{Serrano2007}
J.~Serrano, A.~Bosak, R.~Arenal, M.~Krisch, K.~Watanabe, T.~Taniguchi,
  H.~Kanda, A.~Rubio, and L.~Wirtz.
\newblock Phys. Rev. Lett. \textbf{98}, 095503 (2007).

\bibitem{Allard2010}
A.~Allard and L.~Wirtz.
\newblock Nano Letters \textbf{10}, 11, 4335 (2010).

\bibitem{Fromm2013}
F.~Fromm, M.~H. Oliveira~Jr, A.~Molina-S\'{a}nchez, M.~Hundhausen, J.~M.~J.
  Lopes, H.~Riechert, L.~Wirtz, and T.~Seyller.
\newblock New Journal of Physics \textbf{15}, 4, 043031 (2013).

\bibitem{Endlich2013}
M.~Endlich, A.~Molina-S\'anchez, L.~Wirtz, and J.~Kr\"oger.
\newblock Phys. Rev. B \textbf{88}, 205403 (2013).

\bibitem{kresse:cms:6}
G.~Kresse and J.~Furthm\"uller.
\newblock Comp. Mat. Science \textbf{6}, 15 (1996).

\bibitem{kresse:prb:54}
G.~Kresse and J.~Furthm\"uller.
\newblock Phys. Rev. B \textbf{54}, 11169 (1996).

\bibitem{bloechl:prb:50}
P.~E. Bl{\"o}chl.
\newblock Phys. Rev. B \textbf{50}, 17953 (1994).

\bibitem{kresse:prb:59}
G.~Kresse and D.~Joubert.
\newblock Phys. Rev. B \textbf{59}, 1758 (1999).

\bibitem{kim:prb:80}
Y.-S. Kim, K.~Hummer, and G.~Kresse.
\newblock Phys. Rev. B \textbf{80}, 035203 (2009).

\bibitem{shishkin:prb:74}
M.~Shishkin and G.~Kresse.
\newblock Phys. Rev. B \textbf{74}, 035101 (2006).

\bibitem{shishkin:prb:75}
M.~Shishkin and G.~Kresse.
\newblock Phys. Rev. B \textbf{75}, 235102 (2007).

\bibitem{shishkin:prl:99}
M.~Shishkin, M.~Marsman, and G.~Kresse.
\newblock Phys. Rev. Lett. \textbf{99}, 246403 (2007).

\bibitem{LDA:prl:45}
D.~M. Ceperley and B.~J. Alder.
\newblock Phys. Rev. Lett. \textbf{45}, 566 (1980).

\bibitem{PBEsol:prl:100}
J.~P. Perdew, A.~Ruzsinszky, G.~I. Csonka, O.~A. Vydrov, G.~E. Scuseria, L.~A.
  Constantin, X.~Zhou, and K.~Burke.
\newblock Phys. Rev. Lett. \textbf{100}, 136406 (2008).

\bibitem{klimes:prb:83}
J.~c.~v. Klime\ifmmode~\check{s}\else \v{s}\fi{}, D.~R. Bowler, and
  A.~Michaelides.
\newblock Phys. Rev. B \textbf{83}, 195131 (2011).

\bibitem{dion:prl:92}
M.~Dion, H.~Rydberg, E.~Schr\"der, D.~C. Langreth, and B.~I. Lundqvist.
\newblock Phys. Rev. Lett. \textbf{92}, 246401 (2004).

\bibitem{murnaghan:pnas:30}
F.~D. Murnaghan.
\newblock Proc. Natl. Acad. Sci. USA \textbf{30}, 244 (1944).

\bibitem{Ferrari2006}
A.~C. Ferrari, J.~C. Meyer, V.~Scardaci, C.~Casiraghi, M.~Lazzeri, F.~Mauri,
  S.~Piscanec, D.~Jiang, K.~S. Novoselov, S.~Roth, and A.~K. Geim.
\newblock Phys. Rev. Lett. \textbf{97}, 187401 (2006).

\bibitem{Graf2007}
D.~Graf, F.~Molitor, K.~Ensslin, C.~Stampfer, A.~Jungen, C.~Hierold, and
  L.~Wirtz.
\newblock Nano Letters \textbf{7}, 2, 238 (2007).
\newblock PMID: 17297984.

\bibitem{Berciaud2009}
S.~Berciaud, S.~Ryu, L.~E. Brus, and T.~F. Heinz.
\newblock Nano Letters \textbf{9}, 1, 346 (2009).
\newblock PMID: 19099462.

\bibitem{Forster2013}
F.~Forster, A.~Molina-Sanchez, S.~Engels, A.~Epping, K.~Watanabe, T.~Taniguchi,
  L.~Wirtz, and C.~Stampfer.
\newblock Phys. Rev. B \textbf{88}, 085419 (2013).

\bibitem{Starodub2011}
E.~Starodub, A.~Bostwick, L.~Moreschini, S.~Nie, F.~E. Gabaly, K.~F. McCarty,
  and E.~Rotenberg.
\newblock Phys. Rev. B \textbf{83}, 125428 (2011).

\bibitem{Thomsen2000}
C.~Thomsen and S.~Reich.
\newblock Phys. Rev. Lett. \textbf{85}, 5214 (2000).

\bibitem{Tan2012}
P.~H. Tan, W.~P. Han, W.~J. Zhao, Z.~H. Wu, K.~Chang, H.~Wang, Y.~F. Wang,
  N.~Bonini, N.~Marzari, N.~Pugno, G.~Savini, A.~Lombardo, and A.~C. Ferrari.
\newblock Nat Mater \textbf{11}, 4, 294 (2012).

\bibitem{Berkdemir2013}
A.~Berkdemir, H.~R. Gutierrez, A.~R. Botello-Mendez, N.~Perea-Lopez, A.~L.
  Elias, C.-I. Chia, B.~Wang, V.~H. Crespi, F.~Lopez-Urias, J.-C. Charlier,
  H.~Terrones, and M.~Terrones.
\newblock Sci. Rep. \textbf{3},  (2013).

\bibitem{Bruesch1982}
P.~Br\"uesch.
\newblock \textit{Phonons: Theory and Experiments I}.
\newblock Springer (1982).

\bibitem{Baroni2001}
S.~Baroni, S.~de~Gironcoli, A.~Dal~Corso, and P.~Giannozzi.
\newblock Rev. Mod. Phys. \textbf{73}, 515 (2001).

\bibitem{Gonze1997}
X.~Gonze and C.~Lee.
\newblock Phys. Rev. B \textbf{55}, 10355 (1997).

\bibitem{Jiang2013}
J.-W. Jiang, H.~S. Park, and T.~Rabczuk.
\newblock Journal of Applied Physics \textbf{114}, 6, 064307 (2013).

\bibitem{Livneh2010}
T.~Livneh and E.~Sterer.
\newblock Phys. Rev. B \textbf{81}, 195209 (2010).

\bibitem{Livneh2014}
T.~Livneh and J.~E. Spanier.
\newblock arXiv preprint arXiv:1408.6748  (2014).

\bibitem{Saito1998}
R.~Saito, G.~Dresselhaus, and M.~S. Dresselhaus.
\newblock \textit{Physical Properties of Carbon Nanotubes}.
\newblock Imperial College Press (1998).

\bibitem{Chen1974}
J.~Chen and C.~Wang.
\newblock Solid State Communications \textbf{14}, 9, 857  (1974).

\bibitem{Stacy1985}
A.~Stacy and D.~Hodul.
\newblock Journal of Physics and Chemistry of Solids \textbf{46}, 4, 405
  (1985).

\bibitem{Ribeiro-Soares2014}
J.~Ribeiro-Soares, R.~M. Almeida, E.~B. Barros, P.~T. Araujo, M.~S.
  Dresselhaus, L.~G. Can\ifmmode~\mbox{\c{c}}\else \c{c}\fi{}ado, and A.~Jorio.
\newblock Phys. Rev. B \textbf{90}, 115438 (2014).

\bibitem{Davydov1969}
A.~Davydov.
\newblock \textit{Theory of Molecular Excitons}.
\newblock MacGraw-Hill (1969).

\bibitem{Dawson1975}
P.~Dawson.
\newblock Journal of Physics and Chemistry of Solids \textbf{36}, 12, 1401
  (1975).

\bibitem{Kuroda1979}
N.~Kuroda and Y.~Nishina.
\newblock Phys. Rev. B \textbf{19}, 1312 (1979).

\bibitem{Ghosh1976}
P.~N. Ghosh.
\newblock Solid State Communications \textbf{19}, 7, 639  (1976).

\bibitem{Ghosh1983}
P.~N. Ghosh and C.~R. Maiti.
\newblock Phys. Rev. B \textbf{28}, 2237 (1983).

\bibitem{Gale1997}
J.~D. Gale.
\newblock J. Chem. Soc.{,} Faraday Trans. \textbf{93}, 629 (1997).

\bibitem{Gale2003}
J.~D. Gale and A.~L. Rohl.
\newblock Molecular Simulation \textbf{29}, 5, 291 (2003).

\bibitem{Michel2012}
K.~H. Michel and B.~Verberck.
\newblock Phys. Rev. B \textbf{85}, 094303 (2012).

\bibitem{Mattheiss1973}
L.~F. Mattheiss.
\newblock Phys. Rev. Lett. \textbf{30}, 784 (1973).

\bibitem{Mattheiss1973a}
L.~F. Mattheiss.
\newblock Phys. Rev. B \textbf{8}, 3719 (1973).

\bibitem{Boker2001}
T.~B\"oker, R.~Severin, A.~M\"uller, C.~Janowitz, R.~Manzke, D.~Vo\ss{},
  P.~Kr\"uger, A.~Mazur, and J.~Pollmann.
\newblock Phys. Rev. B \textbf{64}, 235305 (2001).

\bibitem{Peelaers2012}
H.~Peelaers and C.~G. Van~de Walle.
\newblock Phys. Rev. B \textbf{86}, 241401 (2012).

\bibitem{Lebegue2009}
S.~Leb\`egue and O.~Eriksson.
\newblock Phys. Rev. B \textbf{79}, 115409 (2009).

\bibitem{Jones1989}
R.~O. Jones and O.~Gunnarsson.
\newblock Rev. Mod. Phys. \textbf{61}, 689 (1989).

\bibitem{becke:jcp:124}
A.~D. Becke and E.~R. Johnson.
\newblock J. Chem. Phys. \textbf{124}, 221101 (2006).

\bibitem{tran:prl:102}
F.~Tran and P.~Blaha.
\newblock Phys. Rev. Lett. \textbf{102}, 226401 (2010).

\bibitem{Kim2010}
Y.-S. Kim, M.~Marsman, G.~Kresse, F.~Tran, and P.~Blaha.
\newblock Phys. Rev. B \textbf{82}, 205212 (2010).

\bibitem{becke:pra:39}
A.~D. Becke and M.~R. Roussel.
\newblock Phys. Rev. A \textbf{39}, 3761 (1989).

\bibitem{heyd:jcp:118}
J.~Heyd, G.~E. Scuseria, and M.~Ernzerhof.
\newblock J. Chem. Phys. \textbf{118}, 8207 (2003).

\bibitem{paier:jcp:122}
J.~Paier, R.~Hirschl, M.~Marsman, and G.~Kresse.
\newblock J. Chem. Phys. \textbf{122}, 234102 (2005).

\bibitem{krukau:jcp:125}
A.~V. Krukau, O.~A. Vydrov, A.~F. Izmaylov, and G.~E. Scuseria.
\newblock J. Chem. Phys. \textbf{125}, 224106 (2006).

\bibitem{schimka:jcp:134}
L.~Schimka, J.~Harl, and G.~Kresse.
\newblock J. Chem. Phys. \textbf{134}, 024116 (2011).

\bibitem{janesko:pccp:11}
B.~G. Janesko, T.~M. Henderson, and G.~E. Scuseria.
\newblock Phys. Chem. Chem. Phys. \textbf{11}, 443 (2009).

\bibitem{paier:jcp:124}
J.~Paier, M.~Marsman, K.~Hummer, G.~Kresse, I.~C. Gerber, and J.~G.
  {\'A}ng{\'a}n.
\newblock J. Chem. Phys. \textbf{124}, 154709 (2006).

\bibitem{paier:jcp:125:erratum}
J.~Paier, M.~Marsman, K.~Hummer, G.~Kresse, I.~C. Gerber, and J.~G.
  {\'A}ng{\'a}n.
\newblock J. Chem. Phys. \textbf{125}, 249901 (2006).

\bibitem{hummer:prb:75}
K.~Hummer, A.~Grüneis, and G.~Kresse.
\newblock Phys. Rev. B \textbf{75}, 195211 (2007).

\bibitem{paier:prb:79}
G.~I. Csonka, J.~P. Perdew, A.~Ruzsinszky, P.~H.~T. Philipsen, S.~Leb\`egue,
  J.~Paier, O.~A. Vydrov, and J.~G. \'Angy\'an.
\newblock Phys. Rev. B \textbf{79}, 155107 (2009).

\bibitem{note:HSE03}
To avoid confusion concerning the terminology of HSE functional, we would like
  to point out that the original HSE03 functional uses two different
  range-separation (\textit{screening}) parameters $\mu$ for the HF part and
  for the DFT part, respectively. In contrast, the improved HSE06 [A. V. Krukau
  et al., J. Chem. Phys. 125, 224106 (2006)] consistently employs a single
  parameter $\mu$=0.11 a.u.$^{\mathrm{-1}}$ ($\mu = 0.2$\AA$^{-1}$) for both
  parts and was used in all presented calculations.

\bibitem{Komsa2012}
H.-P. Komsa and A.~V. Krasheninnikov.
\newblock Phys. Rev. B \textbf{86}, 241201 (2012).

\bibitem{Ramasubramaniam2012}
A.~Ramasubramaniam.
\newblock Phys. Rev. B \textbf{86}, 115409 (2012).

\bibitem{Yun2012}
W.~S. Yun, S.~W. Han, S.~C. Hong, I.~G. Kim, and J.~D. Lee.
\newblock Phys. Rev. B \textbf{85}, 033305 (2012).

\bibitem{Cheiwchanchamnangij2012}
T.~Cheiwchanchamnangij and W.~R.~L. Lambrecht.
\newblock Phys. Rev. B \textbf{85}, 205302 (2012).

\bibitem{Shi2013}
H.~Shi, H.~Pan, Y.-W. Zhang, and B.~I. Yakobson.
\newblock Phys. Rev. B \textbf{87}, 15, 155304 (2013).

\bibitem{Molina-Sanchez2013}
A.~Molina-S\'anchez, D.~Sangalli, K.~Hummer, A.~Marini, and L.~Wirtz.
\newblock Phys. Rev. B \textbf{88}, 045412 (2013).

\bibitem{zhu:prb:84}
Z.~Y. Zhu, Y.~C. Cheng, and U.~Schwingenschl\"ogl.
\newblock Phys. Rev. B \textbf{84}, 153402 (2011).

\bibitem{Kuc2015}
A.~Kuc and T.~Heine.
\newblock Chem. Soc. Rev. \textbf{44}, 2603 (2015).

\bibitem{Jin2013}
W.~Jin, P.-C. Yeh, N.~Zaki, D.~Zhang, J.~T. Sadowski, A.~Al-Mahboob, A.~M.
  van~der Zande, D.~A. Chenet, J.~I. Dadap, I.~P. Herman, P.~Sutter, J.~Hone,
  and R.~M. Osgood.
\newblock Phys. Rev. Lett. \textbf{111}, 106801 (2013).

\bibitem{Wirtz2006}
L.~Wirtz, A.~Marini, and A.~Rubio.
\newblock Phys. Rev. Lett. \textbf{96}, 126104 (2006).

\bibitem{Liu2013}
G.-B. Liu, W.-Y. Shan, Y.~Yao, W.~Yao, and D.~Xiao.
\newblock Phys. Rev. B \textbf{88}, 8, 085433 (2013).

\bibitem{Liu2015}
G.-B. Liu, D.~Xiao, Y.~Yao, X.~Xu, and W.~Yao.
\newblock Chem. Soc. Rev. \textbf{44}, 2643 (2015).

\bibitem{Cappelluti2013}
E.~Cappelluti, R.~Rold\'an, J.~A. Silva-Guill\'en, P.~Ordej\'on, and F.~Guinea.
\newblock Phys. Rev. B \textbf{88}, 075409 (2013).

\bibitem{Sangalli2012}
D.~Sangalli, A.~Marini, and A.~Debernardi.
\newblock Phys. Rev. B \textbf{86}, 125139 (2012).

\bibitem{mostofi:cpc:178}
A.~A. Mostofi, J.~R. Yates, Y.~S. Lee, I.~Souza, D.~Vanderbilt, and N.~Marzari.
\newblock Comput. Phys. Commun. \textbf{178}, 685 (2008).

\bibitem{vaspwannier}
C.~Franchini, R.~Kov\'{a}\v{c}ik, M.~Marsman, S.~S. Murthy, J.~He, C.~Ederer,
  and G.~Kresse.
\newblock J. of Phys.: Cond. Mat. \textbf{24}, 235602 (2012).

\bibitem{Qiu2013}
D.~Y. Qiu, F.~H. da~Jornada, and S.~G. Louie.
\newblock Phys. Rev. Lett. \textbf{111}, 216805 (2013).

\bibitem{bronsema:zaac:540}
K.~D. Bronsema, J.~L.~D. Boer, and F.~Jellinek.
\newblock Z. anorg. allg. Chem. \textbf{540/541}, 15 (1986).

\bibitem{Jiang2012}
H.~Jiang.
\newblock J. Phys. Chem. C \textbf{116}, 7664 (2012).

\bibitem{Frey1998}
G.~L. Frey, S.~Elani, M.~Homyonfer, Y.~Feldman, and R.~Tenne.
\newblock Phys. Rev. B \textbf{57}, 6666 (1998).

\bibitem{Liang2013}
Y.~Liang, S.~Huang, R.~Soklaski, and L.~Yang.
\newblock Applied Physics Letters \textbf{103}, 042106 (2013).

\bibitem{Huser2013}
F.~H\"user, T.~Olsen, and K.~S. Thygesen.
\newblock Phys. Rev. B \textbf{88}, 245309 (2013).

\bibitem{Caruso2012}
F.~Caruso, P.~Rinke, X.~Ren, M.~Scheffler, and A.~Rubio.
\newblock Phys. Rev. B \textbf{86}, 081102 (2012).

\bibitem{Ellis2011}
J.~K. Ellis, M.~J. Lucero, and G.~E. Scuseria.
\newblock Applied Physics Letters \textbf{99}, 26, 261908 (2011).

\bibitem{Slater1954}
J.~C. Slater and G.~F. Koster.
\newblock Phys. Rev. \textbf{94}, 1498 (1954).

\bibitem{Enderlein1997}
R.~Enderlein and N.~J.~M. Horing.
\newblock \textit{Fundamentals of Semiconductor Physics and Devices}.
\newblock World Scientific Pub Co Inc (1997).

\bibitem{Zahid2013}
F.~Zahid, L.~Liu, Y.~Zhu, J.~Wang, and H.~Guo.
\newblock AIP Advances \textbf{3}, 5, 052111 (2013).

\bibitem{Neville1976}
R.~A. Neville and B.~L. Evans.
\newblock physica status solidi (b) \textbf{73}, 2, 597 (1976).

\bibitem{Coehoorn1987}
R.~Coehoorn, C.~Haas, and R.~A. de~Groot.
\newblock Phys. Rev. B \textbf{35}, 6203 (1987).

\bibitem{Coehoorn1987a}
R.~Coehoorn, C.~Haas, J.~Dijkstra, C.~J.~F. Flipse, R.~A. de~Groot, and
  A.~Wold.
\newblock Phys. Rev. B \textbf{35}, 6195 (1987).

\bibitem{Shi2013a}
H.~Shi, R.~Yan, S.~Bertolazzi, J.~Brivio, B.~Gao, A.~Kis, D.~Jena, H.~G. Xing,
  and L.~Huang.
\newblock ACS Nano \textbf{7}, 2, 1072 (2013).

\bibitem{Arnaud2006}
B.~Arnaud, S.~Leb\`egue, P.~Rabiller, and M.~Alouani.
\newblock Phys. Rev. Lett. \textbf{96}, 026402 (2006).

\bibitem{Wirtz_comment}
L.~Wirtz, A.~Marini, M.~Gr\"uning, C.~Attaccalite, G.~Kresse, and A.~Rubio.
\newblock Phys. Rev. Lett. \textbf{100}, 189701 (2008).

\bibitem{Komsa2013}
H.-P. Komsa and A.~V. Krasheninnikov.
\newblock Phys. Rev. B \textbf{88}, 085318 (2013).

\bibitem{Berghauser2014}
G.~Bergh\"auser and E.~Malic.
\newblock Phys. Rev. B \textbf{89}, 125309 (2014).

\bibitem{Onida2002}
G.~Onida, L.~Reining, and A.~Rubio.
\newblock Rev. Mod. Phys. \textbf{74}, 601 (2002).

\bibitem{Rohlfing2000}
M.~Rohlfing and S.~G. Louie.
\newblock Phys. Rev. B \textbf{62}, 4927 (2000).

\bibitem{Strinati1982}
G.~Strinati.
\newblock Phys. Rev. Lett. \textbf{49}, 1519 (1982).

\bibitem{Strinati1984}
G.~Strinati.
\newblock Phys. Rev. B \textbf{29}, 5718 (1984).

\bibitem{Marini2009}
A.~Marini, C.~Hogan, M.~Grüning, and D.~Varsano.
\newblock Computer Physics Communications \textbf{180}, 8, 1392  (2009).

\bibitem{Marini2008}
A.~Marini.
\newblock Phys. Rev. Lett. \textbf{101}, 106405 (2008).

\bibitem{Deslippe2012}
J.~Deslippe, G.~Samsonidze, D.~A. Strubbe, M.~Jain, M.~L. Cohen, and S.~G.
  Louie.
\newblock Computer Physics Communications \textbf{183}, 6, 1269  (2012).

\bibitem{Hueser2013}
F.~H\"user, T.~Olsen, and K.~S. Thygesen.
\newblock Phys. Rev. B \textbf{88}, 245309 (2013).

\bibitem{Rozzi2006}
C.~A. Rozzi, D.~Varsano, A.~Marini, E.~K.~U. Gross, and A.~Rubio.
\newblock Phys. Rev. B \textbf{73}, 205119 (2006).

\bibitem{Galambosi2011}
S.~Galambosi, L.~Wirtz, J.~A. Soininen, J.~Serrano, A.~Marini, K.~Watanabe,
  T.~Taniguchi, S.~Huotari, A.~Rubio, and K.~H\"am\"al\"ainen.
\newblock Phys. Rev. B \textbf{83}, 081413 (2011).

\bibitem{VanHove1953}
L.~Van~Hove.
\newblock Phys. Rev. \textbf{89}, 1189 (1953).

\bibitem{cardona}
P.~Y. Yu and M.~Cardona.
\newblock \textit{Fundamentals of {S}emiconductors}.
\newblock Springer (1999).

\bibitem{Riefer2011}
A.~Riefer, F.~Fuchs, C.~R\"odl, A.~Schleife, F.~Bechstedt, and R.~Goldhahn.
\newblock Phys. Rev. B \textbf{84}, 075218 (2011).

\bibitem{Klots2014}
A.~R. Klots, A.~K.~M. Newaz, B.~Wang, D.~Prasai, H.~Krzyzanowska, J.~Lin,
  D.~Caudel, N.~J. Ghimire, J.~Yan, B.~L. Ivanov, K.~A. Velizhanin, A.~Burger,
  D.~G. Mandrus, N.~H. Tolk, S.~T. Pantelides, and K.~I. Bolotin.
\newblock Sci. Rep. \textbf{4},  (2014).

\bibitem{Mertens2014}
J.~Mertens, Y.~Shi, A.~Molina-S\'{a}nchez, L.~Wirtz, H.~Y. Yang, and J.~J.
  Baumberg.
\newblock Applied Physics Letters \textbf{104}, 19, 191105 (2014).

\bibitem{Chernikov2014}
A.~Chernikov, T.~C. Berkelbach, H.~M. Hill, A.~Rigosi, Y.~Li, O.~B. Aslan,
  D.~R. Reichman, M.~S. Hybertsen, and T.~F. Heinz.
\newblock Phys. Rev. Lett. \textbf{113}, 076802 (2014).

\bibitem{Antoci1972}
S.~Antoci, P.~Camagni, A.~Manara, and A.~Stella.
\newblock Journal of Physics and Chemistry of Solids \textbf{33}, 6, 1177
  (1972).

\bibitem{Ramirez-Torres2014}
A.~Ramirez-Torres, V.~Turkowski, and T.~S. Rahman.
\newblock Phys. Rev. B \textbf{90}, 085419 (2014).

\bibitem{Yuan2014}
H.~Yuan, X.~Wang, B.~Lian, H.~Zhang, X.~Fang, B.~Shen, G.~Xu, Y.~Xu, S.-C.
  Zhang, H.~Y. Hwang, and Y.~Cui.
\newblock Nat Nano \textbf{advance online publication},  (2014).

\bibitem{Chiu2014}
M.-H. Chiu, M.-Y. Li, W.~Zhang, W.-T. Hsu, W.-H. Chang, M.~Terrones,
  H.~Terrones, and L.-J. Li.
\newblock ACS Nano \textbf{8}, 9, 9649 (2014).
\newblock PMID: 25196077.

\bibitem{Kozawa2014}
D.~Kozawa, R.~Kumar, A.~Carvalho, K.~Kumar~Amara, W.~Zhao, S.~Wang, M.~Toh,
  R.~M. Ribeiro, A.~H. Castro~Neto, K.~Matsuda, and G.~Eda.
\newblock Nat Commun \textbf{5},  (2014).

\bibitem{Koch2004}
H.~Koch, S. W. ;~Haug.
\newblock \textit{Quantum Theory of the Optical and Electronic Properties of
  Semiconductors}.
\newblock World Scientific (2004).

\bibitem{Ugeda2014}
M.~M. Ugeda, A.~J. Bradley, S.-F. Shi, F.~H. da~Jornada, Y.~Zhang, D.~Y. Qiu,
  W.~Ruan, S.-K. Mo, Z.~Hussain, Z.-X. Shen, F.~Wang, S.~G. Louie, and M.~F.
  Crommie.
\newblock Nat Mater \textbf{advance online publication},  (2014).

\bibitem{Mak2014}
K.~F. Mak, K.~L. McGill, J.~Park, and P.~L. McEuen.
\newblock Science \textbf{344}, 6191, 1489 (2014).

\bibitem{Mai2014}
C.~Mai, A.~Barrette, Y.~Yu, Y.~G. Semenov, K.~W. Kim, L.~Cao, and K.~Gundogdu.
\newblock Nano Letters \textbf{14}, 1, 202 (2014).

\bibitem{Jones2013}
A.~M. Jones, H.~Yu, N.~J. Ghimire, S.~Wu, G.~Aivazian, J.~S. Ross, B.~Zhao,
  J.~Yan, D.~G. Mandrus, D.~Xiao, W.~Yao, and X.~Xu.
\newblock Nat Nano \textbf{8}, 9, 634 (2013).

\bibitem{Wang2013}
Q.~Wang, S.~Ge, X.~Li, J.~Qiu, Y.~Ji, J.~Feng, and D.~Sun.
\newblock ACS Nano \textbf{7}, 12, 11087 (2013).

\bibitem{Cui2014}
Q.~Cui, F.~Ceballos, N.~Kumar, and H.~Zhao.
\newblock ACS Nano \textbf{8}, 3, 2970 (2014).

\bibitem{Lagarde2014}
D.~Lagarde, L.~Bouet, X.~Marie, C.~R. Zhu, B.~L. Liu, T.~Amand, P.~H. Tan, and
  B.~Urbaszek.
\newblock Phys. Rev. Lett. \textbf{112}, 047401 (2014).

\end{thebibliography}
\end{document}